\documentclass[twocolumn,tighten]{aastex631}
\usepackage{CJKutf8}
\usepackage{amsmath}
\usepackage{enumitem}
\usepackage{booktabs,pbox}
\usepackage{bm}
\usepackage{colortbl,multirow}

\graphicspath{{./}{./figures/}{../figures/}}

\newcommand{\ualphaCO}{\mbox{$\rm M_\odot~pc^{-2}\ (K~km~s^{-1})^{-1}$}}

\newcommand{\sdmol}{\ensuremath{\langle \Sigma_{\rm mol}^{\rm cloud} \rangle}}
\newcommand{\sigmol}{\ensuremath{\langle \sigma_{\rm mol}^{\rm cloud} \rangle}}
\newcommand{\tauff}{\ensuremath{\langle \tau_{\rm ff}^{\rm cloud} \rangle}}
\newcommand{\avir}{\ensuremath{\langle \alpha_{\rm vir}^{\rm cloud} \rangle}}

\newcommand{\taudep}{\ensuremath{\tau_{\rm dep}^{\rm mol}}}
\newcommand{\eff}{\ensuremath{\epsilon_{\rm ff}^{\rm mol}}}

\shorttitle{cloud-scale gas properties, \taudep , and \eff }
\shortauthors{Leroy, Sun, Meidt et al.}

\begin{document}

\title{Cloud-scale gas properties, depletion times, and star formation efficiency per free-fall time in PHANGS--ALMA}

\suppressAffiliations

\newcommand{\Princeton}{\affiliation{Department of Astrophysical Sciences, Princeton University, 4 Ivy Lane, Princeton, NJ 08544, USA}}

\newcommand{\McMaster}{\affiliation{Department of Physics and Astronomy, McMaster University, 1280 Main Street West, Hamilton, ON L8S 4M1, Canada}}

\newcommand{\CITA}{\affiliation{Canadian Institute for Theoretical Astrophysics (CITA), University of Toronto, 60 St George Street, Toronto, ON M5S 3H8, Canada}}

\newcommand{\OSU}{\affiliation{Department of Astronomy, The Ohio State University, 140 West 18th Avenue, Columbus, OH 43210, USA}}

\newcommand{\CCAPP}{\affiliation{Center for Cosmology and Astroparticle Physics (CCAPP), 191 West Woodruff Avenue, Columbus, OH 43210, USA}}

\newcommand{\Alberta}{\affiliation{Department of Physics, University of Alberta, Edmonton, AB T6G 2E1, Canada}}

\newcommand{\ANU}{\affiliation{Research School of Astronomy and Astrophysics, Australian National University, Canberra, ACT 2611, Australia}}

\newcommand{\Arizona}{\affiliation{Steward Observatory, University of Arizona, 933 North Cherry Avenue, Tucson, AZ 85721, USA}}

\newcommand{\ASIAA}{\affiliation{Institute of Astronomy and Astrophysics, Academia Sinica, No. 1, Sec. 4, Roosevelt Road, Taipei 106319, Taiwan}}

\newcommand{\ASTROThreeD}{\affiliation{ARC Centre of Excellence for All Sky Astrophysics in 3 Dimensions (ASTRO 3D), Australia}}

\newcommand{\Bonn}{\affiliation{Argelander-Institut f\"ur Astronomie, Universit\"at Bonn, Auf dem H\"ugel 71, 53121 Bonn, Germany}}

\newcommand{\Carnegie}{\affiliation{Observatories of the Carnegie Institution for Science, 813 Santa Barbara Street, Pasadena, CA 91101, USA}}

\newcommand{\CCA}{\affiliation{Center for Computational Astrophysics, Flatiron Institute, 162 Fifth Avenue, New York, NY 10010, USA}}

\newcommand{\CfA}{\affiliation{Center for Astrophysics $\mid$ Harvard \& Smithsonian, 60 Garden Street, Cambridge, MA 02138, USA}}

\newcommand{\CITEVA}{\affiliation{Centro de Astronomía (CITEVA), Universidad de Antofagasta, Avenida Angamos 601, Antofagasta, Chile}}

\newcommand{\CNRS}{\affiliation{CNRS, IRAP, 9 Av. du Colonel Roche, BP 44346, F-31028 Toulouse cedex 4, France}}

\newcommand{\Connecticut}{\affiliation{Department of Physics, University of Connecticut, 196A Auditorium Road, Storrs, CT 06269, USA}}

\newcommand{\COOL}{\affiliation{Cosmic Origins Of Life (COOL) Research DAO, coolresearch.io}}

\newcommand{\EPFL}{\affiliation{Institute of Physics, Laboratory for galaxy evolution and spectral modelling, EPFL, Observatoire de Sauverny, Chemin Pegais 51, 1290 Versoix, Switzerland}}

\newcommand{\ESO}{\affiliation{European Southern Observatory, Karl-Schwarzschild Stra{\ss}e 2, D-85748 Garching bei M\"{u}nchen, Germany}}

\newcommand{\Gent}{\affiliation{Sterrenkundig Observatorium, Universiteit Gent, Krijgslaan 281 S9, B-9000 Gent, Belgium}}

\newcommand{\Hawaii}{\affiliation{Institute for Astronomy, University of Hawaii, 2680 Woodlawn Drive, Honolulu, HI 96822, USA}}

\newcommand{\Heidelberg}{\affiliation{Astronomisches Rechen-Institut, Zentrum f\"{u}r Astronomie der Universit\"{a}t Heidelberg, M\"{o}nchhofstra\ss e 12-14, D-69120 Heidelberg, Germany}}

\newcommand{\IAC}{\affiliation{Instituto de Astrof\'isica de Canarias, C/ V\'ia L\'actea s/n, E-38205, La Laguna, Spain}}

\newcommand{\ICRAR}{\affiliation{International Centre for Radio Astronomy Research, University of Western Australia, 35 Stirling Highway, Crawley, WA 6009, Australia}}

\newcommand{\INAF}{\affiliation{INAF -- Osservatorio Astrofisico di Arcetri, Largo E. Fermi 5, I-50157, Firenze, Italy}}

\newcommand{\IPAC}{\affiliation{Caltech-IPAC, 1200 E. California Blvd. Pasadena, CA 91125, USA}}

\newcommand{\IPARC}{\affiliation{Instituto de F\'{\i}sica de Part\'{\i}culas y del Cosmos IPARCOS, Facultad de Ciencias F\'{\i}sicas, Universidad Complutense de Madrid, E-28040, Spain}}

\newcommand{\IRAM}{\affiliation{Institut de Radioastronomie Millim\'etrique (IRAM), 300 Rue de la Piscine, F-38406 Saint Martin d'H\`eres, France}}

\newcommand{\ITA}{\affiliation{Universit\"{a}t Heidelberg, Zentrum f\"{u}r Astronomie, Institut f\"{u}r Theoretische Astrophysik, Albert-Ueberle-Str 2, D-69120 Heidelberg, Germany}}

\newcommand{\IWR}{\affiliation{Universit\"{a}t Heidelberg, Interdisziplin\"{a}res Zentrum f\"{u}r Wissenschaftliches Rechnen, Im Neuenheimer Feld 205, D-69120 Heidelberg, Germany}}

\newcommand{\JHU}{\affiliation{Department of Physics and Astronomy, The Johns Hopkins University, Baltimore, MD 21218, USA}}

\newcommand{\Kansas}{\affiliation{Department of Physics and Astronomy, University of Kansas, 1251 Wescoe Hall Drive, Lawrence, KS 66045, USA}}

\newcommand{\LAM}{\affiliation{Aix Marseille Univ, CNRS, CNES, LAM (Laboratoire d’Astrophysique de Marseille), Marseille, France}}

\newcommand{\Leiden}{\affiliation{Leiden Observatory, Leiden University, P.O. Box 9513, 2300 RA Leiden, The Netherlands}}

\newcommand{\Liverpool}{\affiliation{Astrophysics Research Institute, Liverpool John Moores University, IC2, Liverpool Science Park, 146 Brownlow Hill, Liverpool L3 5RF, UK}}

\newcommand{\Lyon}{\affiliation{Univ Lyon, Univ Lyon 1, ENS de Lyon, CNRS, Centre de Recherche Astrophysique de Lyon UMR5574, F-69230 Saint-Genis-Laval, France}}

\newcommand{\Maryland}{\affiliation{Department of Astronomy and Joint Space-Science Institute, University of Maryland, 4296 Stadium Drive, College Park, MD 20742, USA}}

\newcommand{\MPE}{\affiliation{Max-Planck-Institut f\"{u}r extraterrestrische Physik, Giessenbachstra{\ss}e 1, D-85748 Garching, Germany}}

\newcommand{\MPIA}{\affiliation{Max-Planck-Institut f\"{u}r Astronomie, K\"{o}nigstuhl 17, D-69117, Heidelberg, Germany}}

\newcommand{\Nagoya}{\affiliation{Department of Physics, Nagoya University, Furo-cho, Chikusa-ku, Nagoya, Aichi 464-8602, Japan}}

\newcommand{\NAOJ}{\affil{National Astronomical Observatory of Japan, 2-21-1 Osawa, Mitaka, Tokyo, 181-8588, Japan}}

\newcommand{\Nichidai}{\affil{Department of Physics, General Studies, College of Engineering, Nihon University, 1 Nakagawara, Tokusada, Tamuramachi, Koriyama, Fukushima, 963-8642, Japan}}

\newcommand{\NRAO}{\affiliation{National Radio Astronomy Observatory, 520 Edgemont Road, Charlottesville, VA 22903, USA}}

\newcommand{\OAN}{\affiliation{Observatorio Astron\'{o}mico Nacional (IGN), C/Alfonso XII, 3, E-28014 Madrid, Spain}}

\newcommand{\ObsParis}{\affiliation{Sorbonne Universit\'{e}, Observatoire de Paris, Universit\'{e} PSL, CNRS, LERMA, F-75014, Paris, France}}

\newcommand{\Oxford}{\affiliation{Sub-department of Astrophysics, Department of Physics, University of Oxford, Keble Road, Oxford OX1 3RH, UK}}

\newcommand{\ROE}{\affiliation{Institute for Astronomy, University of Edinburgh, Royal Observatory, Blackford Hill, Edinburgh EH9 3HJ, UK}}

\newcommand{\Rutgers}{\affiliation{Department of Physics and Astronomy, Rutgers, the State University of New Jersey, 136 Frelinghuysen Road, Piscataway, NJ 08854, USA}}

\newcommand{\STScI}{\affiliation{Space Telescope Science Institute, 3700 San Martin Drive, Baltimore, MD 21218, USA}}

\newcommand{\STScIESA}{\affiliation{AURA for the European Space Agency (ESA), Space Telescope Science Institute, 3700 San Martin Drive, Baltimore, MD 21218, USA}}

\newcommand{\Surrey}{\affiliation{Department of Physics, University of Surrey, Guildford GU2 7XH, UK}}

\newcommand{\Sydney}{\affiliation{Sydney Institute for Astronomy, School of Physics A28, The University of Sydney, NSW 2006, Australia}}

\newcommand{\TAPIR}{\affil{TAPIR, California Institute of Technology, Pasadena, CA 91125, USA}}

\newcommand{\Tamkang}{\affiliation{Department of Physics, Tamkang University, No.151, Yingzhuan Rd., Tamsui Dist., New Taipei City 251301, Taiwan}}

\newcommand{\Toulouse}{\affiliation{Universit\'{e} de Toulouse, UPS-OMP, IRAP, F-31028 Toulouse cedex 4, France}}

\newcommand{\Toledo}{\affiliation{University of Toledo, 2801 W. Bancroft St., Mail Stop 111, Toledo, OH 43606, USA}}

\newcommand{\UChile}{\affiliation{Departamento de Astronom\'{i}a, Universidad de Chile, Camino del Observatorio 1515, Las Condes, Santiago, Chile}}

\newcommand{\UCM}{\affiliation{Departamento de F\'{\i}sica de la Tierra y Astrof\'{\i}sica, Universidad Complutense de Madrid, E-28040, Spain}}

\newcommand{\UCSD}{\affiliation{Center for Astrophysics and Space Sciences, Department of Physics,  University of California, San Diego, 9500 Gilman Drive, La Jolla, CA 92093, USA}}

\newcommand{\ULL}{\affiliation{Departamento de Astrof\'isica, Universidad de La Laguna, Av. del Astrof\'isico Francisco S\'anchez s/n, E-38206, La Laguna, Spain}}

\newcommand{\UMass}{\affiliation{University of Massachusetts—Amherst, 710 North Pleasant Street, Amherst, MA 01003, USA}}

\newcommand{\UVa}{\affiliation{University of Virginia, 530 McCormick Road, Charlottesville, VA 22904, USA}}

\newcommand{\Wyoming}{\affiliation{Department of Physics and Astronomy, University of Wyoming, Laramie, WY 82071, USA}}

\newcommand{\Zurich}{\affiliation{Department of Astrophysics, University of Z\"urich, Winterthurerstrasse 190, 8057 Z\"urich, Switzerland}}

\newcommand{\Lund}{\affiliation{Lund Observatory, Division of Astrophysics, Department of Physics, Lund University, Box 43, SE-221 00 Lund, Sweden}}

\newcommand{\Rad}{\affiliation{Radcliffe Institute for Advanced Studies at Harvard University, 10 Garden Street, Cambridge, MA 02138, U.S.A.}}

\newcommand{\Shizuoka}{\affil{Faculty of Global Interdisciplinary Science and Innovation, Shizuoka University, 836 Ohya, Suruga-ku, Shizuoka 422-8529, Japan}}

\newcommand{\Michigan}{\affiliation{Department of Astronomy, University of Michigan, Ann Arbor, MI 48109, USA}}


\author[0000-0002-2545-1700]{Adam~K.~Leroy}
\OSU
\CCAPP

\author[0000-0003-0378-4667]{Jiayi~Sun \begin{CJK*}{UTF8}{gbsn}(孙嘉懿)\end{CJK*}}
\altaffiliation{NASA Hubble Fellow}
\Princeton

\author[0000-0002-6118-4048]{Sharon~Meidt}
\Gent

\author[0000-0002-4287-1088]{Oscar Agertz}
\Lund

\author[0000-0003-2551-7148]{I-Da Chiang \begin{CJK*}{UTF8}{bkai}(江宜達)\end{CJK*}}\ASIAA

\author[0000-0001-6119-9883]{Jindra Gensior}\ROE\Zurich

\author[0000-0001-6708-1317]{Simon C.~O.\ Glover}
\ITA

\author[0000-0001-9852-9954]{Oleg Y. Gnedin}
\Michigan

\author[0000-0002-9181-1161]{Annie~Hughes}
\CNRS

\author[0000-0002-3933-7677]{Eva~Schinnerer}
\MPIA

\author[0000-0003-0410-4504]{Ashley~T.~Barnes} 
\ESO

\author[0000-0003-0166-9745]{Frank Bigiel}
\Bonn

\author[0000-0002-5480-5686]{Alberto D. Bolatto}
\Maryland

\author[0000-0001-6498-2945]{Dario Colombo}
\Bonn

\author[0000-0002-8760-6157]{Jakob den Brok}
\CfA

\author[0000-0002-5635-5180]{M\'elanie~Chevance}
\ITA
\COOL

\author[0000-0001-8241-7704]{Ryan~Chown}
\OSU

\author[0000-0002-1185-2810]{Cosima Eibensteiner}
\altaffiliation{Jansky Fellow of the National Radio Astronomy Observatory}
\NRAO

\author[0009-0001-8660-9962]{Damian R. Gleis}
\MPIA

\author[0000-0002-3247-5321]{Kathryn~Grasha}
\altaffiliation{ARC DECRA Fellow}
\ANU

\author[0000-0001-9656-7682]{Jonathan D. Henshaw} 
\Liverpool
\MPIA

\author[0000-0002-0560-3172]{Ralf S.\ Klessen}\ITA\IWR\CfA\Rad

\author[0000-0001-9605-780X]{Eric W. Koch}
\CfA

\author[0000-0002-0119-1115]{Elias~K.~Oakes}
\Connecticut

\author[0000-0002-1370-6964]{Hsi-An Pan}
\Tamkang

\author[0000-0002-0472-1011]{Miguel Querejeta}
\OAN

\author[0000-0002-5204-2259]{Erik Rosolowsky}
\Alberta

\author[0000-0002-2501-9328]{Toshiki Saito}\Shizuoka

\author[0000-0002-4378-8534]{Karin~Sandstrom}
\UCSD

\author[0000-0002-4781-7291]{Sumit K. Sarbadhicary}
\OSU
\CCAPP
\affiliation{Department of Physics, The Ohio State University, Columbus, Ohio 43210, USA}

\author[0000-0003-4209-1599]{Yu-Hsuan Teng}
\Maryland

\author[0000-0003-1242-505X]{Antonio Usero}
\OAN

\author{Dyas Utomo}
\OSU
\NRAO

\author[0000-0002-0012-2142]{Thomas G. Williams}
\Oxford

\correspondingauthor{Adam K. Leroy}
\email{leroy.42@osu.edu}

\begin{abstract}
We compare measurements of star formation efficiency to cloud-scale gas properties across the PHANGS--ALMA sample. Dividing $67$ galaxies into $1.5$~kpc scale regions, we calculate the molecular gas depletion time $\taudep = \Sigma_{\rm mol} / \Sigma_{\rm SFR}$ and the star formation efficiency per free-fall time $\eff = \tau_{\rm ff} / \taudep$ for each region. Then we test how \taudep\ and \eff\ vary as functions of the regional mass-weighted mean molecular gas properties on cloud scales (60--150~pc): gas surface density, \sdmol , velocity dispersion, \sigmol , virial parameter, \avir , and gravitational free-fall time, \tauff . \tauff\ and \taudep\ correlate positively, consistent with the expectation that gas density plays a key role in setting the rate of star formation. Our fiducial measurements suggest $\taudep \propto \tauff^{0.5}$ and $\eff \approx 0.39\%$, though the exact numbers depend on the adopted fitting methods. We also observe anti-correlations between \taudep\ and \sdmol\ and between \taudep\ and \sigmol . All three correlations may reflect the same underlying link between density and star formation efficiency combined with systematic variations in the degree to which self-gravity binds molecular gas in galaxies. We highlight the \taudep--\sigmol\ relation because of the lower degree of correlation between the axes. Contrary to theoretical expectations, we observe a weak anti-correlation between \taudep\ and \avir\ and no significant correlation between \eff\ and \avir . Our results depend sensitively on the adopted CO-to-H$_2$ conversion factor, with corrections for excitation and emissivity effects in inner galaxies playing an important role. We emphasize that our simple methodology and clean selection allow easy comparison to numerical simulations and highlight this as a logical next direction.
\end{abstract}

\section{Introduction}
\label{sec:intro}

Stars form from clouds of molecular gas, with star formation and molecular gas well-correlated across nearby galaxies \citep[e.g.,][]{WONG02SFGAS,LEROY08SFGAS,BIGIEL08SFGAS,SCHRUBA11SFGAS,KENNICUTT12REVIEW}. Thanks to new surveys by ALMA, we now know that the ${\approx} 100$~pc ``cloud-scale'' properties of molecular gas --- including surface density, velocity dispersion, and dynamical state --- vary systematically, responding to and reflecting their galactic environment \citep[e.g.,][]{SUN18CLOUDS,SCHRUBA19GMCS,SUN22CLOUDS,SCHINNERER24REVIEW}. So far, observations only weakly constrain how these variations in cloud-scale gas properties affect the process of star formation. In other words, a major open question in the field of star formation remains: ``How do the properties of the parent molecular clouds affect the efficiency and rate of star formation?'' 

The initial properties of a cloud of gas \textit{should} affect the rate at which stars form. In a given environment and at fixed mass, denser clouds will be more tightly bound by self-gravity. A cloud with a higher mean density, $\rho^{\rm cloud}_{\rm mol}$, will also have a shorter corresponding cloud-scale gravitational free-fall time, $\tau_{\rm ff}^{\rm cloud} = \sqrt{3 \pi / (32 G \rho_{\rm mol}^{\rm cloud})}$ and might be expected to collapse to form stars faster, all other things being equal. For motions dominated by supersonic, isothermal turbulence, the cloud-scale velocity dispersion, which reflects the Mach number via $\mathcal{M} \propto \sigma_{\rm mol}^{\rm cloud}$, relates to the width of the density distribution \citep[e.g.,][]{VAZQUEZ94TURB,PADOAN02TURB}. Meanwhile, the density distribution sets the fraction of high density, gravitationally bound gas \citep[e.g.,][]{KRUMHOLZ05EFF}. The densest molecular gas is observed to be the immediate site for star formation and to correlate with tracers of star formation in the Milky Way and other galaxies \citep[][]{GAO04DENSE,LADA10DENSE,GARCIABURILLO12DENSE,EVANS14DENSE,JIMENEZ19DENSE}. The balance between kinetic and gravitational energy, often expressed as the virial parameter, $\alpha_{\rm vir} \equiv 2 K / U$, will also affect the ability of gas to collapse and form stars. In both analytic theory and numerical simulations, $\alpha_{\rm vir}$ plays a main role in setting $\epsilon_{\rm ff}$ when all other factors are equal \citep[][]{KRUMHOLZ05EFF,PADOAN12EFF,FEDERRATH13EFF,KIM21FEEDBACKSIMS}, though models in which star formation is dynamically triggered \citep[e.g.,][]{FUKUI21COLLIDE} or in which clouds exist in a state of approximate free-fall \citep[e.g.,][]{VAZQUEZ24EFF} also remain viable.

Observationally, this picture is less clear. Despite decades of studies of molecular cloud properties, there are relatively few quantitative, large-scale comparisons between the efficiency of star formation and the cloud-scale properties of molecular gas. The studies that do exist mainly concentrate on a single region or galaxy, and often focus on a few specific prototype spiral galaxies \citep[e.g., M51, M83, IC 342;][]{MEIDT13SFGAS,LEROY17SFGAS,HIROTA18GMCS,QUEREJETA23SFGAS} with \citet{SCHRUBA19GMCS} and \citet{SUN22CLOUDS} almost the only two studies that compare cloud properties, $\taudep \equiv M_{\rm mol}/{\rm SFR}$, and $\epsilon_{\rm ff} \equiv \tau_{\rm ff} / \taudep$ across significant samples of galaxies and environments. Overall, these works find mixed or weak evidence for any general correlation between $\epsilon_{\rm ff}$ and $\alpha_{\rm vir}$ \citep[][]{LEROY17SFGAS,SCHRUBA19GMCS}. Even the relationship of \taudep\ to cloud-scale mean density, expressed as $\tau_{\rm ff}$, often appears weaker than the theoretically expected $\taudep \propto \tau_{\rm ff}$ or even non-existent \citep[][]{LEROY17SFGAS,SUN23SFGAS,QUEREJETA23SFGAS}. 

Milky Way studies of local clouds demonstrate a clear connection between the distribution of gas density in a cloud and star formation activity \citep[][]{KAINULAINEN09DENSE,LADA10DENSE,LADA12DENSE,EVANS14DENSE}. However, most of these studies target a narrow range of environments with limited statistical power, partially because of the difficulty of accessing the spatially or time-averaged efficiency of star formation from our position within the Milky Way. In one of the most general studies to date, \citet{EVANS21GMCS} consider clouds across the whole Milky Way disk and do find evidence of a correlation between $\alpha_{\rm vir}$ and $\epsilon_{\rm ff}$. They also show that this conclusion depends on methodology, particularly the approach used to estimate molecular gas mass.

In this paper, we attempt to take a next logical step addressing this question by comparing star formation activity to cloud-scale gas properties using the largest homogeneous cloud-scale data set on molecular gas in galaxies, PHANGS--ALMA \citep{PHANGSALMA21SURVEY}. Specifically, we compare the molecular gas depletion time ($\taudep \equiv M_{\rm mol}/{\rm SFR}$), and the star formation efficiency per free fall time ($\epsilon_{\rm ff} \equiv \tau_{\rm ff}/\taudep$)  to the mass-weighted mean molecular gas properties on cloud scales. This paper builds on previous work using PHANGS--ALMA and acts as a direct follow up to \citet{SUN22CLOUDS}, but represents the first systematic attempt to correlate the cloud-scale gas properties with $\taudep$ and $\epsilon_{\rm ff}$ across the whole data set.

We adopt the cross-scale methodology developed in \citet{LEROY16GMCS}, \citet{LEROY17SFGAS}, \citet{SCHRUBA19GMCS}, \citet{SUN20PRESS} and \citet{SUN22CLOUDS}. This approach, described in \S \ref{sec:methods}, breaks galaxies into $\sim$kpc sized regions and summarizes the smaller-scale molecular gas properties via mass-weighted averages. These properties include mass surface density (\sdmol ), velocity dispersion (\sigmol ), and the dynamical state of the gas expressed via the virial parameter (\avir ). We compare these gas properties to the regional mean molecular gas depletion time ($\taudep$) and star formation efficiency per gravitational free fall time ($\eff$). By working with these kpc-sized regions, we average out the time evolution of individual star-forming regions, which causes systematic changes in their apparent $\taudep$ and $\eff$ \citep[see][]{SCHRUBA10SFGAS,LEROY13SFGAS,SEMENOV18GMCS,SEMENOV21GMCS,KRUIJSSEN19TIMES,CHEVANCE20TIMES,KIM22TIMES}. In practice, estimating $\Sigma_{\rm SFR}$ at this resolution also allows us to reduce stochasticity and utilize infrared (IR) data with limited $\approx 15''$ angular resolution. This approach should be simple to reproduce in numerical simulations and easy to apply to other data sets.

We present our core results in \S \ref{sec:results}. We find that the estimated gravitational free-fall time and molecular gas depletion time show the expected correlation, but the details depend sensitively on the adopted CO-to-H$_2$ conversion factor. This same relationship manifests as anti-correlations between \taudep\ and the cloud-scale surface density and velocity dispersion. On the other hand, our results comparing the virial parameter to \taudep\ and the star formation efficiency per free fall time seem to suggest that star formation becomes more efficient when the virial parameter is higher. This is opposite to theoretical expectations for turbulent gas clouds, which predict more efficient star formation for lower virial parameter.

In discussing the results, we suggest a path forward on this topic. We highlight rigorous comparison between observations and numerical simulations as an immediately possible, highly promising next step. In this regard, we emphasize that our measurements are easy to reproduce in simulations of galaxies, and we give a concrete checklist to do so in Appendix \ref{sec:stepbystep}. On intermediate timescales, we also discuss the need for next-generation data sets that cover diverse environments with much higher physical resolution, high completeness, and reliable cloud-scale indicators of star formation activity.

\section{Data and methods}
\label{sec:methods}

\begin{deluxetable*}{cl}[t!]
\tabletypesize{\small}
\tablecaption{Experiment Summary \label{tab:experiment}}
\tablewidth{0pt}
\tablehead{
}
\startdata
\hline
\multicolumn{2}{c}{\textbf{Mass-weighted region-averaged cloud-scale gas properties}} \\
\multicolumn{2}{c}{(mass-weighted averages estimated in each 
$1.5$~kpc diameter region based on CO~(2-1) emission)} \\
\hline
$\left< \Sigma_{\rm mol}^{\rm cloud} \right>$ & Cloud-scale molecular gas mass surface density in M$_\odot$~pc$^{-2}$ . \\
& Estimated from CO~(2-1) line intensity using $\alpha_{\rm CO}^{2-1}$ from \citet{SCHINNERER24REVIEW}. \\
& Corrected for completeness and inclination following \citet{SUN22CLOUDS}. \\
$\left< \tau_{\rm ff}^{\rm cloud} \right>$ & Gravitational free-fall time in yr. \\
& Calculated via $\tau_{\rm ff} = \sqrt{3 \pi / (32 G \rho_{\rm mol}^{\rm cloud})}$ and $\rho = 3 M_{\rm mol}^{\rm beam} / (4 \pi R_{\rm pix}^3$). \\
& $R_{\rm pix} = {\rm min}\!\left[ \ell / 2, \left( \ell^2 H_\mathrm{mol} / (8 \cos i) \right)^{1/3} \right] $ following \citet{SUN22CLOUDS}. \\
$\left< \sigma_{\rm mol}^{\rm cloud} \right>$ & Cloud-scale velocity dispersion in km~s$^{-1}$. \\
& Calculated via the CO~(2-1) line effective width and corrected for inclination and completeness. \\
$\left< \alpha_{\rm vir}^{\rm cloud} \right>$ & Dimensionless virial parameter $\alpha_{\rm vir} = 2~K/U$ considering only gas self-gravity. \\
& Calculated via $\alpha_{\rm vir}^{\rm cloud} = 5 \sigma^2_{\rm mol} R_{\rm pix} / (f G M_{\rm mol}^{\rm beam})$. \\
\hline
\multicolumn{2}{c}{\textbf{Region-averaged quantities related to star formation efficiency}} \\
\multicolumn{2}{c}{(calculated by averaging all emission in each large 
$1.5$~kpc diameter region)} \\
\hline
$\Sigma_{\rm mol}^{\rm kpc}$ & Mean molecular gas surface density over each 1.5~kpc region in M$_\odot$~pc$^{-2}$. \\
$\Sigma_{\rm SFR}^{\rm kpc}$ & Mean star formation rate surface density over each 1.5~kpc region in M$_\odot$~yr$^{-1}$~kpc$^{-2}$. \\
$\tau_{\rm dep}^{\rm mol}$ & Molecular gas depletion time in yr, defined as $\tau_{\rm dep}^{\rm mol} = \Sigma_{\rm mol}^{\rm kpc} / \Sigma_{\rm SFR}^{\rm kpc}$. \\
$\epsilon_{\rm ff}^{\rm mol}$ & Dimensionless star formation efficiency per free fall time, defined as $\epsilon_{\rm ff}^{\rm mol} = \left< \tau_{\rm ff}^{\rm cloud} \right> / \tau_{\rm dep}^{\rm mol}$
\tablenotemark{a}. \\
\hline
\multicolumn{2}{c}{\textbf{CO-to-H$_2$ conversion factor}} \\
\multicolumn{2}{c}{(following \citealt{SCHINNERER24REVIEW} and based on references given in text)} \\
\hline
$\alpha_{\rm CO}^{2-1}$ & Calculated via $\alpha_{\rm CO,MW}^{1-0} \times f(Z) \times g(\Sigma_\star) \times R_{21}(\Sigma_{\rm SFR})^{-1}$ \\
$\alpha_{\rm CO,MW}^{1-0}$ & $=4.35\;\ualphaCO$; Milky Way CO~(1-0) conversion factor  \citep{BOLATTO13REVIEW} \\
$f (Z)$ & $=\left(Z / Z_\odot \right)^{-1.5}$ ; metallicity-dependent ``CO-dark'' term \\
$g (\Sigma_\star)$ & $=\left( {\rm max}\!\left( \Sigma_\star , 100~{\rm M}_\odot~{\rm pc}^{-2} \right)\;/\;100~{\rm M}_\odot~{\rm pc}^{-2} \right)^{-0.25}$; ``starburst'' emissivity term following \citet{CHIANG24XCO} \\
$R_{21} (\Sigma_{\rm SFR})$ & $=0.65 \left( \Sigma_{\rm SFR}\;/\;0.018~{\rm M}_\odot~{\rm yr}^{-1}~{\rm kpc}^{-2} \right)^{0.125}$ with min 0.35, max 1.0; CO line excitation term \\
\hline
\multicolumn{2}{c}{\textbf{Data selection}} \\
\multicolumn{2}{c}{(requiring cloud-scale CO flux completeness $f_{\rm comp} > 0.5$, and $\sdmol > 20$~M$_\odot$~pc$^{-2}$)} \\
\multicolumn{2}{c}{At $\ell = 150$~pc, this selection scheme yields 841 independent, 1.5~kpc hexagonal regions in 67 galaxies} 
\enddata
\tablenotetext{a}{\eff\ combines region-averaged properties and mass-weighted cloud-scale properties. See \S \ref{sec:effvsprops} for discussion.}
\tablecomments{Calculations and database described in detail in \citet{SUN22CLOUDS} and \citet{SUN23SFGAS}. Conversion factor treatment described in \citet{SCHINNERER24REVIEW}. $i$ refers to galaxy inclination, $\ell$ to cloud scale data resolution (beam FWHM), $H_\mathrm{mol}$ to the assumed FWHM vertical thickness of the molecular layer, $M_{\rm mol}^{\rm beam} = \Sigma_{\rm mol}^{\rm cloud} A_{\rm beam}$ with $A_{\rm beam} = \pi \theta^2 / (4 \ln 2)$.}
\end{deluxetable*}

We measure how star formation efficiency correlates with diagnostics that trace the physical state of molecular gas across local galaxies. To do this, we compare cloud-scale ($\ell=150$~pc FWHM resolution) gas properties to the mean molecular gas depletion time over $1.5$~kpc diameter hexagonal regions, which collectively cover the area mapped by the PHANGS--ALMA survey \citep{PHANGSALMA21SURVEY}. PHANGS--ALMA targeted nearby ($d \lesssim 20$~Mpc), low to moderately inclined ($i \lesssim 75^\circ$), relatively massive ($\log_{10} M_\star / {\rm M}_\odot \gtrsim 9.75$, star-forming ($\log_{10} M_\star / {\rm SFR} > -11$~yr$^{-1}$) galaxies and offers a representative view of massive galaxies on or near the star-forming main sequence.

We conduct our analysis using the high-level data products (``the PHANGS mega-tables'') assembled by \citet{SUN22CLOUDS,SUN23SFGAS}. They break galaxies into individual sub-galactic regions and then measure a rich set of galaxy structural, stellar, and ISM properties in each region. Here we consider their main data product, which breaks 80 massive, star-forming disk galaxies from PHANGS--ALMA into individual $1.5$~kpc hexagonal regions. Within each region, the ``mega-tables'' record star formation activity, the total molecular gas mass, and the mass-weighted mean averages of cloud-scale gas properties. We summarize the quantities relevant to our analysis below and in Table \ref{tab:experiment} and refer to \citet{SUN22CLOUDS} and \citet{SUN23SFGAS} for detailed descriptions of how all other quantities are calculated. Appendix \ref{sec:stepbystep} gives a step-by-step guide to reproduce our measurements, e.g., to compare numerical simulations to PHANGS--ALMA.

The core data sets aggregated by \citet{SUN22CLOUDS} that contribute to our measurements are: the PHANGS--ALMA CO~(2-1) maps \citep{PHANGSALMA21SURVEY}; ground-based narrowband H$\alpha$ maps \citep[A.~Razza et al.\ in preparation, see][]{SCHINNERER19TIMES,PAN22TIMES,QUEREJETA24SFGAS}; GALEX \citep{GALEX05} ultraviolet images compiled by \citet{LEROY19Z0MGS}; WISE \citep{WISE10} 3.4--22$\mu$m images compiled by \citet{LEROY19Z0MGS}; and IRAC $3.6\mu$m images \citep{S4GSURVEY,QUEREJETA21}. We adopt distances and orientation parameters for all PHANGS--ALMA targets as measured and compiled by \citet{LANG20KINEMATICS,ANAND21DISTANCES,PHANGSALMA21SURVEY}.

\subsection{Cloud-scale gas properties}
\label{sec:gasprops}

\begin{figure}[t!]
\centering
\includegraphics[width=\columnwidth]{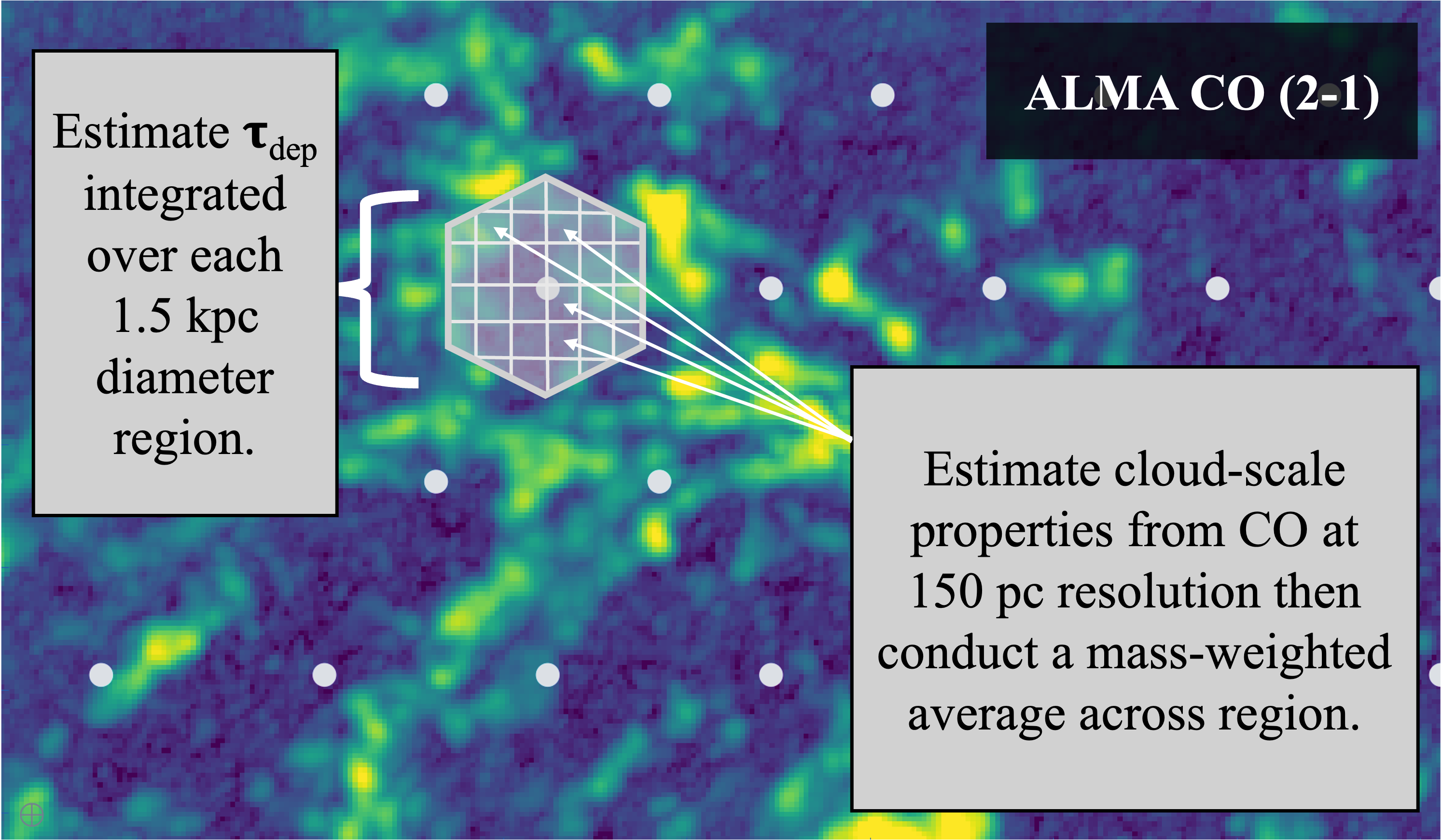}
\caption{Illustration of our experiment over a region in a CO~(2-1) peak intensity map from PHANGS--ALMA. We cover each galaxy with $1.5$~kpc diameter hexes. Within each region we measure the average $\tau_{\rm dep}^{\rm mol}$. We also measure the surface density, line width, gravitational free-fall time, and virial parameter at higher $150$~pc resolution. Then within each region, we weight these by mass and calculate the mass-weighted average $150$~pc value for each property, which we write $\langle X^{\rm cloud} \rangle$. See Table \ref{tab:experiment} and \citet{SUN22CLOUDS,SUN23SFGAS}.
\label{fig:experiment}
}
\end{figure}

To assess the physical state of molecular gas, we use cloud-scale (physical beam size $\ell = 150$~pc) imaging\footnote{Following, e.g., \citet{SCHINNERER24REVIEW} we refer to observations that match the beam to the size of a massive molecular cloud as ``cloud-scale'' imaging, in contrast to ``resolved cloud'' imaging or ``resolved galaxy'' imaging.} of the CO~(2-1) line, which traces the bulk of the molecular gas. Within each region of each galaxy, we measure four gas properties for each individual line of sight where CO is detected. Then, we weight each line of sight by the local gas surface density to derive mass-weighted, region-averaged molecular gas properties. This mass-weighting scheme, illustrated in Fig.~\ref{fig:experiment}, is intended to yield estimates of the physical conditions that set the star formation rate per unit gas, i.e., \taudep , in the region. It has been described in detail by \citet{LEROY16GMCS,LEROY17SFGAS,UTOMO18EFF} and \citet{SUN22CLOUDS}. We refer to quantities computed via this cross-scale averaging as $\left< X^{\rm cloud} \right>$, indicating original measurements on cloud scales followed by weighted averaging over the 1.5~kpc hexagonal regions.

We focus on four gas properties, summarized in Table \ref{tab:experiment}, that should be relevant to the efficiency of star formation:

\begin{enumerate}
\item \textit{Cloud-scale molecular gas surface density ($\Sigma_{\rm mol}^{\rm cloud}$).} The molecular gas mass per unit area relates to the density of the gas, its gravitational potential, and its shielding from external radiation.

\item \textit{Gravitational free-fall time ($\tau_{\rm ff}^{\rm cloud}$).} Based on $\Sigma_{\rm mol}^{\rm cloud}$ and an estimated line-of-sight depth for the gas (see discussion of $R_{\rm pix}$ below), \citet{SUN22CLOUDS} estimate the mass volume density, $\rho_{\rm mol}^{\rm cloud}$, and the corresponding gravitational free fall time $\tau_{\rm ff}^{\rm cloud}$. Note that the region averaging for $\tau_{\rm ff}^{\rm cloud}$ follows a weighting scheme that ensures $\taudep \propto \left< \tau_{\rm ff}^{\rm cloud} \right>$ for a fixed $\epsilon_{\rm ff}^{\rm mol}$ \citep[see][]{SUN22CLOUDS}.

\item \textit{Cloud-scale molecular gas velocity dispersion ($\sigma_{\rm mol}^{\rm cloud}$).} Within individual clouds, turbulence is expected to make a significant contribution to the CO line width \citep[although at these scales, we may also worry about blending of multiple distinct components, bulk motions due to the galactic potential, or collapse of clouds, e.g.,][]{IBANEZMEJIA16GMCS,MEIDT18GMCS,HENSHAW20SCOUSE}. In cases where the line width is dominated by supersonic, isothermal turbulence, the gas velocity dispersion (reflected by the line width) should be proportional to the Mach number of the turbulence \citep[e.g., see][]{MACLOW04REVIEW}. This, in turn, relates to the expected width of the gas density distribution \citep[e.g.,][]{PADOAN02TURB}. All other things being equal, a broader density distribution should lead to more dense, potentially star-forming gas.

\item \textit{Virial parameter ($\alpha_{\rm vir}^{\rm cloud}$).}  The virial parameter $\alpha_{\rm vir} = 2 K / U \propto \sigma_{\rm mol}^2 / (\Sigma_{\rm mol}~R_{\rm cloud})$ expresses the ratio of kinetic energy to potential energy associated with self-gravity in the gas \citep[see][though see discussion for alternate interpretations in \citealt{BALLESTEROS06GMCS}]{MCKEE92GMCS,BERTOLDI92GMCS}. Gas with a lower virial parameter should be prone to collapse and therefore more susceptible to star formation. Many theories of turbulence-regulated star formation predict an anti-correlation between $\alpha_{\rm vir}^{\rm cloud}$ and $\epsilon_{\rm ff}^{\rm mol}$ \citep[e.g.,][]{PADOAN11EFF,FEDERRATH12EFF}, and a similar relationship is also seen in simulations \citep{PADOAN12EFF,KIM21FEEDBACKSIMS}. As with $\tau_{\rm ff}^{\rm cloud}$, $\alpha_{\rm vir}^{\rm cloud}$ requires adopting a relevant size scale, $R_{\rm pix}$.

\end{enumerate}

\noindent For both $\tau_{\rm ff}^{\rm cloud}$ and $\alpha_{\rm vir}^{\rm cloud}$ we require an adopted size scale (see Table \ref{tab:experiment}). Following \citet{SUN22CLOUDS}, we take the relevant scale to be $R_{\rm pix} \equiv (\ell^2 H_{\rm mol} / (8 \cos i))^{1/3}$ at the fiducial $\ell=150$~pc resolution, with $H_{\rm mol}=100$~pc. This assumes that the mass measured in each beam is spread over the entire beam and over a line-of-sight depth of $H_\mathrm{mol}/\cos{i}$, with $H_\mathrm{mol} = 100$~pc the adopted molecular gas disk scale height \citep[e.g.,][]{HEYER15REVIEW}. In a similar analysis targeting M51, \citet{LEROY17SFGAS} explored the difference between fixed and dynamical estimates of $H$ and found qualitatively consistent results between the two approaches. Moreover, \citet{SUN22CLOUDS} showed a good overall match between results for an object oriented analysis, where size is measured from morphological analysis, and this sightline based approach. For more information on $R_{\rm pix}$, as well as the definition, calculation, and physical meaning of these and other quantities, see discussions in \citet{SUN20GMCS}, \citet{SUN22CLOUDS}, \citet{ROSOLOWSKY21GMCS}, and \citet{SCHINNERER24REVIEW}.

The main difference here compared to \citet{SUN22CLOUDS} or \citet{ROSOLOWSKY21GMCS} is that we adopt the new CO-to-H$_2$ conversion factor $\alpha_{\rm CO}^{2-1}$ recommended by \citet[][]{SCHINNERER24REVIEW}\footnote{Our $\alpha_{\rm CO}$ estimates are calculated region-by-region, so when we conduct region-averages we assume a fixed $\alpha_{\rm CO}$ within each region.}. This prescription, which we reproduce in Table \ref{tab:experiment}, attempts to account for excitation variations, emissivity variations, and metallicity effects based on synthesizing work by \citet{BOLATTO13REVIEW,ACCURSO17XCO,GONG20XCO,HU22XCO,SUN20PRESS,SUN22CLOUDS,DENBROK21LINES,LEROY22LINES,TENG23XCO,CHIANG24XCO}. In practice, it resembles a refined version of the $\alpha_{\rm CO}$ recommended by \citet{BOLATTO13REVIEW}. To assess the impact of this choice, we also run our analysis using a fixed Galactic $\alpha_{\rm CO}$. We discuss that the choice of $\alpha_{\rm CO}$ exerts a large impact on our results in \S \ref{sec:alphaco}.

\subsection{Star formation rate per unit gas}
\label{sec:sfe}

We compare region-averaged cloud-scale gas properties to two measures of the efficiency with which molecular gas forms stars. The molecular gas depletion time,

\begin{equation}
\label{eq:tdep}
\taudep = \frac{\Sigma_{\rm mol}^{\rm kpc}}{\Sigma_{\rm SFR}^{\rm kpc}} ,
\end{equation}

\noindent expresses the time needed for current star formation to consume the molecular gas reservoir without replenishment. Here $\Sigma_{\rm mol}^{\rm kpc}$ and $\Sigma_{\rm SFR}^{\rm kpc}$ are the mean molecular gas and star formation rate surface densities over the whole large 
$1.5$-kpc averaging aperture.\footnote{Specifically, as in \citet{SUN23SFGAS}, we convolve both $\Sigma_{\rm mol}$ and $\Sigma_{\rm SFR}$ to share a matched Gaussian kernel with FWHM equal to the hex diameter and then sample the value of those convolved maps at the hex center.} Although \taudep\ is formally a timescale, the physical meaning here is that it expresses the star formation rate per unit molecular gas mass. That is, $(\tau_{\rm dep}^{\rm mol})^{-1}$ is the ``specific'' star formation rate of molecular gas, so that a short $\tau_{\rm dep}^{\rm mol}$ indicates a high rate of star formation per unit molecular gas.

We also compute the star formation efficiency per free fall time,

\begin{equation}
\label{eq:eff}
\epsilon_{\rm ff}^{\rm mol} = \frac{\left< \tau_{\rm ff}^{\rm cloud} \right>}{\taudep} ,
\end{equation}

\noindent which contrasts $\tau_{\rm dep}^{\rm mol}$ with the gravitational free-fall time, $\tau_{\rm ff}^{\rm cloud}$, estimated from the cloud-scale CO imaging. $\epsilon_{\rm ff}^{\rm mol}$ expresses the efficiency of star formation relative to uninhibited gravitational collapse. This is viewed as a theoretically important quantity \citep[e.g.,][]{MCKEE07REVIEW,FEDERRATH12EFF,FEDERRATH13EFF,KRUMHOLZ14REVIEW} and is often implemented as a controlling parameter in sub-grid models for star formation in numerical simulations \citep[e.g.,][]{KATZ92SIMS,HOPKINS11SIMS,KIM14SIMS,AGERTZ21SIMS,SEMENOV21GMCS}. Nonetheless, we caution that comparing $\epsilon_{\rm ff}^{\rm mol}$ to cloud-scale gas properties unavoidably involves plotting highly correlated quantities against one another, because they all depend on the same cloud-scale CO imaging data.

As with the cloud-scale gas properties, we estimate $\Sigma_{\rm mol}$ from PHANGS--ALMA CO~(2-1) imaging using our adopted CO-to-H$_2$ conversion factor. At the $1.5$~kpc resolution where we calculate $\Sigma_{\rm mol}$ and $\tau_{\rm dep}^{\rm mol}$, PHANGS--ALMA yields high significance CO detection for almost every region, and CO flux completeness at low resolution plays little role in our data selection. By default, we estimate $\Sigma_{\rm SFR}$ combining H$\alpha$ narrowband measurements and WISE $22\mu$m mid-infrared emission. These measurements follow \citet{BELFIORE23SFRLARGE} as implemented by \citet{SUN23SFGAS} using narrowband H$\alpha$ maps from A.~Razza et al.\ (in preparation) and $22\mu$m WISE4 maps from \citet{LEROY19Z0MGS}. When H$\alpha$ maps are not available, we instead combine GALEX FUV emission and WISE 22$\mu$m emission, and in a handful of cases we use only WISE $22\mu$m emission. Both \citet{SUN23SFGAS} and \citet{QUEREJETA24SFGAS} have both considered star formation scaling relations in this data set and shown that using narrowband H$\alpha$ and/or H$\alpha$+22$\mu$m matches the results using the ``gold standard'' VLT/MUSE maps to $\approx 0.1$~dex accuracy (though see discussion about next steps in \S\ref{sec:nextsteps}).

\subsection{Data selection and completeness}
\label{sec:selection}

\begin{figure}[t!]
\centering
\includegraphics{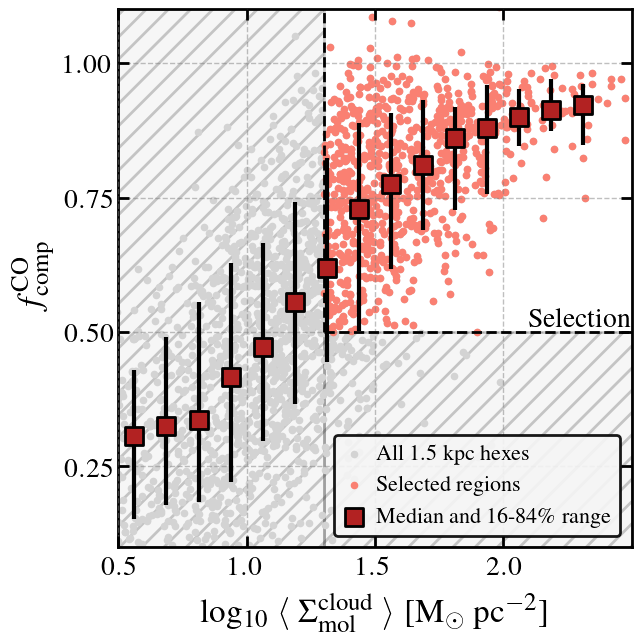}
\caption{The fraction of total CO~(2-1) emission detected at high signal to noise in each aperture, $f_{\rm comp}^{\rm CO}$, as a function of the mass-weighted mean cloud-scale surface density, \sdmol . The binned data show the median and $16-84\%$ range in $f_{\rm comp}^{\rm CO}$ considering all regions. Completeness rises as a function of the mass-weighted mean cloud-scale surface density, with scatter reflecting variation in distance to the target and data quality. We analyze regions with $f_{\rm comp}^{\rm CO} > 0.5$ and $\sdmol > 20$~M$_\odot$~pc$^{-2}$.
\label{fig:completeness}
}
\end{figure}

\begin{figure*}[t!]
\centering
\includegraphics[height=3in]{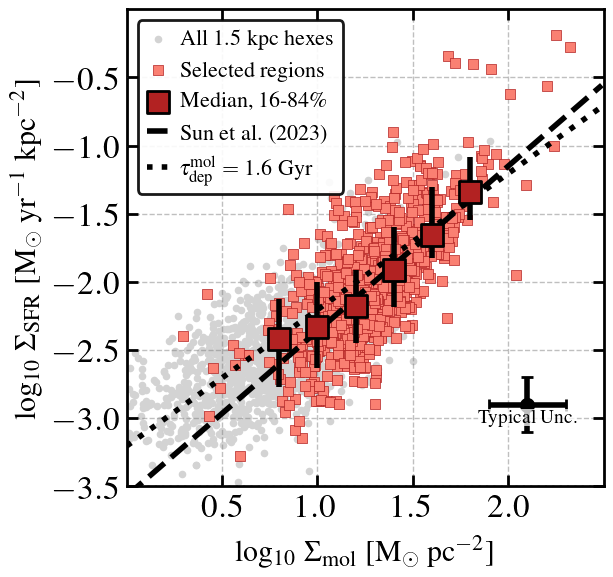}
\includegraphics[height=3in]{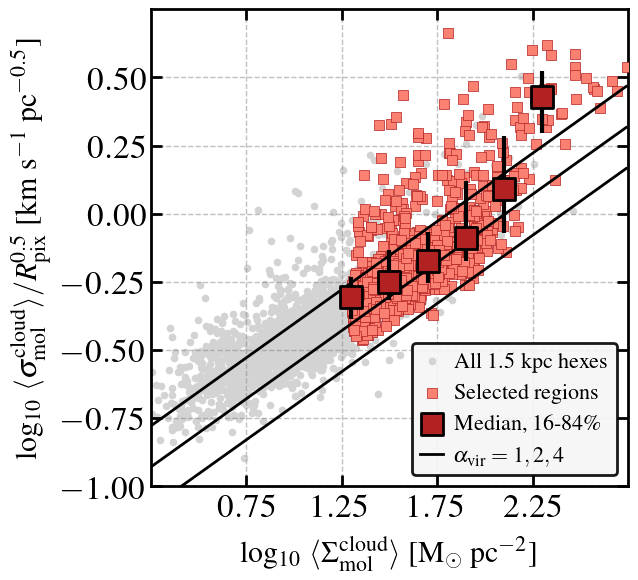}
\caption{Analyzed region in key parameter spaces. \textit{Left:} $\Sigma_{\rm SFR}{-}\Sigma_{\rm mol}$ space, i.e., the resolved molecular Kennicutt-Schmidt relation: gray points show all $1.5$~kpc hexes in the \citet{SUN22CLOUDS} database. Red points show regions selected for our analysis based on their completeness and cloud-scale surface density (\S \ref{sec:selection}, Fig.~\ref{fig:completeness}, Tab.~\ref{tab:experiment}). Dark points with error bars show the median and 16{-}84\% range $\Sigma_{\rm SFR}$ in bins of $\Sigma_{\rm mol}$. Our selection imposes a bias towards molecular-gas rich regions, but within the covered part of parameter space our analyzed regions reflect the overall $\Sigma_{\rm SFR}{-}\Sigma_{\rm mol}$ scaling seen in local galaxies \citep[][]{SUN23SFGAS}. \textit{Right:} Cloud-scale mean $\sigmol / R_{\rm pix}^{0.5}$ as a function of \sdmol\ (the ``Heyer-Keto relation'' capturing the mean dynamical state of molecular clouds) for the same regions, using the same symbols as the left panel. Selected data show $\approx 1$~dex dynamic range in region-averaged gas and SFR surface density, $> 1.5$~dex range in \sdmol , and $\approx 1$~dex span in $\sigmol / R_{\rm pix}^{0.5}$, and they reflect the overall set of high surface density regions found in nearby, star-forming galaxies well.
\label{fig:scaling}
}
\end{figure*}

Measuring $\Sigma_{\rm mol}^{\rm cloud}$, $\sigma_{\rm mol}^{\rm cloud}$, and $\alpha_{\rm vir}^{\rm cloud}$ requires detecting and characterizing CO~(2-1) emission over individual high-resolution lines of sight. In particular, robust detections are required to estimate the velocity dispersion, $\sigma_{\rm mol}$, and virial parameter, $\alpha_{\rm vir}$. PHANGS--ALMA has sensitivity to detect individual $\sim 10^{5}$~M$_\odot$ GMCs in each $60{-}150$~pc beam \citep[see details in][]{PHANGSALMA21SURVEY}. However, over many regions much of the CO flux appears associated with lines of sight that have lower surface brightness. This means only a fraction of the total CO flux in each 1.5~kpc region will enter our high-resolution cloud-scale gas property measurements. \citet{SUN22CLOUDS} calculate this CO flux ``completeness,'' 

\begin{equation}
\label{eq:fcomp}
f_{\rm comp}^{\rm CO} \equiv \frac{\sum_{\rm high~S/N~mask} \I_{\rm CO}^{\rm cloud}}{\sum_{\rm full~region} I_{\rm CO}^{\rm cloud}}~,
\end{equation}

\noindent by comparing the flux associated with individually detected high signal-to-noise sightlines (the numerator) to the total CO~(2-1) flux\footnote{This total flux is well-measured by summing the PHANGS--ALMA data, which include short- and zero-spacing data, without any aggressive high resolution masking.} (the denominator). A high $f_{\rm comp}^{\rm CO}$ means that our cloud-scale measurements capture the physical conditions of the \textit{bulk} molecular gas, which is relevant to the overall \taudep\ in the region. Conversely, a low $f_{\rm comp}^{\rm CO}$ indicates that we merely sample the ``tip of the iceberg'' and that our cloud-scale measurements are not representative and may have little relevance to region-averaged star formation activity. 

\citet{SUN22CLOUDS} correct the measured cloud-scale surface density and $\tauff$ in each hexagonal region to account for the effects of completeness by assuming lognormal probability distribution function for the surface density, in good agreement with observations (see that paper). However, this implies that in regions with very low $f_{\rm comp}$, the measured cloud properties will represent a rather aggressive extrapolation. With this in mind, we consider a region appropriate for the current analysis if the cloud-scale CO~(2-1) completeness $f_{\rm comp}^{\rm CO} > 50\%$.

Figure \ref{fig:completeness} shows $f_{\rm comp}^{\rm CO}$ in PHANGS--ALMA as a function of \sdmol . Based on this figure we impose a \sdmol\ cut at $20$~M$_\odot$~pc$^{-2}$ in addition to requiring $f_{\rm comp}^{\rm CO} > 50\%$. This yields a uniform data set and avoids the case where a few high completeness galaxies dominate low \sdmol\ bins and drive our results on their own. Table \ref{tab:experiment} reports the number of hexagonal regions and galaxies that survive these cuts. Applying this criteria to the $\ell = 150$~pc resolution PHANGS--ALMA data set in \citet{SUN22CLOUDS}, we select $841$ regions in $67$ galaxies.

\subsection{Correlation analysis, binning, fitting, and uncertainties}
\label{sec:fits}

While parametrized fitting and correlation analyses provide useful insights, we would like to first emphasize the region-by-region measurements as our key products. These represent our best estimates of \taudep , \eff , and gas properties with a well-understood selection function for a representative set of nearby galaxies. We present the full measurement set in Table \ref{tab:data}. Appendix \ref{sec:stepbystep} summarizes this section in a simple, step by step guide that can be used to reproduce matched measurements from simulations or other data sets.

We treat the cloud-scale gas properties as independent variables and measure how \taudep\ and \eff\ (the $y$ axis) vary as a function of these mass-weighted average gas properties (the $x$ axis). We measure Spearman's rank correlation relating \taudep\ and \eff\ to each property and fit $y$ as a function of $x$ for each gas property using the functional form

\begin{equation}
\label{eq:line}
y = m \left( x - x_0 \right) + b
\end{equation}

\noindent where $y$ refers to either $\log_{10} \taudep$ or $\log_{10} \eff$, $x$ to the $\log_{10}$ of one of the gas properties that we consider. We calculate $x_0$ as the median of the $x$ data, then maximize the log-likelihood to infer $m$ and $b$. We also include an intrinsic scatter term in the fit, but this tends to be very small given our adopted uncertainty estimates (see below). We therefore report $\sigma$, the robustly estimated rms residual of the data about each fit. 

Because we average over large regions selected to have high \sdmol , the formal statistical uncertainty associated with any individual measurement is small. Instead, systematic effects dominate our uncertainties. In deriving the calibrations that we use for $\Sigma_{\rm SFR}$, \citet{BELFIORE23SFRLARGE} found $\approx 0.15$~dex point-to-point scatter between the estimators that we use and ``gold standard'' extinction corrected H$\alpha$ estimates from MUSE (\citealt{LEROY19Z0MGS} found similar scatter comparing to \citealt{GSWLC16}). Our $\alpha_{\rm CO}$ estimate heavily leverages the work of \citet{CHIANG24XCO} who find $\approx 0.2$~dex scatter between their dust-based $\alpha_{\rm CO}^{2-1}$ and the fit to local conditions that we employ. Combining the uncertainties, we consider any individual \taudep\ estimate to be uncertain by $\approx 0.3$~dex due to uncertainties in our ability to estimate star formation rates and molecular gas masses. We consider molecular gas surface densities to be uncertain by $0.2$~dex at any scale, and $\tau_{\rm ff} \propto \Sigma_{\rm mol}^{-0.5}$ to be uncertain by $0.1$~dex. The main systematic uncertainty on the mass-weighted line width, \sigmol , is related to measurement technique and correction for sensitivity biases \citep[see][]{HENSHAW16GMCS,KOCH18LINEWIDTH,ROSOLOWSKY21GMCS}; we consider these \sigmol\ values to be uncertain by $\approx 0.15$~dex. When we include these uncertainty estimates in the fitting described above, we find little or no need for an intrinsic scatter term. This may imply that the estimates are slightly conservative, i.e., too large, but they do appear grounded in the current literature.

We observe significant galaxy-to-galaxy scatter in our data, meaning that individual galaxies trace out offset relationships in the \taudep\ vs.\ gas properties parameter space. This situation is common in resolved studies of star formation scaling relations \citep[e.g., see][]{LEROY13SFGAS} and likely reflects a mixture of physical scatter due to unplotted additional parameters and/or systematic uncertainties that operate galaxy by galaxy (e.g., calibration uncertainties, uncertainties in the adopted $\alpha_{\rm CO}$, SFR calibration uncertainties, etc.). We account for this in two ways. First we estimate uncertainties on our fits using a ``per galaxy'' bootstrapping approach, where we repeatedly re-draw $N$ galaxies from the parent sample and refit the relation between \taudep\ and gas properties. Second, we construct a version of our plots in which we first normalize the \taudep\ or \eff\ for each region by the mean \taudep\ or \eff\ for the galaxy in which that region resides and then analyze the normalized results. We calculate the mean \taudep\ or \eff\ used for normalization by a molecular mass ($\Sigma_{\rm mol}^{\rm kpc}$) weighted average over the selected regions for that galaxy.

For each variable pair $(x,y)$ we distill the median and 16{-}84\% percentile in $y$ as a function of $x$. To do this, we apply two dimensional gaussian kernel density estimation (KDE) to the data, allowing the bandwidth to be chosen using the default ``Scott's rule'' implement in \texttt{scipy} (this typically yields bandwidths of $\lesssim 0.3$~dex). We evaluate the density on a regularly spaced grid spanning the 1{-}99\% range of the data in $x$ and the full range of the data in $y$. Then, we calculate the median and 16{-}84\% range in $y$ for each value of $x$ in this grid. The results visually match the spine and span of the data well. We fit lines to the median trends, but found them to be similar to those found fitting the data and only report the fits to the data.

We emphasize that the choice to fit $y$ (i.e., \taudep\ or \eff ) while treating $x$ (cloud-scale gas properties) as an independent variable has a significant impact on our results. Our data show significant scatter compared to the dynamic range in both $x$ and $y$, and many of the correlations that we examine are relatively weak. Fitting $x$ as a function of $y$ or conducting a bivariate fit would yield different results. Our adopted scheme is motivated by the desire to address the question ``Given some cloud-scale gas properties, how well can we predict \taudep ?'' (see \S \ref{sec:intro}). Our selection (\S \ref{sec:selection}) also imposes a strong selection in $x$, which would need to be carefully accounted for in any bivariate fit \citep[see, e.g.,][]{LEROY13SFGAS} but does not affect our adopted approach. The key point for the reader is that rigorous comparison to our fitting results should consider the same formulation of dependent and independent variable.

Finally, note that we examine many cases with correlated axes. In some cases this correlation is explicit, e.g., $\eff \propto \langle \left( \Sigma_{\rm mol}^{\rm cloud} \right)^{-0.5} \rangle$. In other cases, e.g., \taudep\ vs. \sdmol , the correlation depends on a cross-scale relationship between $\Sigma_{\rm mol}$ and \sdmol . In many cases our adopted $\alpha_{\rm CO}$ affects both axes. We indicate our expectations for covariance with a vector in each plot along with the typical uncertainties.

\begin{deluxetable*}{lccccccccccccc}[t!]
\tabletypesize{\footnotesize}
\tablecaption{Measurements for individual regions at $\ell = 150$~pc in 1.5~kpc diameter regions \label{tab:data}}
\tablewidth{0pt}
\tablehead{
\colhead{Galaxy} & 
\colhead{Radius} & 
\colhead{$\tau_{\rm dep}^{\rm mol}$} & 
\colhead{$\frac{\tau_{\rm dep}^{\rm mol}}{\left< \tau_{\rm dep}^{\rm mol} \right>_{\rm gal}}$} & 
\colhead{$\alpha_{\rm CO}^{2-1}$} & 
\colhead{$i$} & 
\colhead{$\log_{10} M_\star$} & 
\colhead{$\log_{10} {\rm SFR}$} & 
\colhead{$\left< \Sigma_{mol}^{\rm cloud} \right>$} & 
\colhead{$\left< \tau_{ff}^{\rm cloud} \right>$} & 
\colhead{$\left< \sigma_{mol}^{\rm cloud} \right>$} & 
\colhead{$\left< \alpha_{vir}^{\rm cloud} \right>$} & 
\colhead{$\epsilon_{\rm ff}^{\rm mol}$} & 
\colhead{$\frac{\epsilon_{\rm ff}^{\rm mol}}{\left< \epsilon_{\rm ff}^{\rm mol} \right>_{\rm gal}}$}
\\
\colhead{} & 
\colhead{(kpc)} & 
\colhead{(Gyr)} & 
\colhead{(norm.)} & 
\colhead{($\frac{{\rm M}_\odot~{\rm pc}^{-2}}{{\rm K~km~s}^{-1}}$)} & 
\colhead{($^\circ$)} & 
\colhead{(M$_\odot$)} & 
\colhead{(M$_\odot$~yr$^{-1}$)} & 
\colhead{(M$_\odot$~pc$^{-2}$)} & 
\colhead{(Myr)} & 
\colhead{(km s$^{-1}$)} & 
\colhead{} & 
\colhead{} & 
\colhead{(norm.)}
}
\startdata
ESO097-013 & 0.0 & 0.11 & 0.35 & 1.73 & 64.3 & 10.5 & 0.6 & 182.0 & 3.57 & 24.5 & 7.38 & 0.0339 & 1.23 \\
ESO097-013 & 1.54 & 0.42 & 1.42 & 3.26 & 64.3 & 10.5 & 0.6 & 25.0 & 9.4 & 8.1 & 4.93 & 0.0223 & 0.81 \\
ESO097-013 & 3.36 & 0.99 & 3.32 & 6.65 & 64.3 & 10.5 & 0.6 & 54.0 & 6.69 & 6.0 & 1.04 & 0.0068 & 0.25 \\
IC1954 & 0.0 & 0.92 & 0.66 & 5.26 & 57.1 & 9.7 & -0.4 & 32.0 & 9.1 & 7.1 & 3.52 & 0.0099 & 1.28 \\
IC1954 & 1.82 & 1.71 & 1.23 & 8.29 & 57.1 & 9.7 & -0.4 & 25.0 & 10.65 & 4.7 & 2.05 & 0.0062 & 0.8 \\
IC1954 & 1.82 & 1.41 & 1.02 & 8.14 & 57.1 & 9.7 & -0.4 & 22.0 & 11.35 & 4.6 & 2.37 & 0.0081 & 1.04 \\
IC1954 & 1.97 & 1.52 & 1.1 & 8.29 & 57.1 & 9.7 & -0.4 & 29.0 & 9.92 & 5.2 & 2.36 & 0.0065 & 0.84 \\
IC1954 & 1.97 & 1.54 & 1.11 & 8.42 & 57.1 & 9.7 & -0.4 & 25.0 & 10.76 & 5.1 & 2.51 & 0.007 & 0.9 \\
IC1954 & 2.76 & 1.44 & 1.04 & 9.43 & 57.1 & 9.7 & -0.4 & 21.0 & 11.57 & 4.6 & 2.3 & 0.008 & 1.03 \\
\enddata
\tablecomments{This is a stub. The full table is available online as part of a machine readable table. Columns: (1) galaxy name, (2) galactocentric radius, (3) molecular gas depletion time, (4) molecular gas depeletion time normalized to galaxy average, (5) CO (2-1) to H$_2$ conversion factor, (6) galaxy inclination, (7) galaxy integrated stellar mass, (8) galaxy integrated star formation rate, region-averaged mass weighted molecular gas: (9) surface density, (10) gravitational free-fall time, (11) line width, and (12) virial parameter, (13) star formation efficiency per free fall time, (14) star formation efficiency per free fall time normalized to galaxy average.}
\end{deluxetable*}

\section{Results}
\label{sec:results}

\begin{deluxetable}{cccccc}[t!]
\tabletypesize{\small}
\tablecaption{$\tau_{\rm dep}^{\rm mol}$ and cloud-scale gas properties \label{tab:tdep}}
\tablewidth{0pt}
\tablehead{
\colhead{$x$} & 
\colhead{Rank. corr.} &
\colhead{$m$} &
\colhead{$x_0$} &
\colhead{$b$} &
\colhead{$\sigma$} 
}
\startdata
\hline
--- &  ---  &  ---  &  ---  &  $ 9.33 $  &  $ 0.3 $  \\ 
\hline
$\log_{10} \left< \tau_{\rm ff}^{\rm cloud} \right>$ &  $ 0.18 $  &  $ 0.51 $  &  $ 6.94 $  &  $ 9.31 $  &  $ 0.28 $  \\ 
(normalized) &  $ 0.28 $  &  $ 0.43 $  &  ---  &  --- &  $ 0.18 $  \\ 
\hline
$\log_{10} \left< \Sigma_{\rm mol}^{\rm cloud} \right>$ &  $ -0.22 $  &  $ -0.3 $  &  $ 1.58 $  &  $ 9.32 $  &  $ 0.28 $  \\ 
(normalized) &  $ -0.28 $  &  $ -0.23 $  &  ---  &  --- &  $ 0.18 $  \\ 
\hline
$\log_{10} \left< \sigma_{\rm mol}^{\rm cloud} \right>$ &  $ -0.31 $  &  $ -0.48 $  &  $ 0.74 $  &  $ 9.32 $  &  $ 0.26 $  \\ 
(normalized) &  $ -0.31 $  &  $ -0.3 $  &  ---  &  --- &  $ 0.18 $  \\ 
\hline
$\log_{10} \left< \alpha_{\rm vir}^{\rm cloud} \right>$ &  $ -0.27 $  &  $ -0.27 $  &  $ 0.3 $  &  $ 9.31 $  &  $ 0.27 $  \\ 
(normalized) &  $ -0.21 $  &  $ -0.11 $  &  ---  &  --- &  $ 0.19 $  \\ 
\hline
\hline
\enddata
\tablecomments{Results correlating $\tau_{\rm dep}^{\rm mol}$ with cloud-scale gas properties for our fiducial sample of regions with high gas surface density. Cloud scale properties measured at 150~pc resolution, and our best estimate $\alpha_{\rm CO}$. Fit has form $y = m (x - x_0) + b$. Quantities defined in Table \ref{tab:experiment}. Normalized results first normalize $\tau_{\rm dep}^{\rm mol}$ by the galaxy-average value.}
\end{deluxetable}

Our selection yields gas properties, molecular gas depletion times, and estimates of the star formation efficiency per free fall time for $841$ independent $1.5$~kpc-sized regions in $67$ galaxies with properties measured at $\ell = 150$~pc resolution (we also make measurements using higher resolution CO data, though for fewer galaxies, and provide these in Appendix \ref{sec:resn}). Table~\ref{tab:experiment} summarizes the selection and other details of the experiment. Our core results are the measurements of and relationships between \taudep , \eff , and cloud-scale gas properties. We present the measurements in Table~\ref{tab:data}, summarize the correlation analyses in Tables~\ref{tab:tdep}, \ref{tab:fixaco}, and \ref{tab:eff}, and illustrate the results in \S~\ref{sec:tdepvstff} through \S~\ref{sec:effvsprops}, but first we show the properties of our selected regions in Figure \ref{fig:scaling}.

The left panel in Figure \ref{fig:scaling} shows the region-averaged $\Sigma_{\rm SFR}$ as a function of $\Sigma_{\rm mol}$ for our working data set. These measurements do not involve cloud-scale information but simply show the resolved molecular Kennicutt-Schmidt relation at 1.5~kpc scales. Our selected regions span more than an order of magnitude in $\Sigma_{\rm SFR}$ and $\Sigma_{\rm mol}$. The selected data agree well with the power-law fit by \citet{SUN23SFGAS}, shown by the dashed line,\footnote{Our CO-to-H$_2$ conversion factor follows the prescription in \citet{SCHINNERER24REVIEW} while \citet{SUN23SFGAS} adopts the prescription by \citet{BOLATTO13REVIEW} among others. Fig.~\ref{fig:scaling} shows that the two yield comparable results here.} in which $\Sigma_{\rm SFR} \propto \Sigma_{\rm mol}^{1.2}$. This slope reflects that the highest surface density regions in our targets also have the shortest $\tau_{\rm dep}^{\rm mol}$.

The right panel in Figure \ref{fig:scaling} presents a complementary view, now with cloud-scale information. It shows mass-weighted, region-averaged, cloud-scale velocity dispersion normalized by the adopted size scale, $\sigmol / R_{\rm pix}^{0.5}$, as a function of the cloud-scale surface density, \sdmol . Our selection also samples $> 1.5$~dex in \sdmol , and $\approx 1$~dex in \sigmol .  On average, our target regions follow the ``Heyer-Keto'' relation of $\sigma_{\rm mol}^{\rm cloud} / R_{\rm pix}^{0.5} \propto \left( \Sigma_{\rm mol}^{\rm cloud}\right)^{0.5}$. This is expected among gas structures with similar dynamical state, i.e., similar $\alpha_{\rm vir}$ values \citep[illustrated by the solid black lines, see][]{HEYER09GMCS,HEYER15REVIEW,SUN18CLOUDS,SUN20GMCS,SCHINNERER24REVIEW}. As a result of this, \sdmol\ and \sigmol\ are significantly correlated, and we will see similar relationships between \taudep\ and these two quantities. 

While \sdmol\ and \sigmol\ show a strong correlation, there is still substantial scatter and significant deviations from the fiducial Heyer-Keto relation. In particular, a subset of regions show enhanced \sigmol\ at fixed \sdmol . The gas in these regions has high $\avir \propto \left( \sigma_{\rm mol}^{\rm cloud}\right)^2 / (R_{\rm pix} \Sigma_{\rm mol}^{\rm cloud})$ and thus appears less bound by self-gravity. These high \avir\ data often come from the inner parts of galaxies and show high \sdmol . There, stellar gravity can more strongly affect the molecular gas dynamical state, and unresolved streaming motions and shadowing of multiple components may also contribute to the observed broad CO line width \citep[][J. Henshaw et al. in prep.]{MEIDT13SFGAS,MEIDT18GMCS,SUN18CLOUDS,SUN20GMCS}. We return to this point in more detail in \S \ref{sec:byrad}.

\subsection{Molecular gas depletion time as a function of gravitational free-fall time}
\label{sec:tdepvstff}

\begin{figure*}[t!]
\centering
\includegraphics{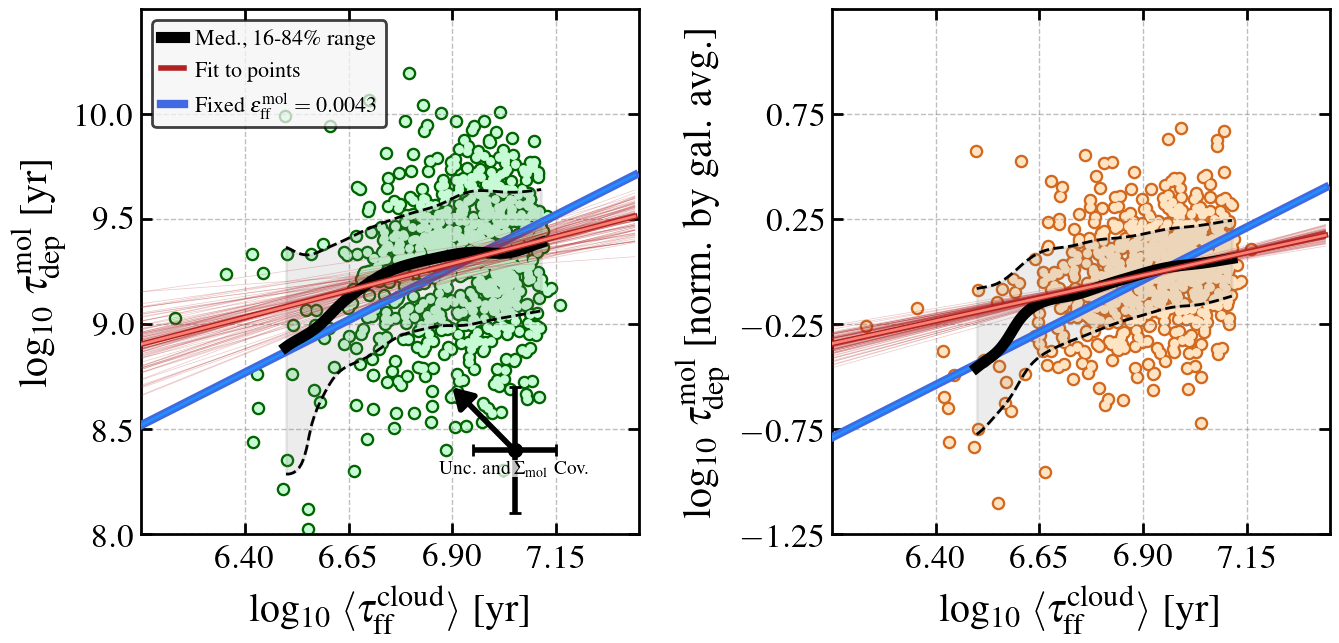}
\caption{Region-averaged molecular gas depletion time, \taudep , as a function of cloud-scale mean free-fall time, $\langle \tau_{\rm ff}^\mathrm{cloud} \rangle$. Points show measurements for individual $1.5$~kpc regions, and the black lines show the median and 16{-}84\% range estimated as described in \S \ref{sec:fits}. The red line shows the best fit to the data (Tab.~\ref{tab:tdep}), with thin traces showing fits to versions of the data set that bootstrap resample the set of included galaxies. For comparison, the blue line shows the expectation for a fixed \eff\ (i.e., $\taudep \propto \langle \tau_{\rm ff}^\mathrm{cloud} \rangle$), and the black error bar indicates typical uncertainties with an arrow showing the sense of correlation between the axes (see \S \ref{sec:methods}). The \textit{left} panel shows results for \taudep\ as measured directly from the data. The \textit{right} panel shows the same data set, but first normalizes each \taudep\ measurements by the mean \taudep\ for that galaxy, thus removing the galaxy-to-galaxy scatter from the left plot. $\taudep$ and $\langle \tau_{\rm ff}^\mathrm{cloud} \rangle$ correlate with each other, but the slope is shallower than that predicted for fixed \eff .
\label{fig:tdepvstff}
}
\end{figure*}

Figure \ref{fig:tdepvstff} plots \taudep\ as a function of the region-averaged cloud-scale gravitational free-fall time, $\tauff$. All other things being equal, higher density gas should form stars faster, and the characteristic time for the gas to collapse under its own gravity, $\tau_{\rm ff} \propto \rho^{-0.5} \propto \left( \Sigma_{\rm mol}^{\rm cloud} \right)^{-0.5}$, has been considered a key timescale in this context.

\begin{figure*}[t!]
\centering
\includegraphics[width=0.465\textwidth]{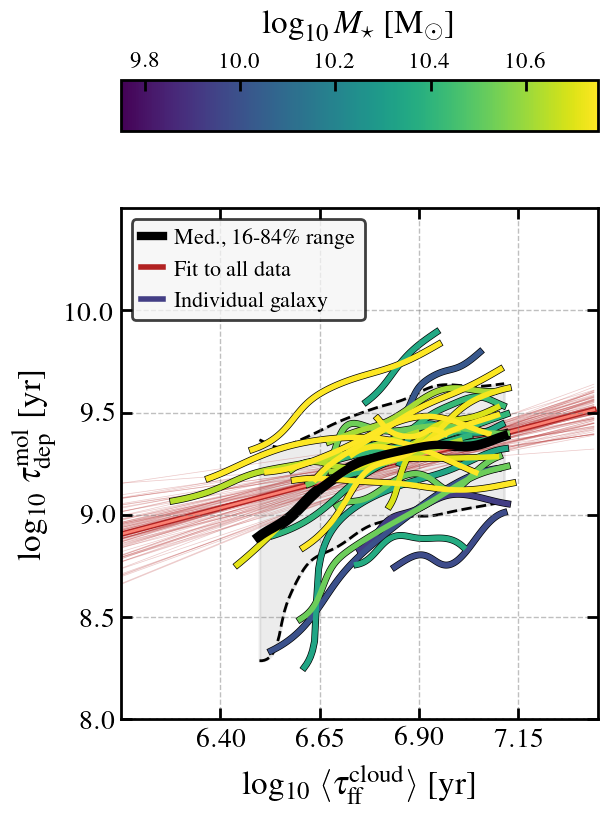}
\includegraphics[width=0.485\textwidth]{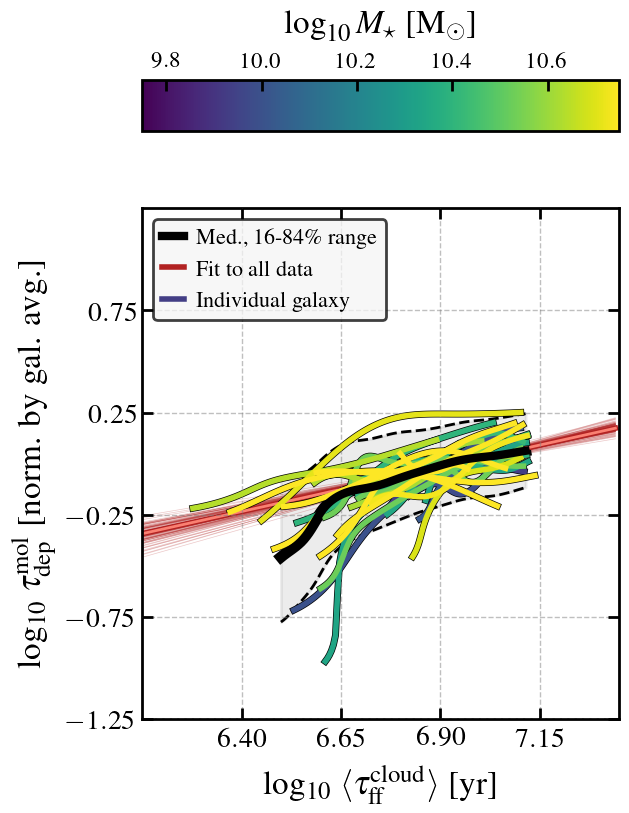}
\caption{Traces of individual galaxies in the same $\taudep {-} \langle \tau_{\rm ff} \rangle$ space shown in Fig.~\ref{fig:tdepvstff}. Each colored line shows the median trend for an individual galaxy, with color indicating the total stellar mass of that galaxy ($M_\star$). The left panel shows these traces in the \taudep\ vs. \tauff\ space. In the right panel, we first normalize \taudep\ by the mean value for each galaxy, thus removing overall galaxy-to-galaxy scatter in \taudep . Individual galaxies show trends with a similar positive correlation to the overall median but significant galaxy-to-galaxy scatter in the slope, the range of \tauff\ covered, and their mean \taudep .
\label{fig:tdepbygal}
}
\end{figure*}

Figure \ref{fig:tdepvstff} does show a weak correlation between \taudep\ and $\langle \tau_{\rm ff}^\mathrm{cloud} \rangle$, with Spearman rank correlation coefficient $0.18$. Gas with higher density (low \tauff) coincides with regions that show higher star formation rate per unit gas (low \taudep ). The blue line showing fixed $\eff$ shows that a single \eff\ can reasonably describe the bulk of the measurements. Most of the data lie within $\pm 0.3$ dex (i.e., a factor of two) of the blue line \citep[consistent with results using the same data set in][]{SUN23SFGAS}. A fixed $\eff$ thus provides a reasonable approximate description of the data. But in detail, $\eff$ tends to be somewhat higher (i.e., the points lie below the blue line) at high $\tauff$, corresponding to low density. This manifests as a best fit \taudep--\tauff\ relation with slope $m = 0.51$ (the red line, Table~\ref{tab:tdep}). This is shallower than the $\taudep \propto \tauff$ expected for fixed \eff , though we reiterate that this specific slope depends on the choice to fit $\taudep$ while treating $\tauff$ as the independent variable (see \S \ref{sec:fits}). 

The right panel in Fig.~\ref{fig:tdepvstff} also shows \taudep\ as a function of \tauff\ but first normalizes \taudep\ by the galaxy-average value. Normalizing leads to an even sharper correlation (Tab.~\ref{tab:tdep}), with the rank correlation coefficient increasing to $0.28$ after normalization. This indicates that $\taudep$ correlates better with $\tauff$ within individual galaxies. The best fit $\taudep{-}\tauff$ slope becomes slightly shallower after normalization (Tab.~\ref{tab:tdep}), $m = 0.43$, but the overall picture appears similar as in the left panel.

Fig.~\ref{fig:tdepbygal} shows why normalizing by galaxy-average $\taudep$ sharpens the $\taudep{-}\tauff$ relation. We plot binned results for individual galaxies before any normalization in the left panel. To a large extent, galaxies trace offset relations in the \taudep--\tauff\ space. This reflects some overall correlation of \taudep\ with host galaxy properties, with a general sense that lower mass galaxies show shorter $\taudep$ (i.e., darker lines tend to appear lower in Fig.~\ref{fig:tdepbygal}). This might reflect real physics, or might reflect that the metallicity-dependence in our $\alpha_{\rm CO}$ prescription remains too weak \citep[because metallicity correlates with $M_\star$;][]{MAIOLINO19REVIEW}. Global variation of $\tau_{\rm dep}^{\rm mol}$ as a function of $M_\star$ has been considered extensively \citep[see, e.g.,][]{SAINTONGE11SFGAS,SAINTONGE17SFGAS,SAINTONGE22REVIEW,SCHRUBA12XCO,LEROY13SFGAS,TACCONI20REVIEW}. For our purposes, the key points are that (1) $\tauff$ variations do not appear to explain all of the observed galaxy-to-galaxy scatter (nor do \sdmol ,  \sigmol , or $\alpha_{\rm vir}$ variations), and (2) \taudep\ correlates with \tauff\ more strongly within individual galaxies. 

In the right panels in Fig.~\ref{fig:tdepvstff} and \ref{fig:tdepbygal}, after normalization most galaxies lie together along a common relation, with individual regions showing rms scatter $\approx 0.2$~dex about the median trend. This represents ${\sim}\sqrt{2}$ times lower scatter than we find before normalizing, implying that about half the total scatter in $\tau_{\rm dep}^{\rm mol}$ can be attributed to galaxy-to-galaxy scatter, while the rest represents scatter within individual galaxies. This resembles results from previous work on $\tau_{\rm dep}^{\rm mol}$ and kpc-scale star formation scaling relations by \citet{LEROY13SFGAS} and \citet{JIMENEZ19DENSE}.

\subsection{Molecular gas depletion time as a function of cloud-scale surface density and line width}
\label{sec:tdepvsprops}

\begin{figure*}
\centering
\includegraphics{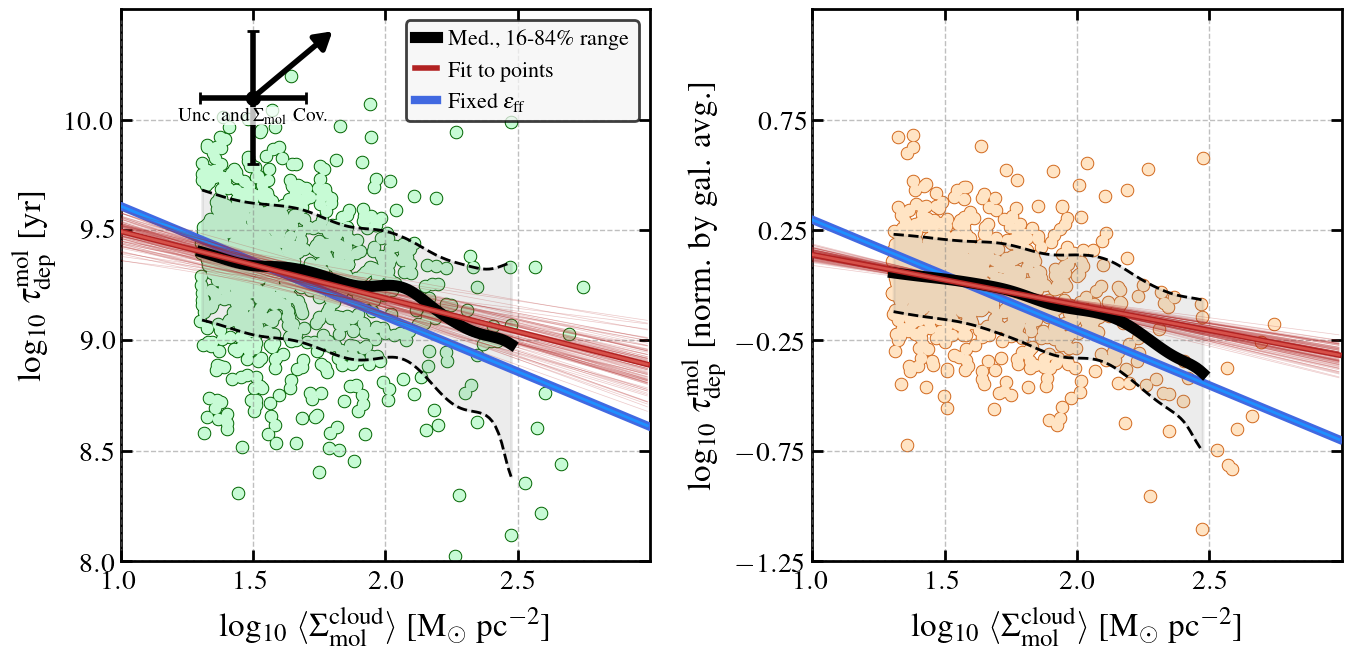}
\includegraphics{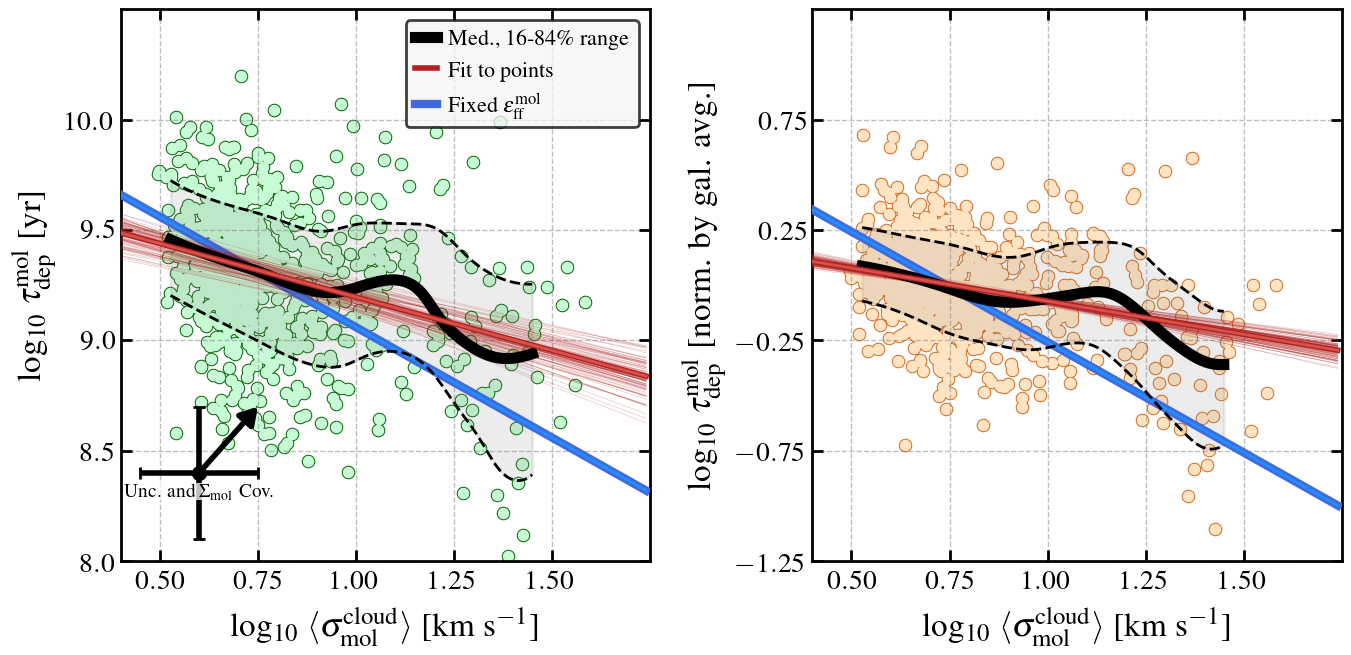}
\caption{\taudep\ as a function of region-averaged cloud-scale surface density (\textit{top}) and line width (\textit{bottom}). As in Fig.~\ref{fig:tdepvstff}, the \textit{left} panels show the measurements of \taudep , while in the \textit{right} panel measurements are normalized by the galaxy-average \taudep , which removes galaxy-to-galaxy scatter in \taudep (see Fig.~\ref{fig:tdepbygal}). Other markings showing the median trend and 16{-}84\% range, best fit, and expectation for fixed \eff , are as in Fig.~\ref{fig:tdepvstff}. \taudep\ anti-correlates with both \sdmol\ and \sigmol (see Table \ref{tab:tdep}), showing a slope shallower than expected for fixed \eff . Given the underlying relation among \sdmol , \sigmol , and $\langle \tau_{\rm ff}^{\rm cloud} \rangle$ these and the plots in Fig.~\ref{fig:tdepvstff} all express highly related trends. We highlight the $\taudep{-}\sigmol$ relation, shown in the bottom row, because it depends less on $\alpha_{\rm CO}$ and potentially resolution compared to other measurements.
\label{fig:tdep_vs_gasprops}
}
\end{figure*}

Fig.~\ref{fig:tdep_vs_gasprops} shows \taudep\ as a function of cloud-scale gas surface density, \sdmol , and velocity dispersion, \sigmol . Tab.~\ref{tab:tdep} reports results for the corresponding correlation analysis. $\taudep$ anti-correlates with both quantities. Surface density and velocity dispersion represent more direct observables than \tauff , and we therefore view these as our most fundamental results. They should be easy to reproduce by applying our methodology to simulations or other data sets (following Appendix \ref{sec:stepbystep}).

To first order, Figs.~\ref{fig:tdepvstff} and \ref{fig:tdep_vs_gasprops} could be viewed as three different manifestations of the same underlying relationship between gas density and star formation efficiency. Because\footnote{Here we discuss general scalings among cloud properties, not our specific observables, and so omit the averaging notation.}
\begin{eqnarray}
\tau_{\rm dep}^{\rm mol} &=& \frac{\tau_{\rm ff}^\mathrm{cloud}}{\eff} \\
\nonumber \tau_{\rm ff}^\mathrm{cloud} &=& \sqrt{\pi / (8 G \Sigma_{\rm mol}^\mathrm{cloud} / R_{\rm cloud})}
\end{eqnarray} 
so that for a fixed \eff\ and size scale, 
\begin{equation}
\label{eq:tdepsdmol}
\taudep \propto \sdmol^{-0.5}
\end{equation} 
The expected anti-correlation with \sigmol\ emerges because for fixed dynamical state, $\alpha_{\rm vir}$ \citep[``the Heyer-Keto relation,'' see][]{HEYER09GMCS,SUN18CLOUDS,ROSOLOWSKY21GMCS,SCHINNERER24REVIEW}, $\Sigma_{\rm mol}$ and $\sigma_{\rm mol}$ are related as
\begin{eqnarray}
\sigma_{\rm mol} &=& \sqrt{\alpha_{\rm vir}~R~\frac{f G \pi \Sigma_{\rm mol}}{5}} 
\end{eqnarray}
with $f$ a geometrical factor related to the cloud density distribution (Table~\ref{tab:experiment}). Since $\sigma_{\rm mol} \propto \sqrt{\Sigma_{\rm mol}}$ at given $\alpha_\mathrm{vir}$ and size scale, we expect
\begin{eqnarray}
\label{eq:tdepsigmol}
\tau_{\rm mol}^{\rm dep} \propto \sigmol^{-1}
\end{eqnarray}
The expected scaling between \sigmol\ and \sdmol\ appears in our data \citep[see Fig.~\ref{fig:scaling} and][]{SUN18CLOUDS,SUN20GMCS,ROSOLOWSKY21GMCS}. We indicate power-laws with the fiducial slopes (Eqs.~\ref{eq:tdepsdmol} and \ref{eq:tdepsigmol}) as blue lines in Fig.~\ref{fig:tdep_vs_gasprops}.

As in Fig.~\ref{fig:tdepvstff}, most of the data lie within $\pm 0.3$~dex of the fiducial relationships, though our fit relations again have shallower slopes compared to the expectations for fixed \eff , fixed dynamical state clouds: $m = -0.3$ for $\taudep {-}\sdmol$ (compared to $-0.5$ expected) and $m = -0.48$ for $\taudep {-}\sigmol$ (compared to $-1$ expected). Again we caution that the adopted methodology affects the precise value of the fit slopes. 

Our results for the $\taudep{-}\sdmol$ relationship match limited previous measurements. In M51, \citet{LEROY17SFGAS} found $\taudep \propto \sdmol ^{-0.14}$ (using fixed $\alpha_{\rm CO}$). Adopting the \citet{BOLATTO13REVIEW} CO-to-H$_2$ prescription, similar to what we use here, the \citet{SUN23SFGAS} fit to the ``free-fall time regulated star formation'' relation yields $\Sigma_{\rm SFR}^{\rm kpc} \propto \left( \Sigma_{\rm mol}^{\rm kpc} / \tauff \right)^{0.75}$, which implies $\taudep \propto \sdmol^{-0.125}$ to $\taudep \propto \sdmol^{-0.375}$ depending on the precise dependence of $\tauff$ on $\Sigma_{\rm mol}^{\rm kpc}$. Our fit value lies intermediate in this regime. They also used the PHANGS--ALMA dataset, so this does not represent an independent measurement, but they did use a different selection and fitting methodology.

We emphasize the \taudep\ vs. \sigmol\ anti-correlation as a useful and new way to approach this topic. This anti-correlation is physically expected for fixed \eff , fixed dynamical state clouds (Eq.~\ref{eq:tdepsigmol}). This first-order expectation does not involve any turbulence physics, it simply reflects that the gas velocity dispersion traces the underlying gravitational potential of the clouds. 

This \taudep\ vs. \sigmol\ relationship is particularly intriguing because  $\alpha_{\rm CO}$ only affects the vertical axis\footnote{Though note that there could be a subtle second-order effect due to variations in $\alpha_{\rm CO}$ within a region, which might affect the weighted averaging scheme.}, so that these plots have less built-in covariance between the axes than most of the other relationships we consider (though see below). Moreover, \sigmol\ may not depend as sensitively on resolution as \sdmol . In the limit of an isolated, low-mass, unresolved cloud, ALMA might still measure an accurate CO line width given enough sensitivity, whereas in this situation \sdmol\ will be strongly affected by beam dilution.

We note that near galaxy centers, the blending of multiple clouds along the line of sight can be a concern, especially at our resolution \citep[][J.~Henshaw et al.\ in preparation]{ROSOLOWSKY21GMCS}. When estimating the CO line width using moment methods, this can lead the moment-based line width to reflect the inter-cloud dispersion rather than the line width of any individual component \citep[][J.~Henshaw et al.\ in preparation]{JEFFRESON22GMCS}. Our line width measurements are not second moments but ``effective widths'' \citep{HEYER01GMCS,LEROY16GMCS,SUN22CLOUDS} derived from the ratio of the line integral to the peak, $\sigma \equiv I_{\rm CO} / \sqrt{2 \pi} I_{\rm pk}$. This has less sensitivity to any inter-cloud dispersion but will still be biased high in case of multi-velocity components. It will be interesting to explore how different line width measures correlate with $\taudep$ and important to match methodology when comparing to our measurements.

\subsection{Molecular gas depletion time as a function of cloud-scale virial parameter}
\label{sec:tdepvsavir}

\begin{figure*}
\centering
\includegraphics{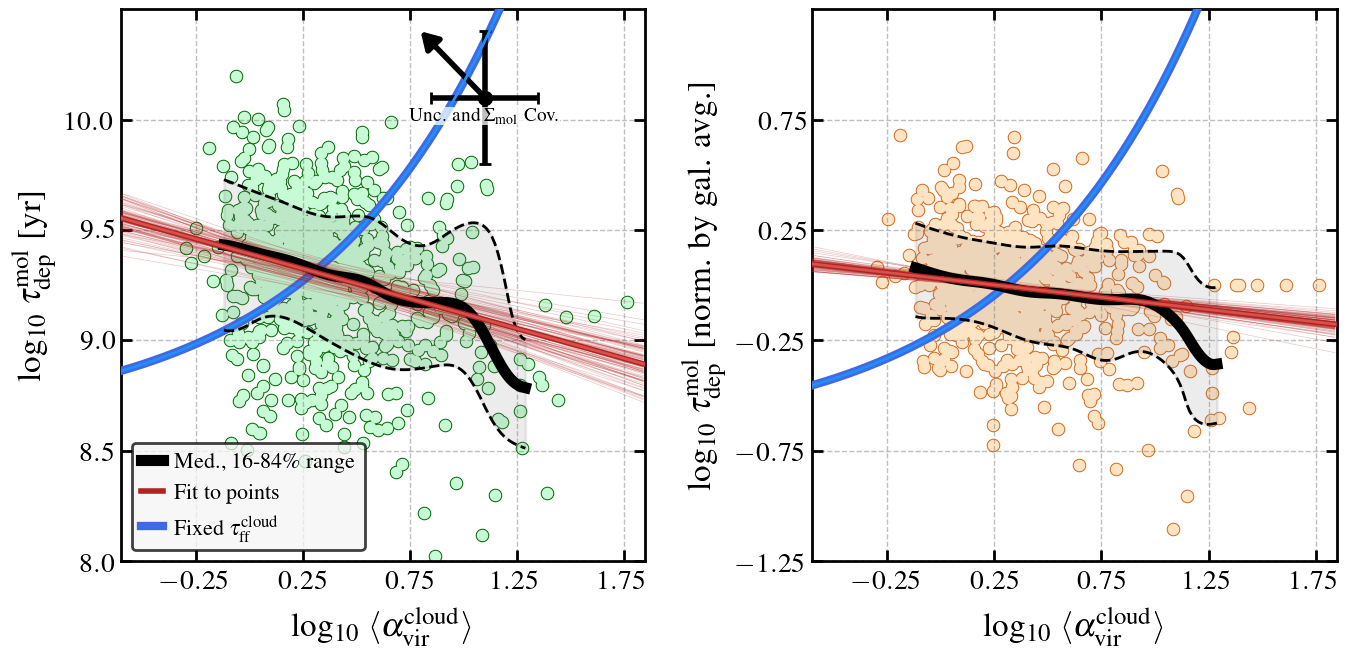}
\caption{\taudep\ as a function of region-averaged cloud-scale virial parameter, \avir . The contents follow Figs.~\ref{fig:tdepvstff} and \ref{fig:tdep_vs_gasprops}. As in those plots, the left panel shows \taudep\ without any normalization, while in the right panel measurements are first normalized by the galaxy-average \taudep . The blue line here shows expectations for fixed $\tau_{\rm ff}^{\rm cloud}$ and varying $\alpha_{\rm vir}^{\rm cloud}$ following \citet{PADOAN12EFF}. The data show an anti-correlation between \taudep\ and \avir . This has the opposite sense compared to the physical expectation that gas more bound by self-gravity (lower $\alpha_{\rm vir}$) should form stars more effectively. The anti-correlation likely reflects that gas in high surface density, inner regions of galaxies is significantly affected by the stellar potential.
\label{fig:tdep_vs_avir}
}
\end{figure*}

Fig.~\ref{fig:tdep_vs_avir} shows \taudep\ as a function of \avir , the region-mean cloud-scale virial parameter. Theoretical models of star formation in turbulent clouds make a clear prediction that, all other things being equal, \taudep\ should correlate with \avir . This reflects that more gravitationally bound gas should form stars more effectively. In Fig.~\ref{fig:tdep_vs_avir} the dashed line shows the expected relationship from \citet{PADOAN12EFF} for fixed $\tauff$; similar predictions would arise from, e.g., \citet{KRUMHOLZ05EFF} or \citet{FEDERRATH12EFF,FEDERRATH13EFF}. Our observations do not show the predicted correlation. In fact, we observe an anti-correlation between \taudep\ and \avir , such that gas less bound by self-gravity appears to form stars more efficiently. We find a best-fit slope of $m = -0.27$, though this drops to $m = -0.11$ after normalize each galaxy by its mean \taudep .

Though contradictory at first glance, this observation fits well with our current understanding of gas properties in the context of galaxy structure. We see that short \taudep\ regions often correspond to galaxy centers, which also show high \avir\ \citep[][]{SUN20GMCS,SUN22CLOUDS}. This high \avir\ can be at least partially attributed to the importance of stellar gravity to molecular gas dynamics in the inner parts of galaxies. Observations suggest that the high stellar and gas mass densities lead to high gas pressures in the inner parts of massive galaxies. This pressure then turns the overwhelming bulk of gas molecular \citep[e.g.,][]{WONG02SFGAS,BLITZ06PRESS,LEROY08SFGAS,EIBENSTEINER2024RMOL}. This can lead to the emergence of a ``diffuse,'' high velocity dispersion medium, in which the gas is bound by the larger-scale galactic potential. This large scale potential is dominated by stellar gravity, not by its own self-gravity. The high stellar densities in the inner regions of galaxies can rival the gas density even within molecular complexes, playing an important role in force balance \citep[e.g.,][]{MEIDT18GMCS,SUN20PRESS,LIU21GMCS}. Moreover, observations suggest that stellar density correlates with molecular gas morphology, addition additional support to the view that stellar gravity exerts an important dynamical effect \citep[][]{LIU21GMCS,DAVIS22GMCS}. 

The denser, inner regions of galaxies show high $\alpha_{\rm vir}$ because, following standard practice for molecular cloud studies, our \avir\ only accounts for gas self-gravity. In cases where the stellar gravity dominates, \avir\ will not capture the true dynamical state of the gas. The expectations mentioned in \S \ref{sec:intro} and \S \ref{sec:methods}, including the blue line from \citet{PADOAN12EFF} in Fig.~\ref{fig:tdep_vs_avir} emerge from theories of turbulent gas-only clouds, and do not typically account for a realistic galactic potential or conditions typical of dense inner galaxies.

Our result has the opposite sense of the trend measured in M51 by \citet{LEROY17SFGAS}, where the expected correlation between \taudep\ and \avir\ was present. Our results finding no overall correlation between \eff\ and \avir\ below do agree with their results in M51 and those of \citet{SCHRUBA19GMCS} studying molecular cloud populations in eight galaxies. M51 may be relatively rare in that the highest \avir\ are driven by streaming motions along the spiral arms, which correspond to regions of suppressed star formation \citep{MEIDT13SFGAS}. This may represent a distinct physical case from the high \avir\ seen in the centers of late type galaxies, which are not generally associated with suppressed star formation. The high $\alpha_{\rm vir}^{\rm cloud}$ found in early-type galaxies, which \textit{are} associated with suppressed star formation may represent yet a third regime \citep{WILLIAMS23GMCS,LU24GMCs}

A related effect where \avir\ misses the full picture has been discussed in the context of \ion{H}{1}-dominated, low surface density parts of galaxies by \citet{SUN18CLOUDS,SUN20GMCS} and \citet{SCHRUBA19GMCS}, following \citet{FIELD11GMCS}. There, atomic gas may make significant contributions to the cloud mass, or molecular gas may represent simply the dense subset of a larger turbulent neutral medium. Our selection largely excludes such regions (see \S \ref{sec:selection} and \S \ref{sec:byrad}), because PHANGS--ALMA lacks the surface brightness sensitivity to achieve high flux completeness at high resolution in such environments. This represents an area for future investigation, and likely relates to why we find results with the opposite direction from the Milky Way work by \citet{EVANS21GMCS}. Their analysis pivots on the behavior of gas at relatively large galactocentric radii and low $\Sigma_{\rm SFR}$ compared to what we study here and, as they discuss, depends sensitively on the metallicity dependence of $\alpha_{\rm CO}$.

\subsection{\texorpdfstring{\taudep }{Depletion time}, gas properties, and galactocentric radius}
\label{sec:byrad}

\begin{figure*}
\centering
\includegraphics{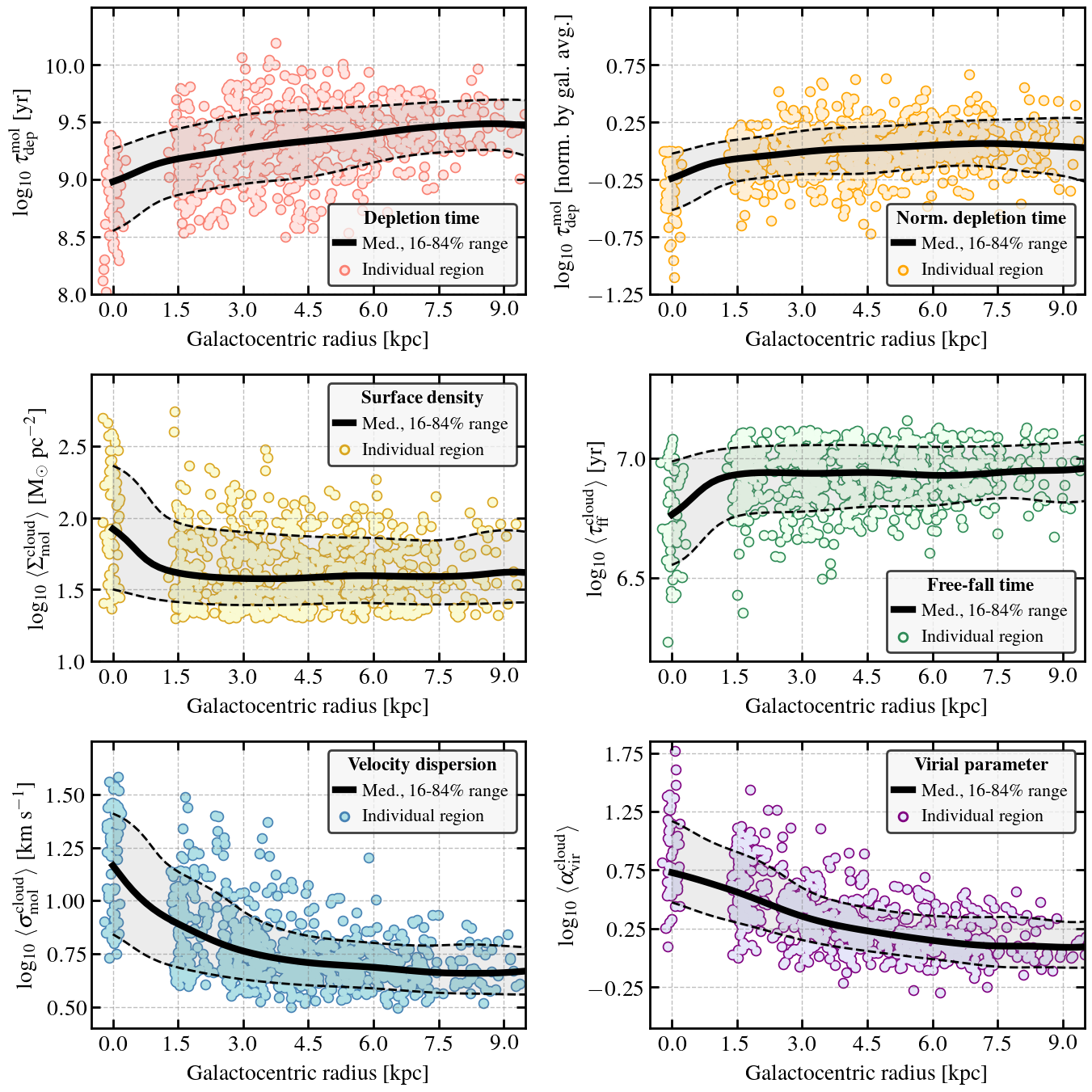}
\caption{Depletion time and mass-weighted mean gas properties as a function of galactocentric radius. Individual points show measurements and central galactocentric radius for individual 1.5~kpc diameter regions (a small amount of scatter has been added to the galactocentric radius for display purposes). The black lines show the median and 16{-}84\% range as a function of galactocentric radius. \textit{Top left:} \taudep\ appears depressed near galaxy centers and shows a modest radial trend, though this appears weaker in the \textit{top right} panel when \taudep\ is first normalized by the galaxy average. \textit{Middle left:} The cloud-scale surface density declines as galactocentric radius increases, with the highest values found in galaxy centers \citep[consistent with][]{SUN18CLOUDS,SUN20GMCS,ROSOLOWSKY21GMCS,SUN22CLOUDS}. The \textit{middle right} panel implies corresponding shorter \tauff . \textit{Bottom left:} The cloud-scale line width, \sigmol , is also highest at galaxy centers and then declines with galactocentric radius. Comparing \sdmol\ and \sigmol, the \textit{bottom right} panel shows that \avir\ is significantly enhanced in galaxy centers and shows a strong gradient as a function of galactocentric radius \citep[see][]{SUN20GMCS}.
\label{fig:gasprops_vs_rad}
}
\end{figure*}

As a general rule, high $\Sigma_{\rm mol}^{\rm kpc}$ can be found in the inner regions of galaxies, with the highest values at galaxy centers \citep[e.g.,][]{YOUNG91SFGAS,YOUNG95SFGAS,REGAN01SFGAS,KUNO07CO,QUEREJETA21}. The same is true for $\Sigma_{\rm mol}^{\rm cloud}$ and $\sigma_{\rm mol}^{\rm cloud}$ \citep[e.g.,][]{JOGEE05SFGAS,SUN20GMCS,SUN22CLOUDS,henshaw2023}. Figure \ref{fig:gasprops_vs_rad} shows the cloud-scale gas properties and \taudep\ as functions of galactocentric radius in our data set. 

All these properties of interest show clear radial trends. The innermost $1.5$~kpc hexes show high \sdmol , high \sigmol , high \avir , and low \taudep . The region-averaged \avir , determined by the ratio of $(\sigma_{\rm mol}^{\rm cloud})^2 / \Sigma_{\rm mol}^{\rm cloud}$, shows a strong radial trend, declining with increasing radius out to radius $> 4$~kpc \citep[something already clear in the beamwise analysis of][also see A.~Hughes et al.\ in preparation]{SUN20GMCS}. In \S \ref{sec:tdepvsprops} and \ref{sec:tdepvsavir} we noted that stellar gravity likely plays an important role in the behavior of \avir . Unresolved streaming or bulk motions may also affect \avir\ \citep{MEIDT13SFGAS,MEIDT18GMCS}, though beam smearing is not expected to exert a large impact at the resolution of PHANGS--ALMA \citep{LANG20KINEMATICS,SUN20GMCS}.

\taudep\ also shows a radial trend with lower values in galaxy centers. After normalization by the galaxy mean \taudep , the strongest effect is lower central \taudep\ compared to the disk average. This radial trend in normalized \taudep , which appears relatively flat with a moderate central depression and large scatter, is consistent with similar measurements for the HERACLES sample by \citet{LEROY13SFGAS} and analysis of the COMING survey by \citet{MURAOKA19COMING}. The low central \taudep\ in PHANGS--ALMA has previously been shown by \citet{TENG24XCO}, who also demonstrate that this result depends critically on the treatment of $\alpha_{\rm CO}$ \citep[see also][]{LEROY13SFGAS,DENBROK23XCO}. We expand on this in \S \ref{sec:alphaco}.

The regularity of the trends in Fig.~\ref{fig:gasprops_vs_rad} reinforce that the cloud-scale gas properties couple to the overall disk structure. Over the last decade, observations have shown that the ${\sim}50{-}150$~pc resolution properties of molecular gas, including $\Sigma_{\rm mol}^{\rm cloud}$ and $P_{\rm int}^{\rm cloud} \propto \Sigma_{\rm mol} \sigma^2_{\rm mol}$, correlate with the large scale properties of the galaxy disk \citep[][]{HUGHES13GMCS,SCHRUBA19GMCS,SUN18CLOUDS,SUN20PRESS,SUN20GMCS,SUN22CLOUDS,SCHINNERER24REVIEW}. \citet{SUN22CLOUDS}, in particular, provide an extensive series of scaling relations that link cloud-scale gas properties, measured using pixel statistics and cloud-finding methods, to the local kpc-scale conditions in the gas disk \citep[reviewed in][]{SCHINNERER24REVIEW}. 

\begin{figure*}
\centering
\includegraphics[width=0.465\textwidth]{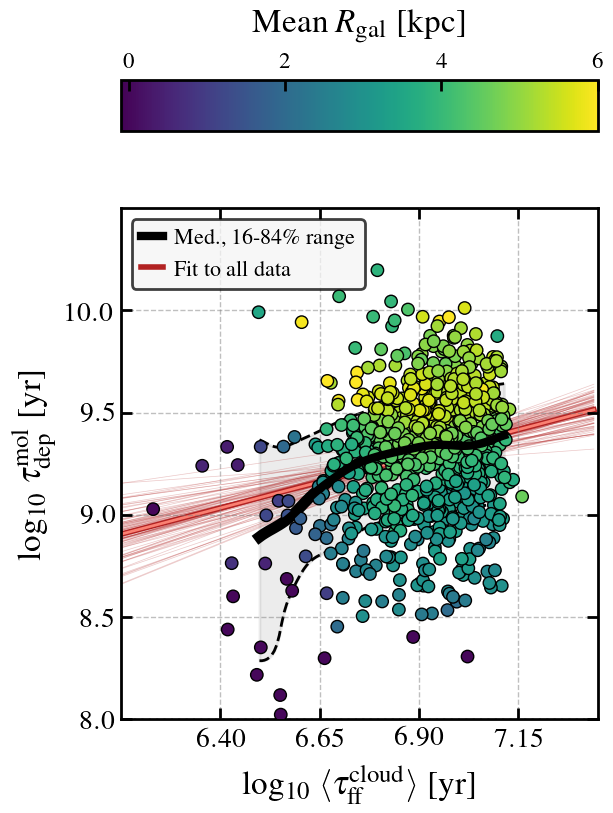}
\includegraphics[width=0.485\textwidth]{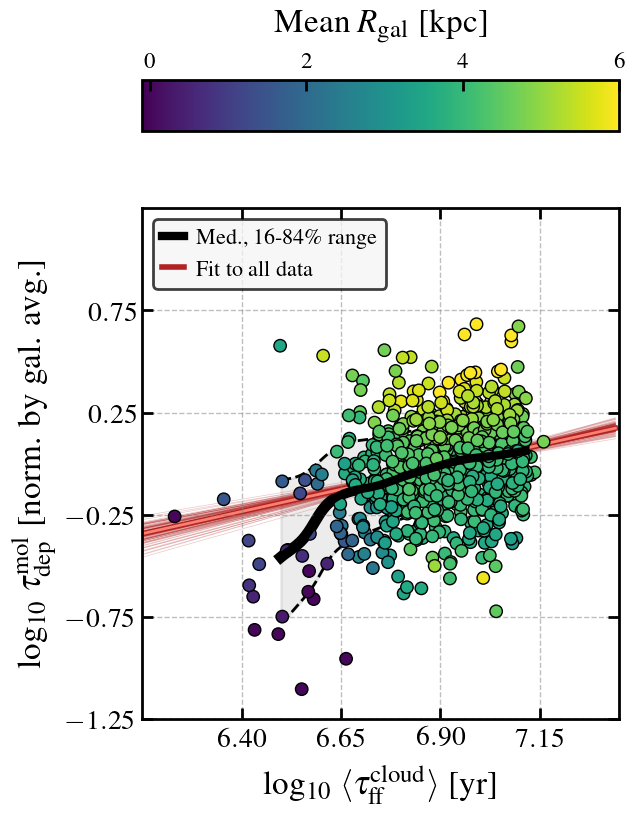}
\caption{As Fig.~\ref{fig:tdepvstff} but indicating the galactocentric radius of data. Each point is colored by the local mean radius among data at similar \tauff\ and \taudep . Long \tauff\ and long \taudep\ both arise preferentially from regions at high galactocentric radius, while denser gas with short depletion times and gravitational free-fall times arises from the central regions. This is particularly clear in the plot with normalized \taudep\ in the right panel. Colors span from galactocentric radius $R_{\rm gal} = 0{-}6$~kpc. 
\label{fig:tdepbyrad}
}
\end{figure*}

This regularity in the variation of gas properties in disks means that the correlations between \taudep\ (or \eff ) and any cloud-scale gas properties almost always have hidden variables related to galactic structure. To show this, Fig.~\ref{fig:tdepbyrad} shows \taudep\ as a function of \tauff\ again (as Fig.~\ref{fig:tdepvstff}), but now coloring the points as a function of galactocentric radius. A clear radial trend is visible in the plot, demonstrating that because of the link between gas properties and $\sim$kpc scale environment in real galaxies, this plot focused on small scale conditions in the gas cannot be easily separated from the rest of galactic structure. We return to this in \S \ref{sec:densdisc}.

Fig \ref{fig:gasprops_vs_rad} and \ref{fig:tdepbyrad} also highlight that the inner regions of galaxies are critical to achieve dynamic range in both \taudep\ and the cloud-scale gas properties. These are the source of our high \sdmol , high \sigmol , high \avir , low \tauff , low \taudep\ points. As noted above, extreme conditions in inner galaxies are well established. In high mass, barred galaxies, the central regions often represent distinct dynamical environments with high surface and volume densities and deep stellar potential wells \citep[e.g.,][]{KORMENDY04REVIEW,MEIDT18GMCS,QUEREJETA21,DAVIS22GMCS,SCHINNERER24REVIEW}. These regions also show enhanced CO emissivity \citep[e.g.,][]{SANDSTROM13XCO,TENG22XCO,TENG23XCO,DENBROK23XCO,CHIANG24XCO}, which manifest as variations in $\alpha_{\rm CO}$. The nature of our sampling scheme, in which each region covers equal area in the galaxy disk, means that these key environments are covered by relatively few data points. It is worth keeping the importance of these scarcer low \taudep , high \sdmol , high \sigmol\ points in mind when examining the plots.

Note that the enhanced densities and shorter \taudep\ conditions observed in galaxy centers refer to the gas, which is often concentrated into bar-fed central molecular zones. Viewed more broadly, massive galaxies often show depressed specific star formation rates, SFR/$M_\star$, in their inner regions \citep[][]{BELFIORE18RPROF,ELLISON18RPROF}. In massive galaxies with bulges, this can reflect the presence of gas-free regions with low $\Sigma_{\rm SFR}$. In follow up work, \citet{PAN24SFGAS} do suggest a longer \taudep\ in the central regions of main sequence galaxies compared to their disks \citep[both][\citealt{LIN22SFGAS}, and \citet{davis2014} also note longer \taudep\ in the inner parts of green valley galaxies, but these are not our focus]{PAN22TIMES}. As discussed in \S \ref{sec:alphaco} and \citet{TENG24XCO} and \citet{LEROY13SFGAS}, the treatment of $\alpha_{\rm CO}$ has a large impact on inferred \taudep\ for galaxy centers and we consider the evidence for low $\alpha_{\rm CO}$ in central molecular zones strong. Still, reconciling these measurements will be an area worth future investigation.

\subsection{The key role of the CO-to-\texorpdfstring{$H_2$}{H2} conversion factor}
\label{sec:alphaco}

\begin{deluxetable}{cccccc}[t!]
\tabletypesize{\small}
\tablecaption{$\tau_{\rm dep}^{\rm mol}$ and cloud-scale gas properties for fixed $\alpha_{\rm CO}$ \label{tab:fixaco}}
\tablewidth{0pt}
\tablehead{
\colhead{$x$} & 
\colhead{Rank. corr.} &
\colhead{$m$} &
\colhead{$x_0$} &
\colhead{$b$} &
\colhead{$\sigma$} 
}
\startdata
\hline
--- &  ---  &  ---  &  ---  &  $ 9.4 $  &  $ 0.28 $  \\ 
$\log_{10} \left< \tau_{\rm ff}^{\rm cloud} \right>$ &  $ -0.18 $  &  $ -0.21 $  &  $ 6.91 $  &  $ 9.37 $  &  $ 0.27 $  \\ 
(normalized) &  $ 0.01 $  &  $ 0.0 $  &  ---  &  --- &  $ 0.16 $  \\ 
\hline
$\log_{10} \left< \Sigma_{\rm mol}^{\rm cloud} \right>$ &  $ 0.13 $  &  $ 0.08 $  &  $ 1.66 $  &  $ 9.38 $  &  $ 0.27 $  \\ 
(normalized) &  $ -0.03 $  &  $ -0.01 $  &  ---  &  --- &  $ 0.16 $  \\ 
\hline
$\log_{10} \left< \sigma_{\rm mol}^{\rm cloud} \right>$ &  $ 0.14 $  &  $ 0.2 $  &  $ 0.74 $  &  $ 9.37 $  &  $ 0.27 $  \\ 
(normalized) &  $ 0.09 $  &  $ 0.07 $  &  ---  &  --- &  $ 0.16 $  \\ 
\hline
$\log_{10} \left< \alpha_{\rm vir}^{\rm cloud} \right>$ &  $ 0.19 $  &  $ 0.1 $  &  $ 0.33 $  &  $ 9.37 $  &  $ 0.26 $  \\ 
(normalized) &  $ 0.18 $  &  $ 0.05 $  &  ---  &  --- &  $ 0.16 $  \\ 
\hline
\hline
\enddata
\tablecomments{Results correlating $\tau_{\rm dep}^{\rm mol}$ with cloud-scale gas properties for our fiducial sample of regions with high gas surface density. Cloud scale properties measured at 150~pc resolution, and a fixed, Milky Way $\alpha_{\rm CO}$. Fit has form $y = m (x - x_0) + b$. Quantities defined in Table \ref{tab:experiment}. Normalized results first normalize $\tau_{\rm dep}^{\rm mol}$ by the galaxy-average value.}
\end{deluxetable}

\begin{figure*}
\centering
\includegraphics[width=0.475\textwidth]{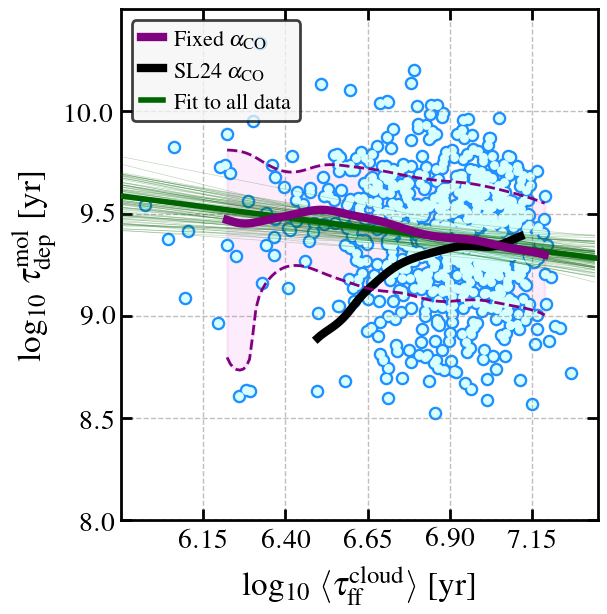}
\includegraphics[width=0.475\textwidth]{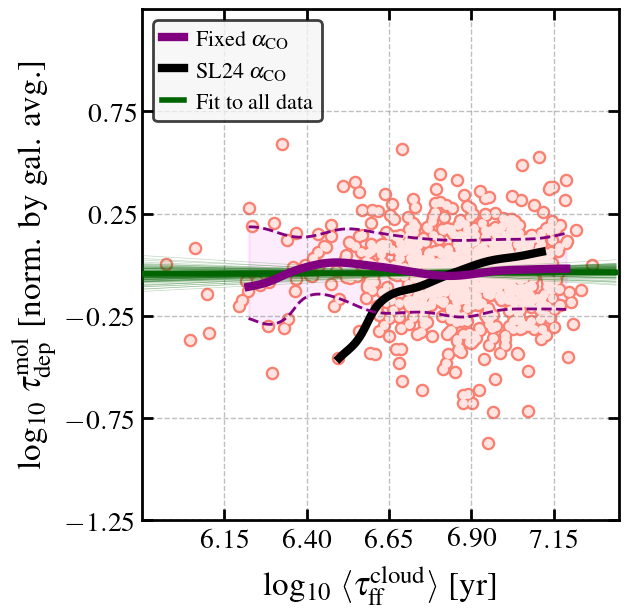}
\caption{As Fig.~\ref{fig:tdepvstff} but now showing \taudep\ and \tauff\ calculated assuming a fixed CO-to-H$_2$ conversion factor, $\alpha_{\rm CO}$. Similar to Fig.~\ref{fig:tdepvstff}, the points show individual regions, the purple lines and shaded region show the median and 16{-}84\% range, and the green line shows the best fit. For comparison, the black line shows the median trend using the variable $\alpha_{\rm CO}$ from Fig.~\ref{fig:tdepvstff}. Adopting a fixed $\alpha_{\rm CO}$ removes any overall correlation between \taudep\ and $\langle \tau_{\rm ff} \rangle$, and even leads to a slight anti-correlation in the left panel. Note that $\alpha_{\rm CO}$ affects both $\taudep$ ($\propto \alpha_{\rm CO}$) and $\tauff$ ($\propto \alpha_{\rm CO}^{-0.5}$), but the effect on \taudep\ is stronger.
\label{fig:tdep_alphaco}
}
\end{figure*}

As described in \S \ref{sec:methods}, we repeat our analysis using a fixed $\alpha_{\rm CO}$ instead of our best estimate. Fig.~\ref{fig:tdep_alphaco} shows these results for the $\taudep{-}\tauff$ relation and Table \ref{tab:fixaco} reports results of the correlation analysis between \taudep\ and all quantities for fixed $\alpha_{\rm CO}$. The fixed $\alpha_{\rm CO}$ leads to a nearly constant \taudep, in contrast to the fiducial results with our best-estimate $\alpha_{\rm CO}$. This, in turn, leads to almost no correlation or even an anti-correlation between \taudep\ and \tauff . Contrasting this with Table~\ref{tab:tdep}, all relationships show weaker correlations and slopes closer to 0.0 at fixed $\alpha_{\rm CO}$. Moreover, the actual sense of the correlation changes in most cases. With fixed $\alpha_{\rm CO}$, \taudep\ appears longer for high \sdmol\ and high \sigmol . The sense of the \avir\ trend also reverses compared to our fiducial case, with higher \avir\ corresponding to higher \taudep , meaning less efficient star formation. All of these trends are weak, however, especially when considering the \taudep\ normalized by the galaxy average.

Our fiducial $\alpha_{\rm CO}$ treatment, adopted from \citet{SCHINNERER24REVIEW}, accounts for variations in excitation and emissivity (sometimes referred to as the ``starburst'' term) and metallicity (the ``CO-dark'' term), similar to \citet{BOLATTO13REVIEW}. For the present work, the excitation and emissivity variations represent the key terms. In our prescription, $\alpha_{\rm CO}^{1-0} \propto \Sigma_\star^{-0.25}$ where $\Sigma_\star > 100$~M$_\odot$~pc$^{-2}$ \citep[based on][]{CHIANG24XCO}, reflecting emissivity variations, and $\alpha_{\rm CO}^{2-1}$ also $\propto \Sigma_{\rm SFR}^{-0.125}$ \citep[based on][and in good agreement with \citealt{KEENAN24LINES}]{LEROY22LINES,DENBROK21LINES,YAJIMA21LINES}, reflecting excitation variations affecting the CO~(2-1)/CO~(1-0) line ratio $R_{21}$. Because our selection focuses on the high surface density parts of galaxies (e.g., Fig.~\ref{fig:scaling}), these terms together significantly affect the estimated molecular gas mass across our sample. The metallicity term, which is critical to integrated galaxy measurements, has a weaker effect here that primarily affects the galaxy-to-galaxy scatter (e.g., Fig.~\ref{fig:tdepbygal}).

The impact of $\alpha_{\rm CO}$ on \taudep\ in galaxy centers has been noted before \citep[][]{LEROY13SFGAS,DENBROK23XCO,TENG23XCO}. In particular, in PHANGS--ALMA, \citet{TENG24XCO} showed that accounting for the emissivity terms in $\alpha_{\rm CO}$ lead to much shorter \taudep\ in galaxy centers, and a much steeper, even multi-valued $\Sigma_{\rm SFR}{-}\Sigma_{\rm mol}$ relationship. \citet{SUN23SFGAS} show a similar effect when applying the \citet{BOLATTO13REVIEW} $\alpha_{\rm CO}$ prescription (which includes an emissivity term) to PHANGS--ALMA. They show that including this term steepens the molecular Kennicutt-Schmidt relation from $\Sigma_{\rm SFR} \propto \Sigma_{\rm mol}^{1.0}$ to $\Sigma_{\rm SFR} \propto \Sigma_{\rm mol}^{1.2}$.

It is worth emphasizing the strong evidence for a variable $\alpha_{\rm CO}$, and specifically for emissivity and excitation variations. Both independent dust-based $\alpha_{\rm CO}$ estimates \citep{SANDSTROM13XCO,CHIANG24XCO} and multi-transition, multi-species spectral line modeling \citep{ISRAEL20XCO,TENG22XCO,TENG23XCO,KEENAN24LINES} agree that the CO emission per unit molecular gas mass is higher in inner galaxies. Similarly, numerous observations agree that high $\Sigma_{\rm SFR}$, inner regions of galaxies show enhanced CO excitation \citep{HERACLES09SURVEY,LEROY13SFGAS,LEROY22LINES,DENBROK21LINES,DENBROK23XCO,YAJIMA21LINES}. All of this matches physical expectations and the sense of variations inferred for merging galaxies with similar or even more extreme conditions \citep[see summary in][]{BOLATTO13REVIEW}.

\subsection{Correlation of 
\texorpdfstring{$\epsilon_{\rm ff}^{\rm mol}$}{the efficiency per free-fall time} with cloud-scale gas properties}
\label{sec:effvsprops}

\begin{deluxetable}{cccccc}[t!]
\tabletypesize{\small}
\tablecaption{$\epsilon_{\rm ff}^{\rm mol}$ and cloud-scale gas properties \label{tab:eff}}
\tablewidth{0pt}
\tablehead{
\colhead{$x$} & 
\colhead{Rank. corr.} &
\colhead{$m$} &
\colhead{$x_0$} &
\colhead{$b$} &
\colhead{$\sigma$} 
}
\startdata
\hline
--- &  ---  &  ---  &  ---  &  $ -2.41 $  &  $ 0.29 $  \\ 
$\log_{10} \left< \tau_{\rm ff}^{\rm cloud} \right>$ &  $ 0.25 $  &  $ 0.46 $  &  $ 6.94 $  &  $ -2.37 $  &  $ 0.28 $  \\ 
(normalized) &  $ 0.11 $  &  $ 0.13 $  &  ---  &  --- &  $ 0.18 $  \\ 
\hline
$\log_{10} \left< \Sigma_{\rm mol}^{\rm cloud} \right>$ &  $ -0.16 $  &  $ -0.15 $  &  $ 1.58 $  &  $ -2.37 $  &  $ 0.29 $  \\ 
(normalized) &  $ -0.14 $  &  $ -0.07 $  &  ---  &  --- &  $ 0.18 $  \\ 
\hline
$\log_{10} \left< \sigma_{\rm mol}^{\rm cloud} \right>$ &  $ 0.01 $  &  $ 0.04 $  &  $ 0.74 $  &  $ -2.38 $  &  $ 0.29 $  \\ 
(normalized) &  $ 0.01 $  &  $ -0.01 $  &  ---  &  --- &  $ 0.18 $  \\ 
\hline
$\log_{10} \left< \alpha_{\rm vir}^{\rm cloud} \right>$ &  $ 0.24 $  &  $ 0.21 $  &  $ 0.3 $  &  $ -2.39 $  &  $ 0.28 $  \\ 
(normalized) &  $ 0.11 $  &  $ 0.04 $  &  ---  &  --- &  $ 0.18 $  \\ 
\hline
\hline
\enddata
\tablecomments{Results correlating $\epsilon_{\rm ff}^{\rm mol}$ with cloud-scale gas properties for our fiducial sample of regions with high gas surface density. Cloud scale properties measured at 150~pc resolution, and our best estimate $\alpha_{\rm CO}$. Fit has form $y = m (x - x_0) + b$. Quantities defined in Table \ref{tab:experiment}. Normalized results first normalize $\epsilon_{\rm ff}^{\rm mol}$ by the galaxy-average value.}
\end{deluxetable}

\begin{figure*}
\centering
\includegraphics{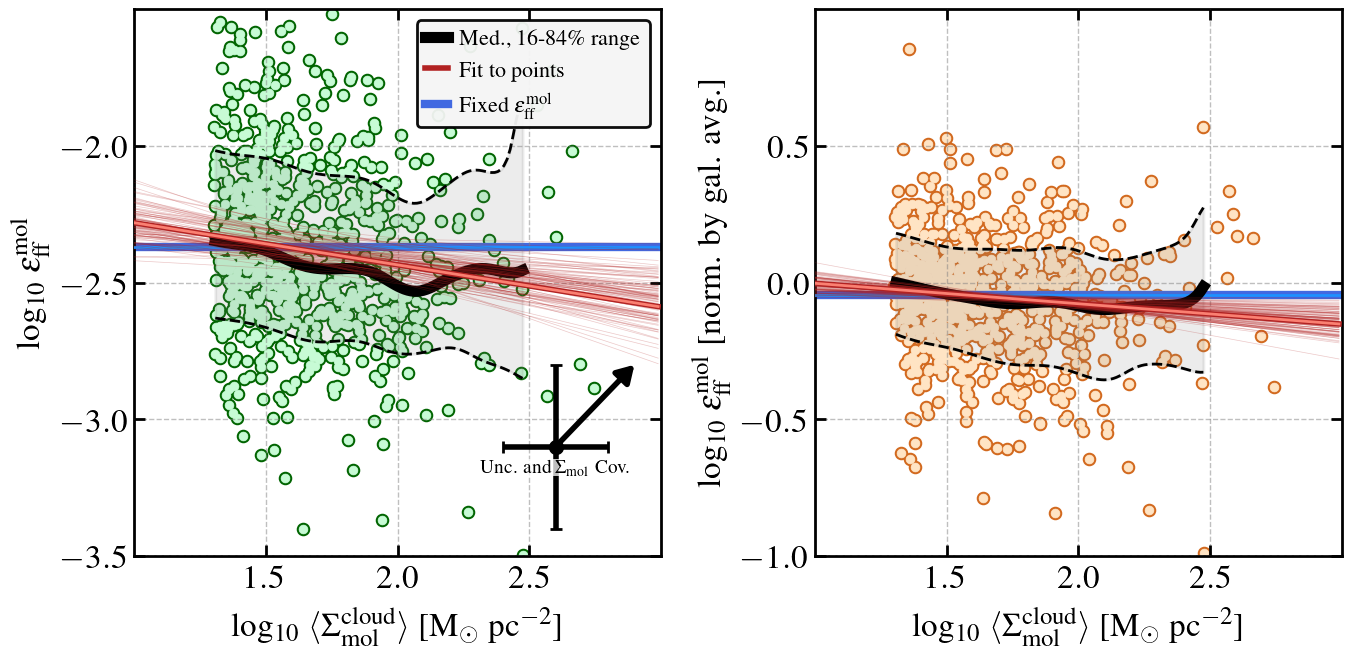}
\includegraphics{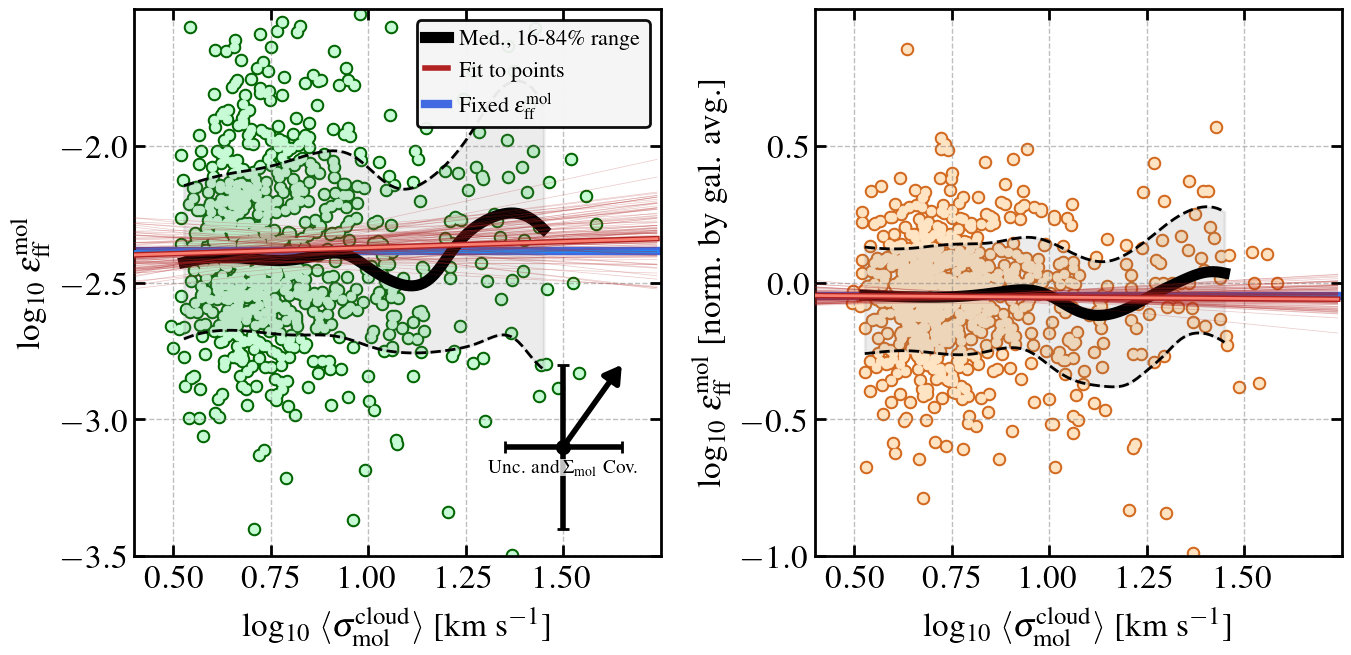}
\caption{Star formation efficiency per free-fall time, \eff\ (Eq.~\ref{eq:eff}), as a function of cloud-scale gas properties, (\textit{top row}) surface density, \sdmol , and (\textit{bottom row}) line width, \sigmol . The points and symbols follow Fig.~\ref{fig:tdepvstff}, \ref{fig:tdep_vs_gasprops}, and \ref{fig:tdep_vs_avir}. Note that by construction the axes here are anti-correlated because \tauff\ is derived from \sdmol\ and is an input to \eff . The correlations between \eff\ and cloud-scale gas properties appear weak or absent (Tab.~\ref{tab:eff}).
\label{fig:eff_vs_gasprops}
}
\end{figure*}

\begin{figure*}
\centering
\includegraphics{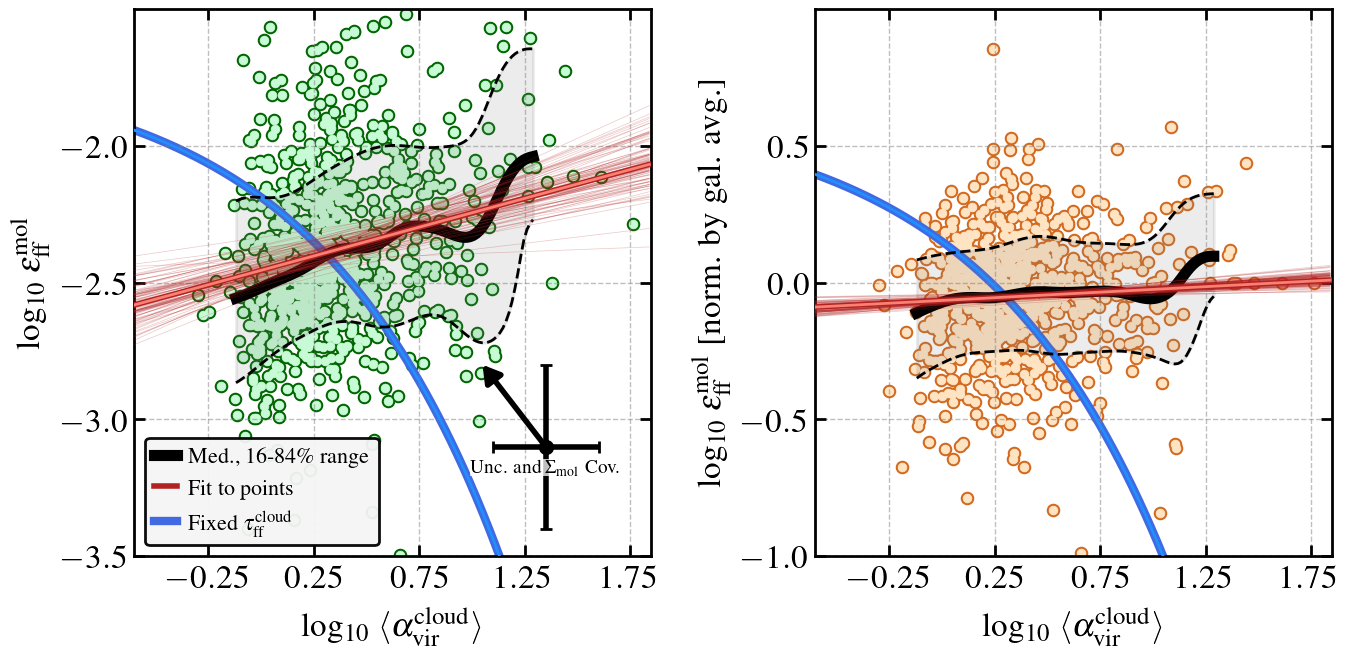}
\caption{Star formation efficiency per free-fall time, \eff\ (Eq.~\ref{eq:eff}), as a function of region-averaged cloud-scale virial parameter \avir . Points and symbols as Fig.~\ref{fig:tdepvstff}, \ref{fig:tdep_vs_gasprops}, and \ref{fig:tdep_vs_avir}. In both panels, we observe \eff\ to weakly correlate with \avir , which goes opposite to the theoretical expectation that more bound gas will form stars more effectively \citep[the curved dashed line shows expectations for fixed $\tauff$ in a simulation of turbulent gas following][]{PADOAN12EFF}. As in Fig.~\ref{fig:eff_vs_gasprops}, the axes are correlated by construction via the dependence of both $\eff$ and $\avir$ on $\sdmol$. The simplest explanation for the observations is that the observed \avir\ reflects motions in the galactic potential rather than probing the fraction of dense, gravitationally bound gas \citep[see][]{MEIDT18GMCS}.
\label{fig:eff_vs_avir}
}
\end{figure*}

We plot our estimated \eff\ as functions of cloud-scale gas properties in Figs.~\ref{fig:eff_vs_gasprops} and \ref{fig:eff_vs_avir} and report results of our correlation analysis in Tab.~\ref{tab:eff}. \citet{PADOAN12EFF}, \citet{FEDERRATH12EFF,FEDERRATH13EFF}, \citet{BURKHART18EFF}, and others all make predictions about how \eff\ should depend on cloud-scale gas properties for star formation in turbulent gas clouds.

Figure~\ref{fig:eff_vs_gasprops} and Tab.~\ref{tab:eff} show almost no relationship between \eff\ and either \sdmol\ or \sigmol, with rank correlation coefficients and slopes both showing very small values. Put in another way, within any reasonable level of uncertainty, a fixed \eff\ with $\pm 0.3$~dex (Table \ref{tab:eff}) scatter appears to be a sufficient description of our present data set. As with the \taudep\ results, this statement depends on our treatment of $\alpha_{\rm CO}$ and data selection, but Fig.~\ref{fig:eff_vs_gasprops} and Tab.~\ref{tab:eff} represent our current best estimates. For reference, the median \eff\ in Table~\ref{tab:eff} corresponds to $0.39\%$ of gas formed into stars per free-fall time, and we measure $\pm 0.29$~dex or a factor of $2$ scatter across the sample. This agrees well with earlier work on PHANGS--ALMA by \citet{UTOMO18EFF} and \citet{SUN23SFGAS}. Indeed, this is almost, though not exactly, the same measurement made in \citet{SUN23SFGAS}.

The dependence of \eff\ on \avir\ has been of interest in the theoretical works mentioned above. These works predict an anti-correlation, such that high \avir\ leads to low \eff . Fig.~\ref{fig:eff_vs_avir} shows this relationship for our data set, with the curved dashed line indicating a simple expectation from \citet{PADOAN12EFF} for the case of fixed \tauff . As with \taudep\ and \avir\ (Fig.~\ref{fig:tdep_vs_avir}), we do not find the predicted theoretical relationship. The data instead appear consistent with a nearly fixed \eff\ over a relatively wide range of \avir , or perhaps even a mild positive correlation between \eff\ and \avir . As discussed in \S \ref{sec:tdepvsavir}, \ref{sec:byrad}, and \S \ref{sec:avirdisc}, this likely reflects that our measured \avir , which only accounts for gas self-gravity, does not capture the true dynamical state of the gas because it neglects stellar gravity.

Finally, we note that though theoretically convenient, treating \eff\ as the dependent variable can be problematic from an observational perspective. This leads to correlated axes, with cloud-scale density or a closely related quantity now entering both axes. And the requirement to detect gas at cloud scales to estimate $\tau_{\rm ff}^{\rm mol}$ (required to calculate $\epsilon_{\rm ff}^{\rm mol}$) means any completeness or surface density cuts will introduce selection effects that affect both the $x$ and $y$ axes, as opposed to mostly only affecting the $x$ axes when considering \taudep . Given that the same physics will manifest in analyses comparing cloud-scale gas properties to \taudep , which already suffer from moderately correlated axes, we recommend those measurements as a cleaner point of comparison.

These correlated axes, along with our fairly narrow dynamic range in cloud properties and selection, explain why the \taudep\ vs. gas property measurements in Table \ref{tab:tdep} do not trivially transform into the \eff\ vs. gas property measurements in Table \ref{tab:eff}. That is, we find almost flat $\eff$ vs. \sdmol\ but the slope of the $\taudep{-}\sdmol$ relation is shallower than the $m=0.5$ expected for fixed \eff\ (and similar for \sigmol ). Again, we recommend to focus on the cleanly reproducible \taudep\ results when comparing to our data (see Appendix \ref{sec:stepbystep}).

\section{Discussion}
\label{sec:discussion}

\subsection{What this measurement is} 

We present a correlation analysis relating cloud-scale gas properties to the molecular gas depletion time across PHANGS--ALMA. We apply a simple selection criteria and use a reproducible methodology that aggregates high-resolution gas properties using a mass-weighting scheme. This yields measurements that average over large enough areas, such that we expect to sample star-forming clouds in all evolutionary stages and access the time-averaged molecular gas depletion time. PHANGS--ALMA does a good job of sampling the massive end of the $z=0$ star-forming main sequence \citep{PHANGSALMA21SURVEY}, so our measurements should represent a definitive view of region-by-region correlations between $\taudep$ and cloud properties for the molecular gas-dominated regions of $z=0$ massive star-forming disk galaxies. 

We highlight our measurements of \taudep\ as a function of \sdmol\ and \sigmol, as these are closest to the empirical data and also capture the key trends. While we provide power-law fits (Tables~\ref{tab:tdep} and \ref{tab:eff}), we encourage direct comparisons to our full measurement set (Table~\ref{tab:data} and Appendix~\ref{sec:resn}) as a more robust approach.

These measurements should be easy to reproduce from observations of other samples of galaxies or numerical simulations. It seems useful to apply this method to numerical simulations that track the time evolution and resolve the small-scale structure of the ISM \citep[e.g.,][]{GRISDALE18,AGERTZ21SIMS,RENAUD21,JEFFRESON22GMCS,JEFFRESON23GMCS,KIM23TIGRESS}. On the one hand, our measurements offer a chance to benchmark such simulations against real disk galaxies. On the other hand, the higher physical detail and access to the time dimension and three dimensional geometry in simulations allow for both a clearer physical interpretation of this type of measurement and the chance to test how different physical prescriptions (e.g., different true \eff , dependence on \avir , dependence on other phases or processes like cloud collisions) would manifest (or not) in our measurements.

\subsection{What this measurement is not} 

We do not measure the initial conditions and final outcomes for individual clouds. Our cloud scale measurements are population averages. These aggregate high-resolution measurements from PHANGS--ALMA but do not distinguish between clouds at different evolutionary stages. Most turbulence-regulated theories of star formation consider how the initial properties of clouds affect the mass of gravitationally bound, collapsing gas \citep{KRUMHOLZ05EFF,PADOAN12EFF,FEDERRATH12EFF,FEDERRATH13EFF}. Using high resolution simulations, \citet{KIM21FEEDBACKSIMS} highlight that feedback from star formation alters the properties of a clouds, including the \avir . Without accounting for the full cloud life cycle, one should not expect the turbulence-regulated models, or any other model focused on initial conditions, to apply rigorously to our measurements.

Indeed, \citet{SEGOVIAOTERO24} demonstrated, using cosmological simulations of galaxy formation, that the underlying ``input'' \eff\ (in their work the \citet[][]{PADOAN12EFF} star formation model was adopted) is challenging to infer observationally. They argued that a mix of cloud evolutionary stages \citep[see also][]{GRISDALE19}, as well as the time lag from the onset of star formation to the resulting tracers of star formation, can wash out correlations with parameters such as \avir\ (see their Figure 4).

Despite this caveat, we emphasize that our presented relationships, their normalization, and their levels of scatter, should still be viewed as benchmarks that numerical simulations should be able to reproduce when our simple methodology is adopted. More, studies of the distribution functions of cloud-scale gas properties in PHANGS, including the gas column distribution function \citep{PATHAK24MIDIR} and beam-by-beam surface density and velocity dispersion measurements \citep{SUN18CLOUDS,SUN20GMCS} find relatively narrow (rms $\sim 0.3{-}0.4$~dex), often lognormal distributions of gas properties in each region. Systematic variations in the mean of these distributions appear mostly correlated with large-scale environment in coherent ways \citep[][and see \S \ref{sec:byrad} and Figures \ref{fig:gasprops_vs_rad} and \ref{fig:tdepbyrad} here]{SUN22CLOUDS}. Given this, we expect that region-to-region variations in the mean cloud-scale gas properties do map fairly directly to variations in the initial conditions for star formation. The impact of feedback seems more likely to represent a correction factor than to completely scramble our results. This would be an excellent topic to investigate further with simulations.

\subsection{Surface density, physical density, and resolution} 
\label{sec:densdisc}

\begin{figure}[t!]
\centering
\includegraphics{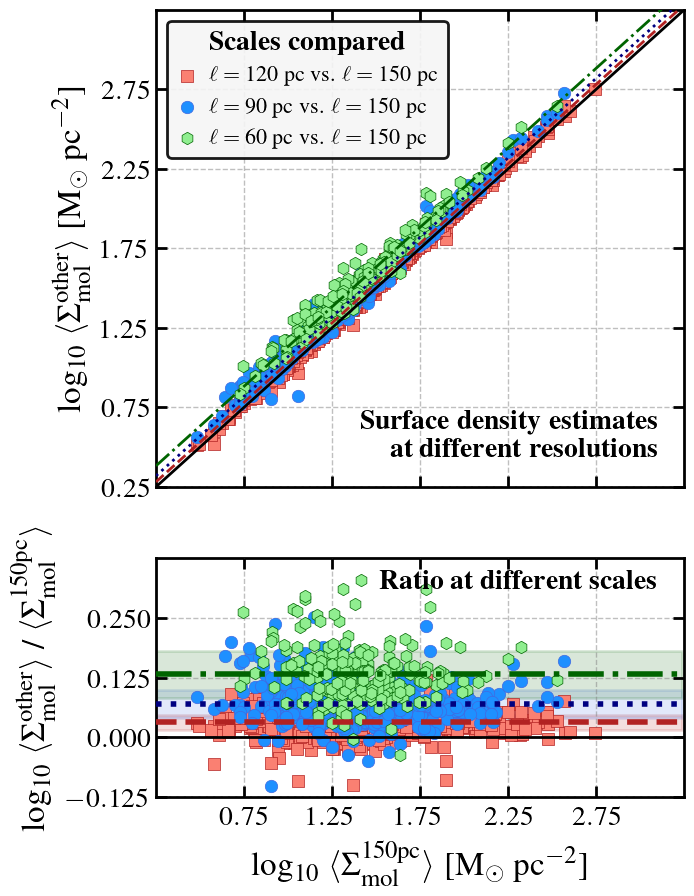}
\caption{Surface density measured at $\ell = 120$, $90$, and $60$~pc resolution compared to our fiducial $\ell = 150$~pc resolution (see also Table \ref{tab:sdresn}). Mass-weighted region-averaged surface densities measured at all four resolutions correlate with one another extremely well, showing high correlation coefficients and little scatter among the ratio of \sdmol\ measured at different scales. As expected, high resolution measurements yield modestly higher surface densities, indicating the presence of beam dilution. But overall the comparison shows no evidence for a preferred scale, and supports that our $\ell = 150$~pc measurements are reasonable indicators of the physical density. Appendix \ref{sec:resn} also shows that the \taudep\ vs. gas properties trends observed at high resolution match our fiducial results.
\label{fig:sdresn}
}
\end{figure}

\begin{deluxetable}{cccc}[t!]
\tabletypesize{\small}
\tablecaption{\sdmol\ and Resolution \label{tab:sdresn}}
\tablewidth{0pt}
\tablehead{
\colhead{Comparison} & 
\colhead{Rank. corr.} &
\colhead{Med. $\log_{10}$ Ratio} &
\colhead{Scatter} \\
\colhead{(vs. $\ell = 150$~pc)} & 
\colhead{} & 
\colhead{(dex)} & 
\colhead{(dex)}
}
\startdata
\hline
$\ell = 120$~pc & $0.998$ & $0.031$ & $0.016$ \\
$\ell = 90$~pc & $0.99$ & $0.069$ & $0.029$ \\
$\ell = 60$~pc & $0.97$ & $0.13$ & $0.050$ \\
\hline
\hline
\enddata
\tablecomments{Correlation between \sdmol\ measured at our fiducial scale, $\ell = 150$~pc resolution (treated as the $x$ variable), and measurements using higher resolution CO~(2-1) data (treated as the $y$ variable) for regions that meet our selection criteria (Fig. \ref{fig:completeness}) Columns report: the nonparameteric rank correlation coefficient between \sdmol\ at the two scales; the median $\log_{10} y/x$ ratio, which captures the amount of beam dilution present; and the scatter in the $\log_{10}$ ratio. See also Fig. \ref{fig:sdresn}.}
\end{deluxetable}

We consider \sdmol\ an observational indicator of the gas volume density and use it to estimate \tauff . For a weakly varying molecular gas scale height \citep[which is the case for the Milky Way;][]{HEYER15REVIEW}, \sdmol\ measures the mean density within the beam. It could be possible that this mean density estimated at 150~pc scales does not indicate the physical density within clouds or the density measured at smaller scales. For example, this could be the case if the molecular ISM consists of otherwise identical small clouds that vary in space density across galaxies \citep[a weak version of this ``universal cloud'' view has been popular, e.g.,][]{BLITZ07GMCS,FUKUI10REVIEW}.

Though our common resolution is $150$~pc, PHANGS--ALMA also represents the largest $120$, $90$, and $60$~pc resolution survey of CO emission from nearby galaxies. In Appendix \ref{sec:resn}, we repeat the analysis comparing \taudep\ to \sdmol\ and \sigmol\ at each of these higher resolutions. Though the numbers of galaxies covered are smaller for these sharper resolutions, they show trends that are consistent with those we observe at 150~pc resolution data (Appendix \ref{sec:resn}).

Figure \ref{fig:sdresn} and Table \ref{tab:sdresn} directly compare \sdmol\ measured at different resolutions in our target regions. The figure shows that the physical resolution has a mild effect on the mass-weighted surface densities, with slightly higher \sdmol\ at higher resolution. This indicates some beam dilution or clumping, which is expected. Beyond the ratio, $\sdmol$ at $150$~pc correlates stunningly well with $\sdmol$ at $120$~pc, $90$~pc, and even $60$~pc resolution. Although many millimeter astronomers hold strongly-felt opinions about the true size of a molecular cloud, as far as we are aware there is no clear evidence for a specific scale where the surface density traces physical density. Instead, our $\ell = 150$~pc data provide an outstanding predictor of the $\ell = 60$~pc \sdmol\ despite $6\times$ difference in beam area. This is in good agreement with previous work examinging the scale dependence of $\Sigma_{\rm mol}$ by \citet{LEROY16GMCS}, \citet{SUN18CLOUDS}, and \citet{SUN20GMCS}.

An important line of evidence reinforcing this view comes from mm-wave spectroscopic tracers of physical density. \cite{GALLAGHER18HCNCLOUD}, \citet{NEUMANN23DENSE}, and \citet{GARCIARODRIGUEZ23DENSE} showed that the HCN/CO ratio measured at $\sim$~kpc scales correlates with \sdmol\ \citep[see also][]{TAFALLA23LINES}. Though the precise density of gas traced by HCN emission is debated, the HCN/CO ratio should be sensitive to physical density \citep[][]{KRUMHOLZ07DENSE,LEROY17DENSE}, and represents one of the most accessible density-sensitive mm-wave line ratios. Reinforcing this view, \citet{JIMENEZ23DENSE} and \citet{STUBER23DENSE} both show that HCN/CO correlates well with the N$_2$H$^+$/CO line ratio, which is widely considered a gold standard tracer of the dense gas fraction in Milky Way studies. We refer the reader to \citet{SCHINNERER24REVIEW} for more discussion.

Together, the consistency of \sdmol -based results across scale and the link between spectroscopic and imaging-based tracers of density give us confidence that \sdmol\ does trace the physical gas density distribution in the beam. The fidelity of the tracer and the correct sub-beam geometry to assume certainly need more study, and these represent important areas for follow up work at higher resolution and simulation-based tests.

\begin{figure*}[t!]
\centering
\includegraphics{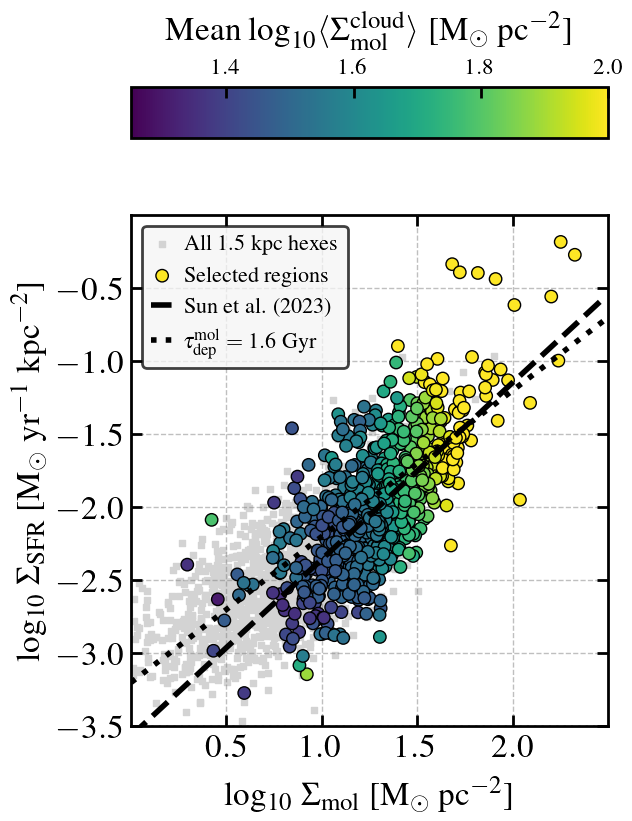}
\includegraphics{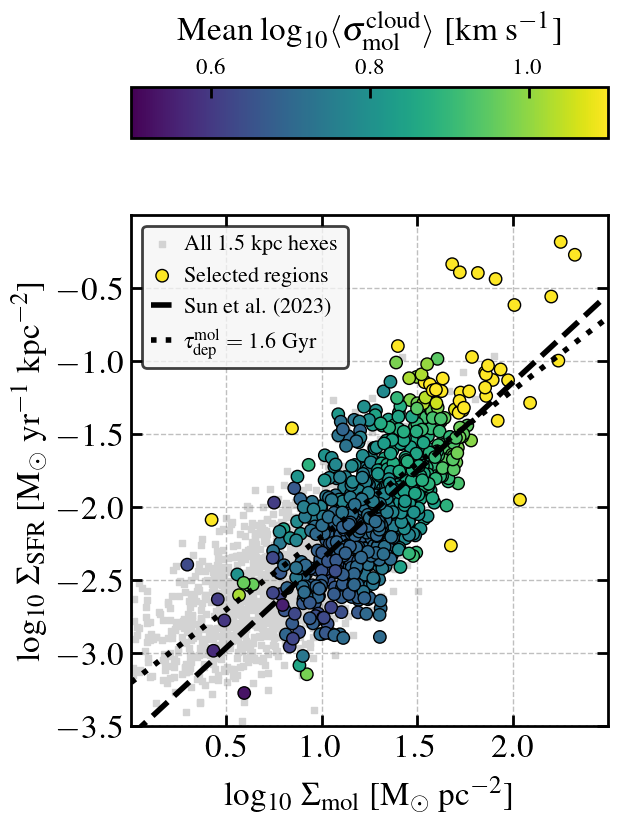}
\caption{The molecular Kennicutt Schmidt relation from Fig.~\ref{fig:scaling} showing only points used in the analysis. Here we color each point by the mean \sdmol\ (left) or \sigmol\ (right) of all points within $\pm 0.1$~dex. The cloud-scale gas properties change in regular ways across $\Sigma_{\rm SFR}^{\rm kpc}{-}\Sigma_{\rm mol}^{\rm kpc}$ space.
\label{fig:turtles}
}
\end{figure*}

We also re-emphasize the point made in \S \ref{sec:byrad}. As demonstrated in \citet{SUN22CLOUDS}, the cloud-scale gas properties vary in a regular way as a  function of $\Sigma_{\rm mol}^{\rm kpc}$, $\Sigma_{\rm SFR}^{\rm kpc}$, the rotation curve and more. Here we observe relationships between \taudep , \sdmol , \sigmol , and \avir , which all then correlate with one another and with these larger-scale conditions. Purely from a data perspective, there is high covariance between these and many other relevant quantities (stellar surface density, the dynamical equilibrium pressure, etc.).  \taudep\ \textit{should} depend on the small scale gas properties, so we expect that the causal relationship likely flows (at least partially) from small to large scales. But a host of hidden variables lurk behind all of these plots.

This also means that as $\Sigma_{\rm mol}^{\rm kpc}$, galactocentric radius, or even global galaxy properties vary, these cloud-scale gas properties will also vary. To illustrate this in a parameter space of broad interest, Fig.~\ref{fig:turtles} shows the $\Sigma_{\rm SFR}^{\rm mol}{-}\Sigma_{\rm mol}^{\rm kpc}$ molecular Kennicutt-Schmidt relation for our target regions (as Fig.~\ref{fig:scaling}) but now coloring by the cloud-scale gas properties. These vary in regular ways across this space, particularly as a function of $\Sigma_{\rm mol}^{\rm kpc}$.

\subsection{The \texorpdfstring{\taudep--\sigmol }{depletion time--velocity dispersion} anti-correlation} 
\label{sec:vdispdisc}

\begin{figure*}[t!]
\centering
\includegraphics{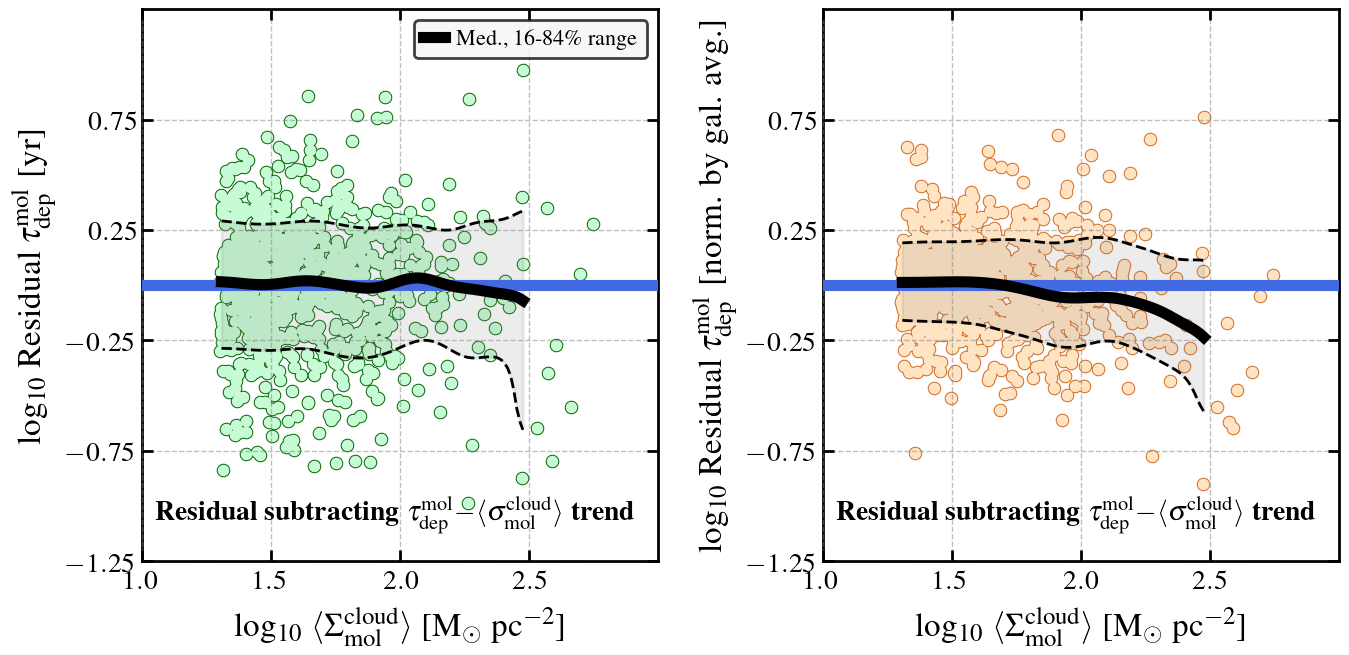}
\caption{Residual molecular gas depletion time, \taudep , as a function of cloud-scale surface density, \sdmol . We calculate the residual by first predicting $\taudep$ from \sigmol . Subtracting (in log space) the trend fit based on line width substantially removes any systematic trend from \taudep\ as a function of surface density. The blue lines indicate a residual of $0$, i.e., no trend. This would be expected if the $\taudep$ vs. \sigmol\ trend expresses the same trend seen for $\taudep$ vs. \sdmol .
\label{fig:vdispresid}
}
\end{figure*}

The anti-correlation between \taudep\ and \sigmol\ mirrors the one relating \taudep\ to \sdmol . As discussed in \S \ref{sec:tdepvsprops}, this anti-correlation would be expected if (1) there is an underlying anti-correlation between density and \taudep\ and (2) clouds show approximately fixed dynamical state. In that case, the line width reflects the potential and so the density of the molecular gas. Then the \taudep--\sdmol\ and \taudep--\sigmol\ relations are manifestations of the same underlying physical correlation. To test this view, Fig.~\ref{fig:vdispresid} shows the residual \taudep\ model as a function of \sdmol\ after subtracting our best-fit \taudep--\sigmol\ relation (Table \ref{tab:tdep}). Except at the very highest \sdmol , most of the \taudep--\sdmol\ trend is removed by subtracting the \taudep--\sigmol\ trend.

This \taudep--\sigmol\ relation is worth more investigation. \sigmol\ has only a weak dependence on $\alpha_{\rm CO}$ (via the weighted average) and may depend less sensitively on resolution than \sdmol . The line width measurement certainly has its own complexities, however. As shown by J.~Henshaw et al.\ (in preparation) studying PHANGS--ALMA, the inner parts of galaxies, which show low \taudep\ also tend to show complex spectral profiles. Our line width measure (effective width) is sensitive to the integrated width of all components, and it will be interesting to explore whether other methods (e.g., second moment, gaussian model fit) show more or less anti-correlation with \taudep . Extending the analysis to well-resolved interacting galaxies, which also tend to show significant spectral complexity \citep[e.g.,][]{BRUNETTI21GMCS,BRUNETTI24GMCS} will provide another interesting counterpoint. Because the gas velocity dispersion may be sensitive to physical conditions even in small, unresolved clouds, this also might offer an interesting direction to study the fainter, smaller clouds in outer galaxy disks.

We have emphasized that the \sigmol\ should trace the gravitational potential and produce the sort of anti-correlation that we see, but the velocity dispersion is often physically interpreted in other ways that may be related to star formation. In the context of turbulence-regulated star formation, the velocity dispersion is often related to the Mach number, which plays an important role in setting the density distribution \citep{PADOAN02TURB,KRUMHOLZ05EFF,FEDERRATH12EFF,FEDERRATH13EFF}. A variety of bulk motions that should influence the line width have also been invoked to explain enhanced or suppressed star formation efficiencies. These include suppression by streaming motions along spiral arms or bars \citep[e.g.,][]{MEIDT13SFGAS}, star formation induced by cloud collisions \citep[e.g.,][]{FUKUI21COLLIDE}, and star formation induced by spiral shocks \citep[see extensive discussion in][]{QUEREJETA24SFGAS}.

Perhaps most importantly, \citet{TENG24XCO} show that the observed CO line width also anti-correlates with $\alpha_{\rm CO}$. They use the \citet{CHIANG24XCO} $\alpha_{\rm CO}$ estimates, which also underpin the emissivity term in the $\alpha_{\rm CO}$ prescription that we use here. \citet{TENG24XCO} argue that the line width serves as a proxy for the inverse of the opacity, and that opacity variations drive a large fraction of the $\alpha_{\rm CO}$ emissivity variations. This agrees well with the spectral line modeling of \citet{TENG22XCO} and \citet{TENG23XCO,DENBROK24XCO}. Since the \taudep\ variations that we observe depend on our adopted $\alpha_{\rm CO}$ (\S \ref{sec:alphaco}, Fig.~\ref{fig:tdep_alphaco}), our measured $\taudep{-}\sigmol$ correlation is formally related to the $\alpha_{\rm CO}{-}\sigmol$ correlation of \citet{TENG24XCO}. It is entirely possible that the line width both traces the gas density (or another physical parameter related to the depletion) \textit{and} drives variations in $\alpha_{\rm CO}$. But at a minimum, the results of \citet{TENG24XCO} are important to bear in mind because they imply a hidden correlation between the \taudep\ and \sigmol\ axes.

\subsection{The lack of a 
\texorpdfstring{\taudep--\avir }{depletion time--virial parameter} correlation}
\label{sec:avirdisc}

Our results disfavor the interpretation that \avir, considering only self-gravity and measured at $60{-}150$~pc scales, indicates the amount of self-gravitating, likely star-forming gas. Instead, the regions that show low \taudep\ also show high \avir\ and tend to occur in the inner parts of galaxies. This agrees with recent results arguing that the broader galactic potential, including significant contribution from stellar gravity, can contribute to the dynamical state of molecular clouds \citep{MEIDT18GMCS,LIU21GMCS}.

In numerical modeling, \citet{SEMENOV17GMCS,SEMENOV18GMCS,SEMENOV21GMCS} find that accounting for the dynamical state of individual parcels of gas is key to reproduce observed \taudep\ and the decorrelation observed between CO and H$\alpha$ at high resolution. \citet{EVANS21GMCS} makes a similar argument for Milky Way clouds. It is tempting to ask whether \avir\ can be salvaged by a more sophisticated formulation that includes information, e.g., on the stellar distribution, rotation curve, or other gas phases. \citet{LIU21GMCS} argue that modeling the large-scale potential, combined with high ($\lesssim 10$~pc) resolution CO imaging may offer a chance to assess the dynamical state on scales that better access the presence of self-gravitating gas. Similarly, \citet{MEIDT18GMCS} argue that extragalactic velocity dispersion measurements can be explained by models that includes the full galactic potential. \citet{SCHRUBA19GMCS} make a similar argument. This may offer a way forward, but lies beyond the scope of this work.

We also note that the $\taudep{-}\avir$ trend almost disappears when the $\taudep{-}\sigmol$ or $\taudep{-}\sdmol$ trends are subtracted from the data, similar to Fig.~\ref{fig:vdispresid}. Given this, an explanation or simulation that matches the other two observed trends will likely capture the impact of dynamical state.

\subsection{Next steps}\label{sec:nextsteps}

As we emphasize above, we consider comparison to numerical simulation a critical next step. PHANGS--ALMA has now measured cloud-scale gas properties across the local galaxy population and found regular patterns and links to \taudep . We have adopted a deliberately simple methodology and tracked completeness, resolution, and other elements needed to make a fair comparison. An ideal next step will be to see how well numerical simulations match the basic distributions in \citet{SUN18CLOUDS,SUN20GMCS} \citep[and soon new dust-based results, e.g.,][]{PATHAK24MIDIR}, the correlations with environment in \citet{SUN22CLOUDS}, and the link to \taudep\ seen here. \citet{DOBBS19GMCS} shows promising results in this direction, finding excellent match between the cloud-scale gas properties measured their dwarf spiral simulation and observations of M33 in \citet{SUN18CLOUDS}. We are eager to see future works comparing more simulations with more representative observational results.

From an observational perspective, ALMA is capable of pushing similar analysis down to $\lesssim 10$~pc resolution over whole galaxies, but this will require a significant commitment. Still, this seems like the qualitative leap in resolution needed to address any ambiguity in the interpretation of \sigmol\ and \avir . Fig.~\ref{fig:sdresn} and Appendix~\ref{sec:resn} suggest that simply increasing the resolution to $\ell \approx 60$~pc is unlikely to yield qualitatively different results. Meanwhile, improving the star formation rate tracer also represents an important next step. We use the best available option for all of PHANGs--ALMA, which is a combination of H$\alpha$ and mid-IR data following \citet{BELFIORE23SFRLARGE}. But this limits the scales over which we can measure \taudep\ and still does not match the quality of an IFU-based H$\alpha$+H$\beta$ estimate \citep{BELFIORE23SFRLARGE}. Coverage of the full sample, e.g., by VLT/MUSE would reduce a lingering source of uncertainty and allow us to average the data in much finer ways, e.g., by constructing averages along likely gas flow lines \citep[e.g.,][]{MEIDT15SFGAS,KODA21SFGAS} in more sharply defined distinct dynamical environments \citep{QUEREJETA21,QUEREJETA24SFGAS}, and so on. They would also allow us to examine statistical distributions of matched resolution SFR and gas tracers rather than relying on region averages for \taudep .

Finally, improved knowledge of $\alpha_{\rm CO}$ remains the single largest uncertainty. The field has made excellent observational and theoretical progress in this area, improving PDR models, building simulations with realistic chemistry and radiative transfer, and constraining line ratio variations, opacity, and expanding estimates based on independent gas tracers like dust. But the uncertainty on $\alpha_{\rm CO}$ in any given environment still dominates the error budget for this entire field.

\section{Summary}
\label{sec:summary}

Using PHANGS--ALMA, we have measured how the molecular gas depletion time, $\taudep$, and the star formation efficiency per free-fall time, $\eff$, correlate with cloud-scale gas properties. Following \citet{SUN22CLOUDS}, we break galaxies into $1.5$~kpc diameter hexagonal regions, calculate the region-mean $\tau_{\rm dep}^{\rm mol}$ and $\epsilon_{\rm ff}^{\rm mol}$, and compare these to the mass-weighted mean molecular gas surface density (\sdmol ), velocity dispersion (\sigmol ), gravitational free-fall time (\tauff ), and virial parameter (\avir ; see Tab.~\ref{tab:experiment}). We measure $150$~pc cloud-scale gas properties, $\tau_{\rm dep}^{\rm mol}$ and $\epsilon_{\rm ff}^{\rm mol}$ for $841$ independent regions in $67$ galaxies. In Appendix~\ref{sec:resn} we also construct higher resolution data sets where we measure the gas properties at $120$, $90$, and $60$~pc resolution. We find substantially similar results varying the resolution, and emphasize that spectroscopic result also indicate that the 150~pc surface density tracks physical density variations across galaxies (\S \ref{sec:densdisc}).

In constructing these measurements, the completeness of the CO data entering the gas property analysis represents an important consideration. We work only with regions where $> 50\%$ of the CO flux is captured by the high-resolution CO map. For PHANGS--ALMA this corresponds approximately to regions with $\sdmol > 20$~M$_\odot$~pc$^{-2}$ (Fig.~\ref{fig:completeness}). This selection yields a set of regions that show good dynamic range in large-scale $\Sigma_{\rm mol}^{\rm kpc}$, $\Sigma_{\rm SFR}^{\rm kpc}$, and cloud-scale gas properties (Fig.~\ref{fig:scaling}). As expected, these regions obey the same large-scale $\Sigma_{\rm SFR}{-}\Sigma_{\rm mol}$ (molecular ``Kennicutt-Schmidt'') and cloud-scale $\sigmol{-}\sdmol$ (``Heyer-Keto'') scalings measured for the overall PHANGS--ALMA data set \citep{SUN20GMCS,SUN22CLOUDS,SUN23SFGAS}.

Analyzing these data, we reach the following main conclusions:

\begin{enumerate}
\item We observe a positive correlation between \taudep\ and $\tauff$ (\S \ref{sec:tdepvstff}, Fig.~\ref{fig:tdepvstff}, Tab.~\ref{tab:tdep}) as expected if density plays an important role in setting the rate at which gas forms stars. This correlation appears stronger after accounting for galaxy-to-galaxy variations in \taudep\ (Figs.~\ref{fig:tdepvstff} and \ref{fig:tdepbygal}). Our best fit line is shallower than the $\taudep \propto \tauff$ relation expected for a fixed efficiency per free fall time, though we emphasize that the fit parameters depend on the adopted methodology. Though \eff\ has been estimated for PHANGS--ALMA before \citep{UTOMO18EFF,SUN22CLOUDS,SUN23SFGAS,SCHINNERER24REVIEW}, this result represents the most direct evidence for an actual correlation between \tauff\ and \taudep\ in our data set.

\item We also observe significant anti-correlations between \taudep\ and \sdmol\ and between \taudep\ and \sigmol\ (Fig.~\ref{fig:tdep_vs_gasprops}, Tab.~\ref{tab:tdep}). Because $\tauff \propto \langle \left(\Sigma_{\rm mol}^{\rm cloud}\right)^{-0.5} \rangle$ and $\sigmol \propto \sdmol^{0.5}$ for clouds with fixed dynamical state (as per Fig.~\ref{fig:scaling}), correlations with this sense could all be expected from a single underlying dependence of \taudep\ on gas volume density. As with \taudep\ and \tauff , the best-fit slopes that we find show shallower slopes than expected for $\taudep \propto \tauff$ and self-gravitating gas. We consider these measurements more empirically robust and the best points for direct comparisons with other observations or simulations.

\item The anti-correlation between \taudep\ and \sigmol\ (Fig.~\ref{fig:eff_vs_gasprops}, Tab.~\ref{tab:tdep}, \S \ref{sec:tdepvsprops}, \ref{sec:vdispdisc}) appears interesting because the $x$-axis does not rely on $\alpha_{\rm CO}$ for the measurement of the cloud-scale gas properties and may be less sensitive to resolution than \sdmol . This may remove some of the built-in correlation present in other scaling relations. We emphasize that this anti-correlation between \taudep\ and \sigmol\ will emerge from clouds with relatively uniform dynamical state with fixed \eff\ regardless of any details related to interstellar turbulence. An important complication is that \sigmol\ may also have a significant impact on $\alpha_{\rm CO}$ \citep{TENG24XCO}, which also affects \taudep .

\item We also observe an anti-correlation between \taudep\ and \avir , and the latter should capture the balance of kinetic and self-gravitational potential energy for the gas (Fig.~\ref{fig:tdep_vs_avir}, Tab.~\ref{fig:tdep_vs_gasprops}). In our measurements, gas that appears less bound by self-gravity forms stars more efficiently, which does not match naive physical expectations. Practically, this reflects that regions with high mean \avir\ often occur in the inner parts of galaxies, where a diffuse, bright molecular medium bound partially by stellar gravity appears common. These regions also appear efficient at forming stars. Both results have significant support from previous studies.

\item Our results depend significantly on the adopted treatment of the CO-to-H$_2$ conversion factor, $\alpha_{\rm CO}$ (Fig.~\ref{fig:tdep_alphaco}, Tab.~\ref{tab:fixaco}, \S \ref{sec:alphaco}). Specifically, the inclusion or omission of a term that accounts for enhanced CO emissivity and excitation in the inner, high surface density parts of galaxies has a large effect on our measurements. When such a term is included, the correlations between \taudep , \tauff , \sdmol , and \sigmol\ discussed above become apparent. When omitting the term, we measure correlations with almost the opposite sense. Including this term represents our best estimate, as these corrections appear necessary based on extensive observational and theoretical work. Despite this previous good work, precise knowledge of $\alpha_{\rm CO}$ almost certainly remains the main obstacle to progress in this field.

\item Recasting our results to consider \eff\ as a function of cloud-scale gas properties, we see almost no correlation between \eff\ and \sdmol\ or \eff\ and \sigmol\ (Fig.~\ref{fig:eff_vs_gasprops}, Tab.~\ref{tab:eff}). Meanwhile, the appear consistent with a nearly fixed \eff over a relatively wide range of \avir, or perhaps even a mild positive correlation between \eff and \avir , in contrast to the theoretically expected anti-correlation for a turbulent cloud of pure gas. Given that these relations induce correlated axes without adding significant information, we suggest to view the measurements comparing \taudep\ to cloud-scale gas properties as more empirically grounded point of comparison.

\item We present quantitative measurements for each region that should be easy to compare to simulations or other samples of galaxies with matched methodology (Table \ref{tab:data}). We implore the community conducting detailed physical simulations of galaxies to attempt such measurements. Applying these methods to cases with well-understood geometry and prescriptions for star formation and stellar feedback represents an important step that will allow us to chart next directions for the field. Appendix \ref{sec:stepbystep} suggests a simple procedure to construct such comparisons.
\end{enumerate}

\noindent These results largely agree with those seen in work on individual galaxies or small samples by \citet{LEROY17SFGAS}, \citet{SCHRUBA19GMCS}, and \citet{QUEREJETA23SFGAS}, but employing a much larger sample.

Finally, a general conclusion from our analysis is that the links between cloud-scale gas properties and $\taudep$ fit naturally into the broader picture expected from simple scaling relations and our knowledge of galaxy structure (Fig.~\ref{fig:tdepbyrad}, \ref{fig:turtles}, Appendix~\ref{sec:resn}). After accounting for $\alpha_{\rm CO}$ variations, $\taudep$ tends to be shorter in high surface density inner regions of galaxies \citep{LEROY13SFGAS,TENG24XCO,SUN23SFGAS}. The cloud-scale \sdmol\ and \sigmol\ and \avir\ also all tend to be higher in these regions, with \sdmol\ and \sigmol\ generally correlating with one another \citep{SUN18CLOUDS,SUN20GMCS,ROSOLOWSKY21GMCS} and the large-scale surface density in the galaxy disk \citep{SUN20PRESS,SUN22CLOUDS}. We thus observe a set of predictable, self-consistent relationships among \taudep , \sdmol , \sigmol, and \tauff .

\section{Acknowledgments}
\label{sec:acknowledgements}

We thank Neal Evans and Davide Elia for pointing out inconsistencies between the data and plots that led to correcting a significant mistake in the manuscript. We thank the anonymous referee for a constructive, timely report.

This work was carried out as part of the PHANGS collaboration.

A.K.L., S.S., and R.C. gratefully acknowledge support from NSF AST AWD 2205628, JWST-GO-02107.009-A, and JWST-GO-03707.001-A. A.K.L. also gratefully acknowledges support by a Humboldt Research Award.

J.S. acknowledges support by the National Aeronautics and Space Administration (NASA) through the NASA Hubble Fellowship grant HST-HF2-51544 awarded by the Space Telescope Science Institute (STScI), which is operated by the Association of Universities for Research in Astronomy, Inc., under contract NAS~5-26555.

M.C.\ gratefully acknowledges funding from the DFG through an Emmy Noether Research Group (grant number CH2137/1-1). COOL Research DAO is a Decentralized Autonomous Organization supporting research in astrophysics aimed at uncovering our cosmic origins.

JG gratefully acknowledges funding via STFC grant ST/Y001133/1 and financial support from the Swiss National Science Foundation (grant no CRSII5\_193826).

HAP acknowledges support from the National Science and Technology Council of Taiwan under grant 110-2112-M-032-020-MY3.

SCOG acknowledges support from the European Research Council via the ERC Synergy Grant ``ECOGAL'' (project ID 855130) and from the German Excellence Strategy via the Heidelberg Cluster of Excellence (EXC 2181 - 390900948) ``STRUCTURES''.

OA acknowledges support from the Knut and Alice Wallenberg Foundation, the Swedish Research Council (grant 2019-04659), and the Swedish National Space Agency (SNSA Dnr 2023-00164).

This paper makes use of the following ALMA data, which have been processed as part of the PHANGS--ALMA CO~(2-1) survey: \\
\noindent ADS/JAO.ALMA\#2012.1.00650.S, \linebreak 
ADS/JAO.ALMA\#2013.1.00803.S, \linebreak 
ADS/JAO.ALMA\#2013.1.01161.S, \linebreak 
ADS/JAO.ALMA\#2015.1.00121.S, \linebreak 
ADS/JAO.ALMA\#2015.1.00782.S, \linebreak 
ADS/JAO.ALMA\#2015.1.00925.S, \linebreak 
ADS/JAO.ALMA\#2015.1.00956.S, \linebreak 
ADS/JAO.ALMA\#2016.1.00386.S, \linebreak 
ADS/JAO.ALMA\#2017.1.00392.S, \linebreak 
ADS/JAO.ALMA\#2017.1.00766.S, \linebreak 
ADS/JAO.ALMA\#2017.1.00886.L, \linebreak 
ADS/JAO.ALMA\#2018.1.01321.S, \linebreak 
ADS/JAO.ALMA\#2018.1.01651.S, \linebreak 
ADS/JAO.ALMA\#2018.A.00062.S, \linebreak 
ADS/JAO.ALMA\#2019.1.01235.S, \linebreak 
ADS/JAO.ALMA\#2019.2.00129.S, \linebreak 
ALMA is a partnership of ESO (representing its member states), NSF (USA), and NINS (Japan), together with NRC (Canada), NSC and ASIAA (Taiwan), and KASI (Republic of Korea), in cooperation with the Republic of Chile. The Joint ALMA Observatory is operated by ESO, AUI/NRAO, and NAOJ. The National Radio Astronomy Observatory is a facility of the National Science Foundation operated under cooperative agreement by Associated Universities, Inc.


\begin{thebibliography}{}
\expandafter\ifx\csname natexlab\endcsname\relax\def\natexlab#1{#1}\fi
\providecommand{\url}[1]{\href{#1}{#1}}
\providecommand{\dodoi}[1]{doi:~\href{http://doi.org/#1}{\nolinkurl{#1}}}
\providecommand{\doeprint}[1]{\href{http://ascl.net/#1}{\nolinkurl{http://ascl.net/#1}}}
\providecommand{\doarXiv}[1]{\href{https://arxiv.org/abs/#1}{\nolinkurl{https://arxiv.org/abs/#1}}}

\bibitem[{{Accurso} {et~al.}(2017){Accurso}, {Saintonge}, {Catinella}, {Cortese}, {Dav{\'e}}, {Dunsheath}, {Genzel}, {Gracia-Carpio}, {Heckman}, {Jimmy}, {Kramer}, {Li}, {Lutz}, {Schiminovich}, {Schuster}, {Sternberg}, {Sturm}, {Tacconi}, {Tran}, \& {Wang}}]{ACCURSO17XCO}
{Accurso}, G., {Saintonge}, A., {Catinella}, B., {et~al.} 2017, \mnras, 470, 4750, \dodoi{10.1093/mnras/stx1556}

\bibitem[{{Agertz} {et~al.}(2021){Agertz}, {Renaud}, {Feltzing}, {Read}, {Ryde}, {Andersson}, {Rey}, {Bensby}, \& {Feuillet}}]{AGERTZ21SIMS}
{Agertz}, O., {Renaud}, F., {Feltzing}, S., {et~al.} 2021, \mnras, 503, 5826, \dodoi{10.1093/mnras/stab322}

\bibitem[{{Anand} {et~al.}(2021){Anand}, {Lee}, {Van Dyk}, {Leroy}, {Rosolowsky}, {Schinnerer}, {Larson}, {Kourkchi}, {Kreckel}, {Scheuermann}, {Rizzi}, {Thilker}, {Tully}, {Bigiel}, {Blanc}, {Boquien}, {Chandar}, {Dale}, {Emsellem}, {Deger}, {Glover}, {Grasha}, {Groves}, {S. Klessen}, {Kruijssen}, {Querejeta}, {S{\'a}nchez-Bl{\'a}zquez}, {Schruba}, {Turner}, {Ubeda}, {Williams}, \& {Whitmore}}]{ANAND21DISTANCES}
{Anand}, G.~S., {Lee}, J.~C., {Van Dyk}, S.~D., {et~al.} 2021, \mnras, 501, 3621, \dodoi{10.1093/mnras/staa3668}

\bibitem[{{Ballesteros-Paredes}(2006)}]{BALLESTEROS06GMCS}
{Ballesteros-Paredes}, J. 2006, \mnras, 372, 443, \dodoi{10.1111/j.1365-2966.2006.10880.x}

\bibitem[{{Belfiore} {et~al.}(2018){Belfiore}, {Maiolino}, {Bundy}, {Masters}, {Bershady}, {Oyarz{\'u}n}, {Lin}, {Cano-Diaz}, {Wake}, {Spindler}, {Thomas}, {Brownstein}, {Drory}, \& {Yan}}]{BELFIORE18RPROF}
{Belfiore}, F., {Maiolino}, R., {Bundy}, K., {et~al.} 2018, \mnras, 477, 3014, \dodoi{10.1093/mnras/sty768}

\bibitem[{{Belfiore} {et~al.}(2023){Belfiore}, {Leroy}, {Sun}, {Barnes}, {Boquien}, {Cao}, {Congiu}, {Dale}, {Egorov}, {Eibensteiner}, {Glover}, {Grasha}, {Groves}, {Klessen}, {Kreckel}, {Neumann}, {Querejeta}, {Sanchez-Blazquez}, {Schinnerer}, \& {Williams}}]{BELFIORE23SFRLARGE}
{Belfiore}, F., {Leroy}, A.~K., {Sun}, J., {et~al.} 2023, \aap, 670, A67, \dodoi{10.1051/0004-6361/202244863}

\bibitem[{{Bertoldi} \& {McKee}(1992)}]{BERTOLDI92GMCS}
{Bertoldi}, F., \& {McKee}, C.~F. 1992, \apj, 395, 140, \dodoi{10.1086/171638}

\bibitem[{{Bigiel} {et~al.}(2008){Bigiel}, {Leroy}, {Walter}, {Brinks}, {de Blok}, {Madore}, \& {Thornley}}]{BIGIEL08SFGAS}
{Bigiel}, F., {Leroy}, A., {Walter}, F., {et~al.} 2008, \aj, 136, 2846, \dodoi{10.1088/0004-6256/136/6/2846}

\bibitem[{{Blitz} {et~al.}(2007){Blitz}, {Fukui}, {Kawamura}, {Leroy}, {Mizuno}, \& {Rosolowsky}}]{BLITZ07GMCS}
{Blitz}, L., {Fukui}, Y., {Kawamura}, A., {et~al.} 2007, in Protostars and Planets V, ed. B.~{Reipurth}, D.~{Jewitt}, \& K.~{Keil}, 81, \dodoi{10.48550/arXiv.astro-ph/0602600}

\bibitem[{{Blitz} \& {Rosolowsky}(2006)}]{BLITZ06PRESS}
{Blitz}, L., \& {Rosolowsky}, E. 2006, \apj, 650, 933, \dodoi{10.1086/505417}

\bibitem[{{Bolatto} {et~al.}(2013){Bolatto}, {Wolfire}, \& {Leroy}}]{BOLATTO13REVIEW}
{Bolatto}, A.~D., {Wolfire}, M., \& {Leroy}, A.~K. 2013, \araa, 51, 207, \dodoi{10.1146/annurev-astro-082812-140944}

\bibitem[{{Brunetti} {et~al.}(2021){Brunetti}, {Wilson}, {Sliwa}, {Schinnerer}, {Aalto}, \& {Peck}}]{BRUNETTI21GMCS}
{Brunetti}, N., {Wilson}, C.~D., {Sliwa}, K., {et~al.} 2021, \mnras, 500, 4730, \dodoi{10.1093/mnras/staa3425}

\bibitem[{{Brunetti} {et~al.}(2024){Brunetti}, {Wilson}, {He}, {Sun}, {Leroy}, {Rosolowsky}, {Bemis}, {Bigiel}, {Groves}, {Saito}, \& {Schinnerer}}]{BRUNETTI24GMCS}
{Brunetti}, N., {Wilson}, C.~D., {He}, H., {et~al.} 2024, \mnras, 530, 597, \dodoi{10.1093/mnras/stae890}

\bibitem[{{Burkhart}(2018)}]{BURKHART18EFF}
{Burkhart}, B. 2018, \apj, 863, 118, \dodoi{10.3847/1538-4357/aad002}

\bibitem[{{Chevance} {et~al.}(2020){Chevance}, {Kruijssen}, {Hygate}, {Schruba}, {Longmore}, {Groves}, {Henshaw}, {Herrera}, {Hughes}, {Jeffreson}, {Lang}, {Leroy}, {Meidt}, {Pety}, {Razza}, {Rosolowsky}, {Schinnerer}, {Bigiel}, {Blanc}, {Emsellem}, {Faesi}, {Glover}, {Haydon}, {Ho}, {Kreckel}, {Lee}, {Liu}, {Querejeta}, {Saito}, {Sun}, {Usero}, \& {Utomo}}]{CHEVANCE20TIMES}
{Chevance}, M., {Kruijssen}, J.~M.~D., {Hygate}, A. P.~S., {et~al.} 2020, \mnras, 493, 2872, \dodoi{10.1093/mnras/stz3525}

\bibitem[{{Chiang} {et~al.}(2024){Chiang}, {Sandstrom}, {Chastenet}, {Bolatto}, {Koch}, {Leroy}, {Sun}, {Teng}, \& {Williams}}]{CHIANG24XCO}
{Chiang}, I.-D., {Sandstrom}, K.~M., {Chastenet}, J., {et~al.} 2024, \apj, 964, 18, \dodoi{10.3847/1538-4357/ad23ed}

\bibitem[{{Davis} {et~al.}(2014){Davis}, {Young}, {Crocker}, {Bureau}, {Blitz}, {Alatalo}, {Emsellem}, {Naab}, {Bayet}, {Bois}, {Bournaud}, {Cappellari}, {Davies}, {de Zeeuw}, {Duc}, {Khochfar}, {Krajnovi{\'c}}, {Kuntschner}, {McDermid}, {Morganti}, {Oosterloo}, {Sarzi}, {Scott}, {Serra}, \& {Weijmans}}]{davis2014}
{Davis}, T.~A., {Young}, L.~M., {Crocker}, A.~F., {et~al.} 2014, \mnras, 444, 3427, \dodoi{10.1093/mnras/stu570}

\bibitem[{{Davis} {et~al.}(2022){Davis}, {Gensior}, {Bureau}, {Cappellari}, {Choi}, {Elford}, {Kruijssen}, {Lelli}, {Liang}, {Liu}, {Ruffa}, {Saito}, {Sarzi}, {Schruba}, \& {Williams}}]{DAVIS22GMCS}
{Davis}, T.~A., {Gensior}, J., {Bureau}, M., {et~al.} 2022, \mnras, 512, 1522, \dodoi{10.1093/mnras/stac600}

\bibitem[{{den Brok} {et~al.}(2024){den Brok}, {Jim{\'e}nez-Donaire}, {Leroy}, {Schinnerer}, {Bigiel}, {Pety}, {Petitpas}, {Usero}, {Teng}, {Humire}, {Koch}, {Rosolowsky}, {Sandstrom}, {Liu}, {Zhang}, {Stuber}, {Chevance}, {Dale}, {Eibensteiner}, {Gali{\'c}}, {Glover}, {Pan}, {Querejeta}, {Smith}, {Williams}, {Wilner}, \& {Zhang}}]{DENBROK24XCO}
{den Brok}, J., {Jim{\'e}nez-Donaire}, M.~J., {Leroy}, A., {et~al.} 2024, arXiv e-prints, arXiv:2410.21399, \dodoi{10.48550/arXiv.2410.21399}

\bibitem[{{den Brok} {et~al.}(2021){den Brok}, {Chatzigiannakis}, {Bigiel}, {Puschnig}, {Barnes}, {Leroy}, {Jim{\'e}nez-Donaire}, {Usero}, {Schinnerer}, {Rosolowsky}, {Faesi}, {Grasha}, {Hughes}, {Kruijssen}, {Liu}, {Neumann}, {Pety}, {Querejeta}, {Saito}, {Schruba}, \& {Stuber}}]{DENBROK21LINES}
{den Brok}, J.~S., {Chatzigiannakis}, D., {Bigiel}, F., {et~al.} 2021, \mnras, 504, 3221, \dodoi{10.1093/mnras/stab859}

\bibitem[{{den Brok} {et~al.}(2023){den Brok}, {Bigiel}, {Chastenet}, {Sandstrom}, {Leroy}, {Usero}, {Schinnerer}, {Rosolowsky}, {Koch}, {Chiang}, {Barnes}, {Puschnig}, {Saito}, {Be{\v{s}}li{\'c}}, {Chevance}, {Dale}, {Eibensteiner}, {Glover}, {Jim{\'e}nez-Donaire}, {Teng}, \& {Williams}}]{DENBROK23XCO}
{den Brok}, J.~S., {Bigiel}, F., {Chastenet}, J., {et~al.} 2023, \aap, 676, A93, \dodoi{10.1051/0004-6361/202245718}

\bibitem[{{Dobbs} {et~al.}(2019){Dobbs}, {Rosolowsky}, {Pettitt}, {Braine}, {Corbelli}, \& {Sun}}]{DOBBS19GMCS}
{Dobbs}, C.~L., {Rosolowsky}, E., {Pettitt}, A.~R., {et~al.} 2019, \mnras, 485, 4997, \dodoi{10.1093/mnras/stz674}

\bibitem[{{Eibensteiner} {et~al.}(2024){Eibensteiner}, {Sun}, {Bigiel}, {Leroy}, {Schinnerer}, {Rosolowsky}, {Kurapati}, {Pisano}, {de Blok}, {Barnes}, {Thorp}, {Colombo}, {Koch}, {Chiang}, {Ostriker}, {Murphy}, {Zabel}, {Laudage}, {Maccagni}, {Healy}, {Sekhar}, {Utomo}, {den Brok}, {Cao}, {Chevance}, {Dale}, {Faesi}, {Glover}, {He}, {Jeffreson}, {Jim{\'e}nez-Donaire}, {Klessen}, {Neumann}, {Pan}, {Pathak}, {Querejeta}, {Teng}, {Usero}, \& {Williams}}]{EIBENSTEINER2024RMOL}
{Eibensteiner}, C., {Sun}, J., {Bigiel}, F., {et~al.} 2024, arXiv e-prints, arXiv:2407.01716, \dodoi{10.48550/arXiv.2407.01716}

\bibitem[{{Ellison} {et~al.}(2018){Ellison}, {S{\'a}nchez}, {Ibarra-Medel}, {Antonio}, {Mendel}, \& {Barrera-Ballesteros}}]{ELLISON18RPROF}
{Ellison}, S.~L., {S{\'a}nchez}, S.~F., {Ibarra-Medel}, H., {et~al.} 2018, \mnras, 474, 2039, \dodoi{10.1093/mnras/stx2882}

\bibitem[{{Evans} {et~al.}(2014){Evans}, {Heiderman}, \& {Vutisalchavakul}}]{EVANS14DENSE}
{Evans}, Neal~J., I., {Heiderman}, A., \& {Vutisalchavakul}, N. 2014, \apj, 782, 114, \dodoi{10.1088/0004-637X/782/2/114}

\bibitem[{{Evans} {et~al.}(2021){Evans}, {Heyer}, {Miville-Desch{\^e}nes}, {Nguyen-Luong}, \& {Merello}}]{EVANS21GMCS}
{Evans}, Neal~J., I., {Heyer}, M., {Miville-Desch{\^e}nes}, M.-A., {Nguyen-Luong}, Q., \& {Merello}, M. 2021, \apj, 920, 126, \dodoi{10.3847/1538-4357/ac1425}

\bibitem[{{Federrath} \& {Klessen}(2012)}]{FEDERRATH12EFF}
{Federrath}, C., \& {Klessen}, R.~S. 2012, \apj, 761, 156, \dodoi{10.1088/0004-637X/761/2/156}

\bibitem[{{Federrath} \& {Klessen}(2013)}]{FEDERRATH13EFF}
---. 2013, \apj, 763, 51, \dodoi{10.1088/0004-637X/763/1/51}

\bibitem[{{Field} {et~al.}(2011){Field}, {Blackman}, \& {Keto}}]{FIELD11GMCS}
{Field}, G.~B., {Blackman}, E.~G., \& {Keto}, E.~R. 2011, \mnras, 416, 710, \dodoi{10.1111/j.1365-2966.2011.19091.x}

\bibitem[{{Fukui} {et~al.}(2021){Fukui}, {Habe}, {Inoue}, {Enokiya}, \& {Tachihara}}]{FUKUI21COLLIDE}
{Fukui}, Y., {Habe}, A., {Inoue}, T., {Enokiya}, R., \& {Tachihara}, K. 2021, \pasj, 73, S1, \dodoi{10.1093/pasj/psaa103}

\bibitem[{{Fukui} \& {Kawamura}(2010)}]{FUKUI10REVIEW}
{Fukui}, Y., \& {Kawamura}, A. 2010, \araa, 48, 547, \dodoi{10.1146/annurev-astro-081309-130854}

\bibitem[{{Gallagher} {et~al.}(2018){Gallagher}, {Leroy}, {Bigiel}, {Cormier}, {Jim{\'e}nez-Donaire}, {Hughes}, {Pety}, {Schinnerer}, {Sun}, {Usero}, {Utomo}, {Bolatto}, {Chevance}, {Faesi}, {Glover}, {Kepley}, {Kruijssen}, {Krumholz}, {Meidt}, {Meier}, {Murphy}, {Querejeta}, {Rosolowsky}, {Saito}, \& {Schruba}}]{GALLAGHER18HCNCLOUD}
{Gallagher}, M.~J., {Leroy}, A.~K., {Bigiel}, F., {et~al.} 2018, \apjl, 868, L38, \dodoi{10.3847/2041-8213/aaf16a}

\bibitem[{{Gao} \& {Solomon}(2004)}]{GAO04DENSE}
{Gao}, Y., \& {Solomon}, P.~M. 2004, \apj, 606, 271, \dodoi{10.1086/382999}

\bibitem[{{Garc{\'\i}a-Burillo} {et~al.}(2012){Garc{\'\i}a-Burillo}, {Usero}, {Alonso-Herrero}, {Graci{\'a}-Carpio}, {Pereira-Santaella}, {Colina}, {Planesas}, \& {Arribas}}]{GARCIABURILLO12DENSE}
{Garc{\'\i}a-Burillo}, S., {Usero}, A., {Alonso-Herrero}, A., {et~al.} 2012, \aap, 539, A8, \dodoi{10.1051/0004-6361/201117838}

\bibitem[{{Garc{\'\i}a-Rodr{\'\i}guez} {et~al.}(2023){Garc{\'\i}a-Rodr{\'\i}guez}, {Usero}, {Leroy}, {Bigiel}, {Jim{\'e}nez-Donaire}, {Liu}, {Querejeta}, {Saito}, {Schinnerer}, {Barnes}, {Belfiore}, {Be{\v{s}}li{\'c}}, {Cao}, {Chevance}, {Dale}, {den Brok}, {Eibensteiner}, {Garc{\'\i}a-Burillo}, {Glover}, {Klessen}, {Pety}, {Puschnig}, {Rosolowsky}, {Sandstrom}, {Sormani}, {Teng}, \& {Williams}}]{GARCIARODRIGUEZ23DENSE}
{Garc{\'\i}a-Rodr{\'\i}guez}, A., {Usero}, A., {Leroy}, A.~K., {et~al.} 2023, \aap, 672, A96, \dodoi{10.1051/0004-6361/202244317}

\bibitem[{{Gong} {et~al.}(2020){Gong}, {Ostriker}, {Kim}, \& {Kim}}]{GONG20XCO}
{Gong}, M., {Ostriker}, E.~C., {Kim}, C.-G., \& {Kim}, J.-G. 2020, \apj, 903, 142, \dodoi{10.3847/1538-4357/abbdab}

\bibitem[{{Grisdale} {et~al.}(2018){Grisdale}, {Agertz}, {Renaud}, \& {Romeo}}]{GRISDALE18}
{Grisdale}, K., {Agertz}, O., {Renaud}, F., \& {Romeo}, A.~B. 2018, \mnras, 479, 3167, \dodoi{10.1093/mnras/sty1595}

\bibitem[{{Grisdale} {et~al.}(2019){Grisdale}, {Agertz}, {Renaud}, {Romeo}, {Devriendt}, \& {Slyz}}]{GRISDALE19}
{Grisdale}, K., {Agertz}, O., {Renaud}, F., {et~al.} 2019, \mnras, 486, 5482, \dodoi{10.1093/mnras/stz1201}

\bibitem[{{Henshaw} {et~al.}(2023){Henshaw}, {Barnes}, {Battersby}, {Ginsburg}, {Sormani}, \& {Walker}}]{henshaw2023}
{Henshaw}, J.~D., {Barnes}, A.~T., {Battersby}, C., {et~al.} 2023, in Astronomical Society of the Pacific Conference Series, Vol. 534, Astronomical Society of the Pacific Conference Series, ed. S.~{Inutsuka}, Y.~{Aikawa}, T.~{Muto}, K.~{Tomida}, \& M.~{Tamura}, 83, \dodoi{10.48550/arXiv.2203.11223}

\bibitem[{{Henshaw} {et~al.}(2016){Henshaw}, {Longmore}, {Kruijssen}, {Davies}, {Bally}, {Barnes}, {Battersby}, {Burton}, {Cunningham}, {Dale}, {Ginsburg}, {Immer}, {Jones}, {Kendrew}, {Mills}, {Molinari}, {Moore}, {Ott}, {Pillai}, {Rathborne}, {Schilke}, {Schmiedeke}, {Testi}, {Walker}, {Walsh}, \& {Zhang}}]{HENSHAW16GMCS}
{Henshaw}, J.~D., {Longmore}, S.~N., {Kruijssen}, J.~M.~D., {et~al.} 2016, \mnras, 457, 2675, \dodoi{10.1093/mnras/stw121}

\bibitem[{{Henshaw} {et~al.}(2020){Henshaw}, {Kruijssen}, {Longmore}, {Riener}, {Leroy}, {Rosolowsky}, {Ginsburg}, {Battersby}, {Chevance}, {Meidt}, {Glover}, {Hughes}, {Kainulainen}, {Klessen}, {Schinnerer}, {Schruba}, {Beuther}, {Bigiel}, {Blanc}, {Emsellem}, {Henning}, {Herrera}, {Koch}, {Pety}, {Ragan}, \& {Sun}}]{HENSHAW20SCOUSE}
{Henshaw}, J.~D., {Kruijssen}, J.~M.~D., {Longmore}, S.~N., {et~al.} 2020, Nature Astronomy, 4, 1064, \dodoi{10.1038/s41550-020-1126-z}

\bibitem[{{Heyer} \& {Dame}(2015)}]{HEYER15REVIEW}
{Heyer}, M., \& {Dame}, T.~M. 2015, \araa, 53, 583, \dodoi{10.1146/annurev-astro-082214-122324}

\bibitem[{{Heyer} {et~al.}(2009){Heyer}, {Krawczyk}, {Duval}, \& {Jackson}}]{HEYER09GMCS}
{Heyer}, M., {Krawczyk}, C., {Duval}, J., \& {Jackson}, J.~M. 2009, \apj, 699, 1092, \dodoi{10.1088/0004-637X/699/2/1092}

\bibitem[{{Heyer} {et~al.}(2001){Heyer}, {Carpenter}, \& {Snell}}]{HEYER01GMCS}
{Heyer}, M.~H., {Carpenter}, J.~M., \& {Snell}, R.~L. 2001, \apj, 551, 852, \dodoi{10.1086/320218}

\bibitem[{{Hirota} {et~al.}(2018){Hirota}, {Egusa}, {Baba}, {Kuno}, {Muraoka}, {Tosaki}, {Miura}, {Nakanishi}, \& {Kawabe}}]{HIROTA18GMCS}
{Hirota}, A., {Egusa}, F., {Baba}, J., {et~al.} 2018, \pasj, 70, 73, \dodoi{10.1093/pasj/psy071}

\bibitem[{{Hopkins} {et~al.}(2011){Hopkins}, {Quataert}, \& {Murray}}]{HOPKINS11SIMS}
{Hopkins}, P.~F., {Quataert}, E., \& {Murray}, N. 2011, \mnras, 417, 950, \dodoi{10.1111/j.1365-2966.2011.19306.x}

\bibitem[{{Hu} {et~al.}(2022){Hu}, {Schruba}, {Sternberg}, \& {van Dishoeck}}]{HU22XCO}
{Hu}, C.-Y., {Schruba}, A., {Sternberg}, A., \& {van Dishoeck}, E.~F. 2022, \apj, 931, 28, \dodoi{10.3847/1538-4357/ac65fd}

\bibitem[{{Hughes} {et~al.}(2013){Hughes}, {Meidt}, {Colombo}, {Schinnerer}, {Pety}, {Leroy}, {Dobbs}, {Garc{\'\i}a-Burillo}, {Thompson}, {Dumas}, {Schuster}, \& {Kramer}}]{HUGHES13GMCS}
{Hughes}, A., {Meidt}, S.~E., {Colombo}, D., {et~al.} 2013, \apj, 779, 46, \dodoi{10.1088/0004-637X/779/1/46}

\bibitem[{{Ib{\'a}{\~n}ez-Mej{\'\i}a} {et~al.}(2016){Ib{\'a}{\~n}ez-Mej{\'\i}a}, {Mac Low}, {Klessen}, \& {Baczynski}}]{IBANEZMEJIA16GMCS}
{Ib{\'a}{\~n}ez-Mej{\'\i}a}, J.~C., {Mac Low}, M.-M., {Klessen}, R.~S., \& {Baczynski}, C. 2016, \apj, 824, 41, \dodoi{10.3847/0004-637X/824/1/41}

\bibitem[{{Israel}(2020)}]{ISRAEL20XCO}
{Israel}, F.~P. 2020, \aap, 635, A131, \dodoi{10.1051/0004-6361/201834198}

\bibitem[{{Jeffreson} {et~al.}(2023){Jeffreson}, {Semenov}, \& {Krumholz}}]{JEFFRESON23GMCS}
{Jeffreson}, S. M.~R., {Semenov}, V.~A., \& {Krumholz}, M.~R. 2023, arXiv e-prints, arXiv:2301.10251, \dodoi{10.48550/arXiv.2301.10251}

\bibitem[{{Jeffreson} {et~al.}(2022){Jeffreson}, {Sun}, \& {Wilson}}]{JEFFRESON22GMCS}
{Jeffreson}, S. M.~R., {Sun}, J., \& {Wilson}, C.~D. 2022, \mnras, 515, 1663, \dodoi{10.1093/mnras/stac1874}

\bibitem[{{Jim{\'e}nez-Donaire} {et~al.}(2019){Jim{\'e}nez-Donaire}, {Bigiel}, {Leroy}, {Usero}, {Cormier}, {Puschnig}, {Gallagher}, {Kepley}, {Bolatto}, {Garc{\'\i}a-Burillo}, {Hughes}, {Kramer}, {Pety}, {Schinnerer}, {Schruba}, {Schuster}, \& {Walter}}]{JIMENEZ19DENSE}
{Jim{\'e}nez-Donaire}, M.~J., {Bigiel}, F., {Leroy}, A.~K., {et~al.} 2019, \apj, 880, 127, \dodoi{10.3847/1538-4357/ab2b95}

\bibitem[{{Jim{\'e}nez-Donaire} {et~al.}(2023){Jim{\'e}nez-Donaire}, {Usero}, {Be{\v{s}}li{\'c}}, {Tafalla}, {Chac{\'o}n-Tanarro}, {Salom{\'e}}, {Eibensteiner}, {Garc{\'\i}a-Rodr{\'\i}guez}, {Hacar}, {Barnes}, {Bigiel}, {Chevance}, {Colombo}, {Dale}, {Davis}, {Glover}, {Kauffmann}, {Klessen}, {Leroy}, {Neumann}, {Pan}, {Pety}, {Querejeta}, {Saito}, {Schinnerer}, {Stuber}, \& {Williams}}]{JIMENEZ23DENSE}
{Jim{\'e}nez-Donaire}, M.~J., {Usero}, A., {Be{\v{s}}li{\'c}}, I., {et~al.} 2023, \aap, 676, L11, \dodoi{10.1051/0004-6361/202347050}

\bibitem[{{Jogee} {et~al.}(2005){Jogee}, {Scoville}, \& {Kenney}}]{JOGEE05SFGAS}
{Jogee}, S., {Scoville}, N., \& {Kenney}, J. D.~P. 2005, \apj, 630, 837, \dodoi{10.1086/432106}

\bibitem[{{Kainulainen} {et~al.}(2009){Kainulainen}, {Beuther}, {Henning}, \& {Plume}}]{KAINULAINEN09DENSE}
{Kainulainen}, J., {Beuther}, H., {Henning}, T., \& {Plume}, R. 2009, \aap, 508, L35, \dodoi{10.1051/0004-6361/200913605}

\bibitem[{{Katz}(1992)}]{KATZ92SIMS}
{Katz}, N. 1992, \apj, 391, 502, \dodoi{10.1086/171366}

\bibitem[{{Keenan} {et~al.}(2024){Keenan}, {Marrone}, \& {Keating}}]{KEENAN24LINES}
{Keenan}, R.~P., {Marrone}, D.~P., \& {Keating}, G.~K. 2024, arXiv e-prints, arXiv:2409.03963, \dodoi{10.48550/arXiv.2409.03963}

\bibitem[{{Kennicutt} \& {Evans}(2012)}]{KENNICUTT12REVIEW}
{Kennicutt}, R.~C., \& {Evans}, N.~J. 2012, \araa, 50, 531, \dodoi{10.1146/annurev-astro-081811-125610}

\bibitem[{{Kim} {et~al.}(2023){Kim}, {Kim}, {Gong}, \& {Ostriker}}]{KIM23TIGRESS}
{Kim}, C.-G., {Kim}, J.-G., {Gong}, M., \& {Ostriker}, E.~C. 2023, \apj, 946, 3, \dodoi{10.3847/1538-4357/acbd3a}

\bibitem[{{Kim} {et~al.}(2022){Kim}, {Chevance}, {Kruijssen}, {Leroy}, {Schruba}, {Barnes}, {Bigiel}, {Blanc}, {Cao}, {Congiu}, {Dale}, {Faesi}, {Glover}, {Grasha}, {Groves}, {Hughes}, {Klessen}, {Kreckel}, {McElroy}, {Pan}, {Pety}, {Querejeta}, {Razza}, {Rosolowsky}, {Saito}, {Schinnerer}, {Sun}, {Tomi{\v{c}}i{\'c}}, {Usero}, \& {Williams}}]{KIM22TIMES}
{Kim}, J., {Chevance}, M., {Kruijssen}, J.~M.~D., {et~al.} 2022, \mnras, 516, 3006, \dodoi{10.1093/mnras/stac2339}

\bibitem[{{Kim} {et~al.}(2021){Kim}, {Ostriker}, \& {Filippova}}]{KIM21FEEDBACKSIMS}
{Kim}, J.-G., {Ostriker}, E.~C., \& {Filippova}, N. 2021, \apj, 911, 128, \dodoi{10.3847/1538-4357/abe934}

\bibitem[{{Kim} {et~al.}(2014){Kim}, {Abel}, {Agertz}, {Bryan}, {Ceverino}, {Christensen}, {Conroy}, {Dekel}, {Gnedin}, {Goldbaum}, {Guedes}, {Hahn}, {Hobbs}, {Hopkins}, {Hummels}, {Iannuzzi}, {Keres}, {Klypin}, {Kravtsov}, {Krumholz}, {Kuhlen}, {Leitner}, {Madau}, {Mayer}, {Moody}, {Nagamine}, {Norman}, {Onorbe}, {O'Shea}, {Pillepich}, {Primack}, {Quinn}, {Read}, {Robertson}, {Rocha}, {Rudd}, {Shen}, {Smith}, {Szalay}, {Teyssier}, {Thompson}, {Todoroki}, {Turk}, {Wadsley}, {Wise}, {Zolotov}, \& {AGORA Collaboration29}}]{KIM14SIMS}
{Kim}, J.-h., {Abel}, T., {Agertz}, O., {et~al.} 2014, \apjs, 210, 14, \dodoi{10.1088/0067-0049/210/1/14}

\bibitem[{{Koch} {et~al.}(2018){Koch}, {Rosolowsky}, \& {Leroy}}]{KOCH18LINEWIDTH}
{Koch}, E., {Rosolowsky}, E., \& {Leroy}, A.~K. 2018, Research Notes of the American Astronomical Society, 2, 220, \dodoi{10.3847/2515-5172/aaf508}

\bibitem[{{Koda}(2021)}]{KODA21SFGAS}
{Koda}, J. 2021, Research Notes of the American Astronomical Society, 5, 222, \dodoi{10.3847/2515-5172/ac2d34}

\bibitem[{{Kormendy} \& {Kennicutt}(2004)}]{KORMENDY04REVIEW}
{Kormendy}, J., \& {Kennicutt}, Robert~C., J. 2004, \araa, 42, 603, \dodoi{10.1146/annurev.astro.42.053102.134024}

\bibitem[{{Kruijssen} {et~al.}(2019){Kruijssen}, {Schruba}, {Chevance}, {Longmore}, {Hygate}, {Haydon}, {McLeod}, {Dalcanton}, {Tacconi}, \& {van Dishoeck}}]{KRUIJSSEN19TIMES}
{Kruijssen}, J.~M.~D., {Schruba}, A., {Chevance}, M., {et~al.} 2019, \nat, 569, 519, \dodoi{10.1038/s41586-019-1194-3}

\bibitem[{{Krumholz}(2014)}]{KRUMHOLZ14REVIEW}
{Krumholz}, M.~R. 2014, \physrep, 539, 49, \dodoi{10.1016/j.physrep.2014.02.001}

\bibitem[{{Krumholz} \& {McKee}(2005)}]{KRUMHOLZ05EFF}
{Krumholz}, M.~R., \& {McKee}, C.~F. 2005, \apj, 630, 250, \dodoi{10.1086/431734}

\bibitem[{{Krumholz} \& {Thompson}(2007)}]{KRUMHOLZ07DENSE}
{Krumholz}, M.~R., \& {Thompson}, T.~A. 2007, \apj, 669, 289, \dodoi{10.1086/521642}

\bibitem[{{Kuno} {et~al.}(2007){Kuno}, {Sato}, {Nakanishi}, {Hirota}, {Tosaki}, {Shioya}, {Sorai}, {Nakai}, {Nishiyama}, \& {Vila-Vilar{\'o}}}]{KUNO07CO}
{Kuno}, N., {Sato}, N., {Nakanishi}, H., {et~al.} 2007, \pasj, 59, 117, \dodoi{10.1093/pasj/59.1.117}

\bibitem[{{Lada} {et~al.}(2012){Lada}, {Forbrich}, {Lombardi}, \& {Alves}}]{LADA12DENSE}
{Lada}, C.~J., {Forbrich}, J., {Lombardi}, M., \& {Alves}, J.~F. 2012, \apj, 745, 190, \dodoi{10.1088/0004-637X/745/2/190}

\bibitem[{{Lada} {et~al.}(2010){Lada}, {Lombardi}, \& {Alves}}]{LADA10DENSE}
{Lada}, C.~J., {Lombardi}, M., \& {Alves}, J.~F. 2010, \apj, 724, 687, \dodoi{10.1088/0004-637X/724/1/687}

\bibitem[{{Lang} {et~al.}(2020){Lang}, {Meidt}, {Rosolowsky}, {Nofech}, {Schinnerer}, {Leroy}, {Emsellem}, {Pessa}, {Glover}, {Groves}, {Hughes}, {Kruijssen}, {Querejeta}, {Schruba}, {Bigiel}, {Blanc}, {Chevance}, {Colombo}, {Faesi}, {Henshaw}, {Herrera}, {Liu}, {Pety}, {Puschnig}, {Saito}, {Sun}, \& {Usero}}]{LANG20KINEMATICS}
{Lang}, P., {Meidt}, S.~E., {Rosolowsky}, E., {et~al.} 2020, \apj, 897, 122, \dodoi{10.3847/1538-4357/ab9953}

\bibitem[{{Leroy} {et~al.}(2008){Leroy}, {Walter}, {Brinks}, {Bigiel}, {de Blok}, {Madore}, \& {Thornley}}]{LEROY08SFGAS}
{Leroy}, A.~K., {Walter}, F., {Brinks}, E., {et~al.} 2008, \aj, 136, 2782, \dodoi{10.1088/0004-6256/136/6/2782}

\bibitem[{{Leroy} {et~al.}(2009){Leroy}, {Walter}, {Bigiel}, {Usero}, {Weiss}, {Brinks}, {de Blok}, {Kennicutt}, {Schuster}, {Kramer}, {Wiesemeyer}, \& {Roussel}}]{HERACLES09SURVEY}
{Leroy}, A.~K., {Walter}, F., {Bigiel}, F., {et~al.} 2009, \aj, 137, 4670, \dodoi{10.1088/0004-6256/137/6/4670}

\bibitem[{{Leroy} {et~al.}(2013){Leroy}, {Walter}, {Sandstrom}, {Schruba}, {Munoz-Mateos}, {Bigiel}, {Bolatto}, {Brinks}, {de Blok}, {Meidt}, {Rix}, {Rosolowsky}, {Schinnerer}, {Schuster}, \& {Usero}}]{LEROY13SFGAS}
{Leroy}, A.~K., {Walter}, F., {Sandstrom}, K., {et~al.} 2013, \aj, 146, 19, \dodoi{10.1088/0004-6256/146/2/19}

\bibitem[{{Leroy} {et~al.}(2016){Leroy}, {Hughes}, {Schruba}, {Rosolowsky}, {Blanc}, {Bolatto}, {Colombo}, {Escala}, {Kramer}, {Kruijssen}, {Meidt}, {Pety}, {Querejeta}, {Sandstrom}, {Schinnerer}, {Sliwa}, \& {Usero}}]{LEROY16GMCS}
{Leroy}, A.~K., {Hughes}, A., {Schruba}, A., {et~al.} 2016, \apj, 831, 16, \dodoi{10.3847/0004-637X/831/1/16}

\bibitem[{{Leroy} {et~al.}(2017{\natexlab{a}}){Leroy}, {Schinnerer}, {Hughes}, {Kruijssen}, {Meidt}, {Schruba}, {Sun}, {Bigiel}, {Aniano}, {Blanc}, {Bolatto}, {Chevance}, {Colombo}, {Gallagher}, {Garcia-Burillo}, {Kramer}, {Querejeta}, {Pety}, {Thompson}, \& {Usero}}]{LEROY17SFGAS}
{Leroy}, A.~K., {Schinnerer}, E., {Hughes}, A., {et~al.} 2017{\natexlab{a}}, \apj, 846, 71, \dodoi{10.3847/1538-4357/aa7fef}

\bibitem[{{Leroy} {et~al.}(2017{\natexlab{b}}){Leroy}, {Usero}, {Schruba}, {Bigiel}, {Kruijssen}, {Kepley}, {Blanc}, {Bolatto}, {Cormier}, {Gallagher}, {Hughes}, {Jim{\'e}nez-Donaire}, {Rosolowsky}, \& {Schinnerer}}]{LEROY17DENSE}
{Leroy}, A.~K., {Usero}, A., {Schruba}, A., {et~al.} 2017{\natexlab{b}}, \apj, 835, 217, \dodoi{10.3847/1538-4357/835/2/217}

\bibitem[{{Leroy} {et~al.}(2019){Leroy}, {Sandstrom}, {Lang}, {Lewis}, {Salim}, {Behrens}, {Chastenet}, {Chiang}, {Gallagher}, {Kessler}, \& {Utomo}}]{LEROY19Z0MGS}
{Leroy}, A.~K., {Sandstrom}, K.~M., {Lang}, D., {et~al.} 2019, \apjs, 244, 24, \dodoi{10.3847/1538-4365/ab3925}

\bibitem[{{Leroy} {et~al.}(2021){Leroy}, {Schinnerer}, {Hughes}, {Rosolowsky}, {Pety}, {Schruba}, {Usero}, {Blanc}, {Chevance}, {Emsellem}, {Faesi}, {Herrera}, {Liu}, {Meidt}, {Querejeta}, {Saito}, {Sandstrom}, {Sun}, {Williams}, {Anand}, {Barnes}, {Behrens}, {Belfiore}, {Benincasa}, {Be{\v{s}}li{\'c}}, {Bigiel}, {Bolatto}, {den Brok}, {Cao}, {Chandar}, {Chastenet}, {Chiang}, {Congiu}, {Dale}, {Deger}, {Eibensteiner}, {Egorov}, {Garc{\'\i}a-Rodr{\'\i}guez}, {Glover}, {Grasha}, {Henshaw}, {Ho}, {Kepley}, {Kim}, {Klessen}, {Kreckel}, {Koch}, {Kruijssen}, {Larson}, {Lee}, {Lopez}, {Machado}, {Mayker}, {McElroy}, {Murphy}, {Ostriker}, {Pan}, {Pessa}, {Puschnig}, {Razza}, {S{\'a}nchez-Bl{\'a}zquez}, {Santoro}, {Sardone}, {Scheuermann}, {Sliwa}, {Sormani}, {Stuber}, {Thilker}, {Turner}, {Utomo}, {Watkins}, \& {Whitmore}}]{PHANGSALMA21SURVEY}
{Leroy}, A.~K., {Schinnerer}, E., {Hughes}, A., {et~al.} 2021, \apjs, 257, 43, \dodoi{10.3847/1538-4365/ac17f3}

\bibitem[{{Leroy} {et~al.}(2022){Leroy}, {Rosolowsky}, {Usero}, {Sandstrom}, {Schinnerer}, {Schruba}, {Bolatto}, {Sun}, {Barnes}, {Belfiore}, {Bigiel}, {den Brok}, {Cao}, {Chiang}, {Chevance}, {Dale}, {Eibensteiner}, {Faesi}, {Glover}, {Hughes}, {Jim{\'e}nez Donaire}, {Klessen}, {Koch}, {Kruijssen}, {Liu}, {Meidt}, {Pan}, {Pety}, {Puschnig}, {Querejeta}, {Saito}, {Sardone}, {Watkins}, {Weiss}, \& {Williams}}]{LEROY22LINES}
{Leroy}, A.~K., {Rosolowsky}, E., {Usero}, A., {et~al.} 2022, \apj, 927, 149, \dodoi{10.3847/1538-4357/ac3490}

\bibitem[{{Lin} {et~al.}(2022){Lin}, {Ellison}, {Pan}, {Thorp}, {Yu}, {Belfiore}, {Hsieh}, {Maiolino}, {Ramya}, {S{\'a}nchez}, \& {Su}}]{LIN22SFGAS}
{Lin}, L., {Ellison}, S.~L., {Pan}, H.-A., {et~al.} 2022, \apj, 926, 175, \dodoi{10.3847/1538-4357/ac4ccc}

\bibitem[{{Liu} {et~al.}(2021){Liu}, {Bureau}, {Blitz}, {Davis}, {Onishi}, {Smith}, {North}, \& {Iguchi}}]{LIU21GMCS}
{Liu}, L., {Bureau}, M., {Blitz}, L., {et~al.} 2021, \mnras, 505, 4048, \dodoi{10.1093/mnras/stab1537}

\bibitem[{{Lu} {et~al.}(2024){Lu}, {Haggard}, {Bureau}, {Gensior}, {Jeffreson}, {Robert}, {Williams}, {Liang}, {Choi}, {Davis}, {Babic}, {Boyce}, {Cheung}, {Drissen}, {Elford}, {Liu}, {Martin}, {Rhea}, {Rousseau-Nepton}, \& {Ruffa}}]{LU24GMCs}
{Lu}, A., {Haggard}, D., {Bureau}, M., {et~al.} 2024, \mnras, 531, 3888, \dodoi{10.1093/mnras/stae1395}

\bibitem[{{Mac Low} \& {Klessen}(2004)}]{MACLOW04REVIEW}
{Mac Low}, M.-M., \& {Klessen}, R.~S. 2004, Reviews of Modern Physics, 76, 125, \dodoi{10.1103/RevModPhys.76.125}

\bibitem[{{Maiolino} \& {Mannucci}(2019)}]{MAIOLINO19REVIEW}
{Maiolino}, R., \& {Mannucci}, F. 2019, \aapr, 27, 3, \dodoi{10.1007/s00159-018-0112-2}

\bibitem[{{Martin} {et~al.}(2005){Martin}, {Fanson}, {Schiminovich}, {Morrissey}, {Friedman}, {Barlow}, {Conrow}, {Grange}, {Jelinsky}, {Milliard}, {Siegmund}, {Bianchi}, {Byun}, {Donas}, {Forster}, {Heckman}, {Lee}, {Madore}, {Malina}, {Neff}, {Rich}, {Small}, {Surber}, {Szalay}, {Welsh}, \& {Wyder}}]{GALEX05}
{Martin}, D.~C., {Fanson}, J., {Schiminovich}, D., {et~al.} 2005, \apjl, 619, L1, \dodoi{10.1086/426387}

\bibitem[{{McKee} \& {Ostriker}(2007)}]{MCKEE07REVIEW}
{McKee}, C.~F., \& {Ostriker}, E.~C. 2007, \araa, 45, 565, \dodoi{10.1146/annurev.astro.45.051806.110602}

\bibitem[{{McKee} \& {Zweibel}(1992)}]{MCKEE92GMCS}
{McKee}, C.~F., \& {Zweibel}, E.~G. 1992, \apj, 399, 551, \dodoi{10.1086/171946}

\bibitem[{{Meidt} {et~al.}(2013){Meidt}, {Schinnerer}, {Garc{\'\i}a-Burillo}, {Hughes}, {Colombo}, {Pety}, {Dobbs}, {Schuster}, {Kramer}, {Leroy}, {Dumas}, \& {Thompson}}]{MEIDT13SFGAS}
{Meidt}, S.~E., {Schinnerer}, E., {Garc{\'\i}a-Burillo}, S., {et~al.} 2013, \apj, 779, 45, \dodoi{10.1088/0004-637X/779/1/45}

\bibitem[{{Meidt} {et~al.}(2015){Meidt}, {Hughes}, {Dobbs}, {Pety}, {Thompson}, {Garc{\'\i}a-Burillo}, {Leroy}, {Schinnerer}, {Colombo}, {Querejeta}, {Kramer}, {Schuster}, \& {Dumas}}]{MEIDT15SFGAS}
{Meidt}, S.~E., {Hughes}, A., {Dobbs}, C.~L., {et~al.} 2015, \apj, 806, 72, \dodoi{10.1088/0004-637X/806/1/72}

\bibitem[{{Meidt} {et~al.}(2018){Meidt}, {Leroy}, {Rosolowsky}, {Kruijssen}, {Schinnerer}, {Schruba}, {Pety}, {Blanc}, {Bigiel}, {Chevance}, {Hughes}, {Querejeta}, \& {Usero}}]{MEIDT18GMCS}
{Meidt}, S.~E., {Leroy}, A.~K., {Rosolowsky}, E., {et~al.} 2018, \apj, 854, 100, \dodoi{10.3847/1538-4357/aaa290}

\bibitem[{{Muraoka} {et~al.}(2019){Muraoka}, {Sorai}, {Miyamoto}, {Yoda}, {Morokuma-Matsui}, {Kobayashi}, {Kuroda}, {Kaneko}, {Kuno}, {Takeuchi}, {Nakanishi}, {Watanabe}, {Tanaka}, {Yasuda}, {Yajima}, {Shibata}, {Salak}, {Espada}, {Matsumoto}, {Noma}, {Kita}, {Komatsuzaki}, {Kajikawa}, {Yashima}, {Pan}, {Oi}, {Seta}, \& {Nakai}}]{MURAOKA19COMING}
{Muraoka}, K., {Sorai}, K., {Miyamoto}, Y., {et~al.} 2019, \pasj, 71, S15, \dodoi{10.1093/pasj/psz015}

\bibitem[{{Neumann} {et~al.}(2023){Neumann}, {Gallagher}, {Bigiel}, {Leroy}, {Barnes}, {Usero}, {den Brok}, {Belfiore}, {Be{\v{s}}li{\'c}}, {Cao}, {Chevance}, {Dale}, {Eibensteiner}, {Glover}, {Grasha}, {Henshaw}, {Jim{\'e}nez-Donaire}, {Klessen}, {Kruijssen}, {Liu}, {Meidt}, {Pety}, {Puschnig}, {Querejeta}, {Rosolowsky}, {Schinnerer}, {Schruba}, {Sormani}, {Sun}, {Teng}, \& {Williams}}]{NEUMANN23DENSE}
{Neumann}, L., {Gallagher}, M.~J., {Bigiel}, F., {et~al.} 2023, \mnras, 521, 3348, \dodoi{10.1093/mnras/stad424}

\bibitem[{{Padoan} {et~al.}(2012){Padoan}, {Haugb{\o}lle}, \& {Nordlund}}]{PADOAN12EFF}
{Padoan}, P., {Haugb{\o}lle}, T., \& {Nordlund}, {\r{A}}. 2012, \apjl, 759, L27, \dodoi{10.1088/2041-8205/759/2/L27}

\bibitem[{{Padoan} \& {Nordlund}(2002)}]{PADOAN02TURB}
{Padoan}, P., \& {Nordlund}, {\r{A}}. 2002, \apj, 576, 870, \dodoi{10.1086/341790}

\bibitem[{{Padoan} \& {Nordlund}(2011)}]{PADOAN11EFF}
---. 2011, \apj, 730, 40, \dodoi{10.1088/0004-637X/730/1/40}

\bibitem[{{Pan} {et~al.}(2022){Pan}, {Schinnerer}, {Hughes}, {Leroy}, {Groves}, {Barnes}, {Belfiore}, {Bigiel}, {Blanc}, {Cao}, {Chevance}, {Congiu}, {Dale}, {Eibensteiner}, {Emsellem}, {Faesi}, {Glover}, {Grasha}, {Herrera}, {Ho}, {Klessen}, {Kruijssen}, {Lang}, {Liu}, {McElroy}, {Meidt}, {Murphy}, {Pety}, {Querejeta}, {Razza}, {Rosolowsky}, {Saito}, {Santoro}, {Schruba}, {Sun}, {Tomi{\v{c}}i{\'c}}, {Usero}, {Utomo}, \& {Williams}}]{PAN22TIMES}
{Pan}, H.-A., {Schinnerer}, E., {Hughes}, A., {et~al.} 2022, \apj, 927, 9, \dodoi{10.3847/1538-4357/ac474f}

\bibitem[{{Pan} {et~al.}(2024){Pan}, {Lin}, {Ellison}, {Thorp}, {S{\'a}nchez}, {Bluck}, {Belfiore}, {Piotrowska}, {Scudder}, \& {Baker}}]{PAN24SFGAS}
{Pan}, H.-A., {Lin}, L., {Ellison}, S.~L., {et~al.} 2024, \apj, 964, 120, \dodoi{10.3847/1538-4357/ad28c1}

\bibitem[{{Pathak} {et~al.}(2024){Pathak}, {Leroy}, {Thompson}, {Lopez}, {Belfiore}, {Boquien}, {Dale}, {Glover}, {Klessen}, {Koch}, {Rosolowsky}, {Sandstrom}, {Schinnerer}, {Smith}, {Sun}, {Sutter}, {Williams}, {Bigiel}, {Cao}, {Chastenet}, {Chevance}, {Chown}, {Emsellem}, {Faesi}, {Larson}, {Lee}, {Meidt}, {Ostriker}, {Ramambason}, {Sarbadhicary}, \& {Thilker}}]{PATHAK24MIDIR}
{Pathak}, D., {Leroy}, A.~K., {Thompson}, T.~A., {et~al.} 2024, \aj, 167, 39, \dodoi{10.3847/1538-3881/ad110d}

\bibitem[{{Querejeta} {et~al.}(2021){Querejeta}, {Schinnerer}, {Meidt}, {Sun}, {Leroy}, {Emsellem}, {Klessen}, {Mu{\~n}oz-Mateos}, {Salo}, {Laurikainen}, {Be{\v{s}}li{\'c}}, {Blanc}, {Chevance}, {Dale}, {Eibensteiner}, {Faesi}, {Garc{\'\i}a-Rodr{\'\i}guez}, {Glover}, {Grasha}, {Henshaw}, {Herrera}, {Hughes}, {Kreckel}, {Kruijssen}, {Liu}, {Murphy}, {Pan}, {Pety}, {Razza}, {Rosolowsky}, {Saito}, {Schruba}, {Usero}, {Watkins}, \& {Williams}}]{QUEREJETA21}
{Querejeta}, M., {Schinnerer}, E., {Meidt}, S., {et~al.} 2021, \aap, 656, A133, \dodoi{10.1051/0004-6361/202140695}

\bibitem[{{Querejeta} {et~al.}(2023){Querejeta}, {Pety}, {Schruba}, {Leroy}, {Herrera}, {Chiang}, {Meidt}, {Rosolowsky}, {Schinnerer}, {Schuster}, {Sun}, {Herrmann}, {Barnes}, {Be{\v{s}}li{\'c}}, {Bigiel}, {Cao}, {Chevance}, {Eibensteiner}, {Emsellem}, {Faesi}, {Hughes}, {Kim}, {Klessen}, {Kreckel}, {Kruijssen}, {Liu}, {Neumayer}, {Pan}, {Saito}, {Sandstrom}, {Teng}, {Usero}, {Williams}, \& {Zakardjian}}]{QUEREJETA23SFGAS}
{Querejeta}, M., {Pety}, J., {Schruba}, A., {et~al.} 2023, \aap, 680, A4, \dodoi{10.1051/0004-6361/202143023}

\bibitem[{{Querejeta} {et~al.}(2024){Querejeta}, {Leroy}, {Meidt}, {Schinnerer}, {Belfiore}, {Emsellem}, {Klessen}, {Sun}, {Sormani}, {Be{\v{s}}lic}, {Cao}, {Chevance}, {Colombo}, {Dale}, {Garc{\'\i}a-Burillo}, {Glover}, {Grasha}, {Groves}, {Koch}, {Neumann}, {Pan}, {Pessa}, {Pety}, {Pinna}, {Ramambason}, {Razza}, {Romanelli}, {Rosolowsky}, {Ruiz-Garc{\'\i}a}, {S{\'a}nchez-Bl{\'a}zquez}, {Smith}, {Stuber}, {Ubeda}, {Usero}, \& {Williams}}]{QUEREJETA24SFGAS}
{Querejeta}, M., {Leroy}, A.~K., {Meidt}, S.~E., {et~al.} 2024, arXiv e-prints, arXiv:2405.05364, \dodoi{10.48550/arXiv.2405.05364}

\bibitem[{{Regan} {et~al.}(2001){Regan}, {Thornley}, {Helfer}, {Sheth}, {Wong}, {Vogel}, {Blitz}, \& {Bock}}]{REGAN01SFGAS}
{Regan}, M.~W., {Thornley}, M.~D., {Helfer}, T.~T., {et~al.} 2001, \apj, 561, 218, \dodoi{10.1086/323221}

\bibitem[{{Renaud} {et~al.}(2021){Renaud}, {Romeo}, \& {Agertz}}]{RENAUD21}
{Renaud}, F., {Romeo}, A.~B., \& {Agertz}, O. 2021, \mnras, 508, 352, \dodoi{10.1093/mnras/stab2604}

\bibitem[{{Rosolowsky} {et~al.}(2021){Rosolowsky}, {Hughes}, {Leroy}, {Sun}, {Querejeta}, {Schruba}, {Usero}, {Herrera}, {Liu}, {Pety}, {Saito}, {Be{\v{s}}li{\'c}}, {Bigiel}, {Blanc}, {Chevance}, {Dale}, {Deger}, {Faesi}, {Glover}, {Henshaw}, {Klessen}, {Kruijssen}, {Larson}, {Lee}, {Meidt}, {Mok}, {Schinnerer}, {Thilker}, \& {Williams}}]{ROSOLOWSKY21GMCS}
{Rosolowsky}, E., {Hughes}, A., {Leroy}, A.~K., {et~al.} 2021, \mnras, 502, 1218, \dodoi{10.1093/mnras/stab085}

\bibitem[{{Saintonge} \& {Catinella}(2022)}]{SAINTONGE22REVIEW}
{Saintonge}, A., \& {Catinella}, B. 2022, \araa, 60, 319, \dodoi{10.1146/annurev-astro-021022-043545}

\bibitem[{{Saintonge} {et~al.}(2011){Saintonge}, {Kauffmann}, {Wang}, {Kramer}, {Tacconi}, {Buchbender}, {Catinella}, {Graci{\'a}-Carpio}, {Cortese}, {Fabello}, {Fu}, {Genzel}, {Giovanelli}, {Guo}, {Haynes}, {Heckman}, {Krumholz}, {Lemonias}, {Li}, {Moran}, {Rodriguez-Fernandez}, {Schiminovich}, {Schuster}, \& {Sievers}}]{SAINTONGE11SFGAS}
{Saintonge}, A., {Kauffmann}, G., {Wang}, J., {et~al.} 2011, \mnras, 415, 61, \dodoi{10.1111/j.1365-2966.2011.18823.x}

\bibitem[{{Saintonge} {et~al.}(2017){Saintonge}, {Catinella}, {Tacconi}, {Kauffmann}, {Genzel}, {Cortese}, {Dav{\'e}}, {Fletcher}, {Graci{\'a}-Carpio}, {Kramer}, {Heckman}, {Janowiecki}, {Lutz}, {Rosario}, {Schiminovich}, {Schuster}, {Wang}, {Wuyts}, {Borthakur}, {Lamperti}, \& {Roberts-Borsani}}]{SAINTONGE17SFGAS}
{Saintonge}, A., {Catinella}, B., {Tacconi}, L.~J., {et~al.} 2017, \apjs, 233, 22, \dodoi{10.3847/1538-4365/aa97e0}

\bibitem[{{Salim} {et~al.}(2016){Salim}, {Lee}, {Janowiecki}, {da Cunha}, {Dickinson}, {Boquien}, {Burgarella}, {Salzer}, \& {Charlot}}]{GSWLC16}
{Salim}, S., {Lee}, J.~C., {Janowiecki}, S., {et~al.} 2016, \apjs, 227, 2, \dodoi{10.3847/0067-0049/227/1/2}

\bibitem[{{Sandstrom} {et~al.}(2013){Sandstrom}, {Leroy}, {Walter}, {Bolatto}, {Croxall}, {Draine}, {Wilson}, {Wolfire}, {Calzetti}, {Kennicutt}, {Aniano}, {Donovan Meyer}, {Usero}, {Bigiel}, {Brinks}, {de Blok}, {Crocker}, {Dale}, {Engelbracht}, {Galametz}, {Groves}, {Hunt}, {Koda}, {Kreckel}, {Linz}, {Meidt}, {Pellegrini}, {Rix}, {Roussel}, {Schinnerer}, {Schruba}, {Schuster}, {Skibba}, {van der Laan}, {Appleton}, {Armus}, {Brandl}, {Gordon}, {Hinz}, {Krause}, {Montiel}, {Sauvage}, {Schmiedeke}, {Smith}, \& {Vigroux}}]{SANDSTROM13XCO}
{Sandstrom}, K.~M., {Leroy}, A.~K., {Walter}, F., {et~al.} 2013, \apj, 777, 5, \dodoi{10.1088/0004-637X/777/1/5}

\bibitem[{{Schinnerer} \& {Leroy}(2024)}]{SCHINNERER24REVIEW}
{Schinnerer}, E., \& {Leroy}, A.~K. 2024, arXiv e-prints, arXiv:2403.19843, \dodoi{10.48550/arXiv.2403.19843}

\bibitem[{{Schinnerer} {et~al.}(2019){Schinnerer}, {Hughes}, {Leroy}, {Groves}, {Blanc}, {Kreckel}, {Bigiel}, {Chevance}, {Dale}, {Emsellem}, {Faesi}, {Glover}, {Grasha}, {Henshaw}, {Hygate}, {Kruijssen}, {Meidt}, {Pety}, {Querejeta}, {Rosolowsky}, {Saito}, {Schruba}, {Sun}, \& {Utomo}}]{SCHINNERER19TIMES}
{Schinnerer}, E., {Hughes}, A., {Leroy}, A., {et~al.} 2019, \apj, 887, 49, \dodoi{10.3847/1538-4357/ab50c2}

\bibitem[{{Schruba} {et~al.}(2019){Schruba}, {Kruijssen}, \& {Leroy}}]{SCHRUBA19GMCS}
{Schruba}, A., {Kruijssen}, J.~M.~D., \& {Leroy}, A.~K. 2019, \apj, 883, 2, \dodoi{10.3847/1538-4357/ab3a43}

\bibitem[{{Schruba} {et~al.}(2010){Schruba}, {Leroy}, {Walter}, {Sandstrom}, \& {Rosolowsky}}]{SCHRUBA10SFGAS}
{Schruba}, A., {Leroy}, A.~K., {Walter}, F., {Sandstrom}, K., \& {Rosolowsky}, E. 2010, \apj, 722, 1699, \dodoi{10.1088/0004-637X/722/2/1699}

\bibitem[{{Schruba} {et~al.}(2011){Schruba}, {Leroy}, {Walter}, {Bigiel}, {Brinks}, {de Blok}, {Dumas}, {Kramer}, {Rosolowsky}, {Sandstrom}, {Schuster}, {Usero}, {Weiss}, \& {Wiesemeyer}}]{SCHRUBA11SFGAS}
{Schruba}, A., {Leroy}, A.~K., {Walter}, F., {et~al.} 2011, \aj, 142, 37, \dodoi{10.1088/0004-6256/142/2/37}

\bibitem[{{Schruba} {et~al.}(2012){Schruba}, {Leroy}, {Walter}, {Bigiel}, {Brinks}, {de Blok}, {Kramer}, {Rosolowsky}, {Sandstrom}, {Schuster}, {Usero}, {Weiss}, \& {Wiesemeyer}}]{SCHRUBA12XCO}
---. 2012, \aj, 143, 138, \dodoi{10.1088/0004-6256/143/6/138}

\bibitem[{{Segovia Otero} {et~al.}(2024){Segovia Otero}, {Agertz}, {Renaud}, {Kraljic}, {Romeo}, \& {Semenov}}]{SEGOVIAOTERO24}
{Segovia Otero}, {\'A}., {Agertz}, O., {Renaud}, F., {et~al.} 2024, arXiv e-prints, arXiv:2410.08266, \dodoi{10.48550/arXiv.2410.08266}

\bibitem[{{Semenov} {et~al.}(2017){Semenov}, {Kravtsov}, \& {Gnedin}}]{SEMENOV17GMCS}
{Semenov}, V.~A., {Kravtsov}, A.~V., \& {Gnedin}, N.~Y. 2017, \apj, 845, 133, \dodoi{10.3847/1538-4357/aa8096}

\bibitem[{{Semenov} {et~al.}(2018){Semenov}, {Kravtsov}, \& {Gnedin}}]{SEMENOV18GMCS}
---. 2018, \apj, 861, 4, \dodoi{10.3847/1538-4357/aac6eb}

\bibitem[{{Semenov} {et~al.}(2021){Semenov}, {Kravtsov}, \& {Gnedin}}]{SEMENOV21GMCS}
---. 2021, \apj, 918, 13, \dodoi{10.3847/1538-4357/ac0a77}

\bibitem[{{Sheth} {et~al.}(2010){Sheth}, {Regan}, {Hinz}, {Gil de Paz}, {Men{\'e}ndez-Delmestre}, {Mu{\~n}oz-Mateos}, {Seibert}, {Kim}, {Laurikainen}, {Salo}, {Gadotti}, {Laine}, {Mizusawa}, {Armus}, {Athanassoula}, {Bosma}, {Buta}, {Capak}, {Jarrett}, {Elmegreen}, {Elmegreen}, {Knapen}, {Koda}, {Helou}, {Ho}, {Madore}, {Masters}, {Mobasher}, {Ogle}, {Peng}, {Schinnerer}, {Surace}, {Zaritsky}, {Comer{\'o}n}, {de Swardt}, {Meidt}, {Kasliwal}, \& {Aravena}}]{S4GSURVEY}
{Sheth}, K., {Regan}, M., {Hinz}, J.~L., {et~al.} 2010, \pasp, 122, 1397, \dodoi{10.1086/657638}

\bibitem[{{Stuber} {et~al.}(2023){Stuber}, {Pety}, {Schinnerer}, {Bigiel}, {Usero}, {Be{\v{s}}li{\'c}}, {Querejeta}, {Jim{\'e}nez-Donaire}, {Leroy}, {den Brok}, {Neumann}, {Eibensteiner}, {Teng}, {Barnes}, {Chevance}, {Colombo}, {Dale}, {Glover}, {Liu}, \& {Pan}}]{STUBER23DENSE}
{Stuber}, S.~K., {Pety}, J., {Schinnerer}, E., {et~al.} 2023, \aap, 680, L20, \dodoi{10.1051/0004-6361/202348205}

\bibitem[{{Sun} {et~al.}(2018){Sun}, {Leroy}, {Schruba}, {Rosolowsky}, {Hughes}, {Kruijssen}, {Meidt}, {Schinnerer}, {Blanc}, {Bigiel}, {Bolatto}, {Chevance}, {Groves}, {Herrera}, {Hygate}, {Pety}, {Querejeta}, {Usero}, \& {Utomo}}]{SUN18CLOUDS}
{Sun}, J., {Leroy}, A.~K., {Schruba}, A., {et~al.} 2018, \apj, 860, 172, \dodoi{10.3847/1538-4357/aac326}

\bibitem[{{Sun} {et~al.}(2020{\natexlab{a}}){Sun}, {Leroy}, {Ostriker}, {Hughes}, {Rosolowsky}, {Schruba}, {Schinnerer}, {Blanc}, {Faesi}, {Kruijssen}, {Meidt}, {Utomo}, {Bigiel}, {Bolatto}, {Chevance}, {Chiang}, {Dale}, {Emsellem}, {Glover}, {Grasha}, {Henshaw}, {Herrera}, {Jimenez-Donaire}, {Lee}, {Pety}, {Querejeta}, {Saito}, {Sandstrom}, \& {Usero}}]{SUN20PRESS}
{Sun}, J., {Leroy}, A.~K., {Ostriker}, E.~C., {et~al.} 2020{\natexlab{a}}, \apj, 892, 148, \dodoi{10.3847/1538-4357/ab781c}

\bibitem[{{Sun} {et~al.}(2020{\natexlab{b}}){Sun}, {Leroy}, {Schinnerer}, {Hughes}, {Rosolowsky}, {Querejeta}, {Schruba}, {Liu}, {Saito}, {Herrera}, {Faesi}, {Usero}, {Pety}, {Kruijssen}, {Ostriker}, {Bigiel}, {Blanc}, {Bolatto}, {Boquien}, {Chevance}, {Dale}, {Deger}, {Emsellem}, {Glover}, {Grasha}, {Groves}, {Henshaw}, {Jimenez-Donaire}, {Kim}, {Klessen}, {Kreckel}, {Lee}, {Meidt}, {Sandstrom}, {Sardone}, {Utomo}, \& {Williams}}]{SUN20GMCS}
{Sun}, J., {Leroy}, A.~K., {Schinnerer}, E., {et~al.} 2020{\natexlab{b}}, \apjl, 901, L8, \dodoi{10.3847/2041-8213/abb3be}

\bibitem[{{Sun} {et~al.}(2022){Sun}, {Leroy}, {Rosolowsky}, {Hughes}, {Schinnerer}, {Schruba}, {Koch}, {Blanc}, {Chiang}, {Groves}, {Liu}, {Meidt}, {Pan}, {Pety}, {Querejeta}, {Saito}, {Sandstrom}, {Sardone}, {Usero}, {Utomo}, {Williams}, {Barnes}, {Benincasa}, {Bigiel}, {Bolatto}, {Boquien}, {Chevance}, {Dale}, {Deger}, {Emsellem}, {Glover}, {Grasha}, {Henshaw}, {Klessen}, {Kreckel}, {Kruijssen}, {Ostriker}, \& {Thilker}}]{SUN22CLOUDS}
{Sun}, J., {Leroy}, A.~K., {Rosolowsky}, E., {et~al.} 2022, \aj, 164, 43, \dodoi{10.3847/1538-3881/ac74bd}

\bibitem[{{Sun} {et~al.}(2023){Sun}, {Leroy}, {Ostriker}, {Meidt}, {Rosolowsky}, {Schinnerer}, {Wilson}, {Utomo}, {Belfiore}, {Blanc}, {Emsellem}, {Faesi}, {Groves}, {Hughes}, {Koch}, {Kreckel}, {Liu}, {Pan}, {Pety}, {Querejeta}, {Razza}, {Saito}, {Sardone}, {Usero}, {Williams}, {Bigiel}, {Bolatto}, {Chevance}, {Dale}, {Gensior}, {Glover}, {Grasha}, {Henshaw}, {Jim{\'e}nez-Donaire}, {Klessen}, {Kruijssen}, {Murphy}, {Neumann}, {Teng}, \& {Thilker}}]{SUN23SFGAS}
{Sun}, J., {Leroy}, A.~K., {Ostriker}, E.~C., {et~al.} 2023, \apjl, 945, L19, \dodoi{10.3847/2041-8213/acbd9c}

\bibitem[{{Tacconi} {et~al.}(2020){Tacconi}, {Genzel}, \& {Sternberg}}]{TACCONI20REVIEW}
{Tacconi}, L.~J., {Genzel}, R., \& {Sternberg}, A. 2020, \araa, 58, 157, \dodoi{10.1146/annurev-astro-082812-141034}

\bibitem[{{Tafalla} {et~al.}(2023){Tafalla}, {Usero}, \& {Hacar}}]{TAFALLA23LINES}
{Tafalla}, M., {Usero}, A., \& {Hacar}, A. 2023, \aap, 679, A112, \dodoi{10.1051/0004-6361/202346136}

\bibitem[{{Teng} {et~al.}(2022){Teng}, {Sandstrom}, {Sun}, {Leroy}, {Johnson}, {Bolatto}, {Kruijssen}, {Schruba}, {Usero}, {Barnes}, {Bigiel}, {Blanc}, {Groves}, {Israel}, {Liu}, {Rosolowsky}, {Schinnerer}, {Smith}, \& {Walter}}]{TENG22XCO}
{Teng}, Y.-H., {Sandstrom}, K.~M., {Sun}, J., {et~al.} 2022, \apj, 925, 72, \dodoi{10.3847/1538-4357/ac382f}

\bibitem[{{Teng} {et~al.}(2023){Teng}, {Sandstrom}, {Sun}, {Gong}, {Bolatto}, {Chiang}, {Leroy}, {Usero}, {Glover}, {Klessen}, {Liu}, {Querejeta}, {Schinnerer}, {Bigiel}, {Cao}, {Chevance}, {Eibensteiner}, {Grasha}, {Israel}, {Murphy}, {Neumann}, {Pan}, {Pinna}, {Sormani}, {Smith}, {Walter}, \& {Williams}}]{TENG23XCO}
---. 2023, \apj, 950, 119, \dodoi{10.3847/1538-4357/accb86}

\bibitem[{{Teng} {et~al.}(2024){Teng}, {Chiang}, {Sandstrom}, {Sun}, {Leroy}, {Bolatto}, {Usero}, {Ostriker}, {Querejeta}, {Chastenet}, {Bigiel}, {Boquien}, {den Brok}, {Cao}, {Chevance}, {Chown}, {Colombo}, {Eibensteiner}, {Glover}, {Grasha}, {Henshaw}, {Jim{\'e}nez-Donaire}, {Liu}, {Murphy}, {Pan}, {Stuber}, \& {Williams}}]{TENG24XCO}
{Teng}, Y.-H., {Chiang}, I.-D., {Sandstrom}, K.~M., {et~al.} 2024, \apj, 961, 42, \dodoi{10.3847/1538-4357/ad10ae}

\bibitem[{{Utomo} {et~al.}(2018){Utomo}, {Sun}, {Leroy}, {Kruijssen}, {Schinnerer}, {Schruba}, {Bigiel}, {Blanc}, {Chevance}, {Emsellem}, {Herrera}, {Hygate}, {Kreckel}, {Ostriker}, {Pety}, {Querejeta}, {Rosolowsky}, {Sandstrom}, \& {Usero}}]{UTOMO18EFF}
{Utomo}, D., {Sun}, J., {Leroy}, A.~K., {et~al.} 2018, \apjl, 861, L18, \dodoi{10.3847/2041-8213/aacf8f}

\bibitem[{{Vazquez-Semadeni}(1994)}]{VAZQUEZ94TURB}
{Vazquez-Semadeni}, E. 1994, \apj, 423, 681, \dodoi{10.1086/173847}

\bibitem[{{V{\'a}zquez-Semadeni} {et~al.}(2024){V{\'a}zquez-Semadeni}, {Palau}, {G{\'o}mez}, {Arroyo-Ch{\'a}vez}, {Alig}, {Ballesteros-Paredes}, {Camacho}, {Gonz{\'a}lez-Samaniego}, \& {Burkert}}]{VAZQUEZ24EFF}
{V{\'a}zquez-Semadeni}, E., {Palau}, A., {G{\'o}mez}, G.~C., {et~al.} 2024, arXiv e-prints, arXiv:2408.10406, \dodoi{10.48550/arXiv.2408.10406}

\bibitem[{{Williams} {et~al.}(2023){Williams}, {Bureau}, {Davis}, {Cappellari}, {Choi}, {Elford}, {Iguchi}, {Gensior}, {Liang}, {Lu}, {Ruffa}, \& {Zhang}}]{WILLIAMS23GMCS}
{Williams}, T.~G., {Bureau}, M., {Davis}, T.~A., {et~al.} 2023, \mnras, \dodoi{10.1093/mnras/stad2455}

\bibitem[{{Wong} \& {Blitz}(2002)}]{WONG02SFGAS}
{Wong}, T., \& {Blitz}, L. 2002, \apj, 569, 157, \dodoi{10.1086/339287}

\bibitem[{{Wright} {et~al.}(2010){Wright}, {Eisenhardt}, {Mainzer}, {Ressler}, {Cutri}, {Jarrett}, {Kirkpatrick}, {Padgett}, {McMillan}, {Skrutskie}, {Stanford}, {Cohen}, {Walker}, {Mather}, {Leisawitz}, {Gautier}, {McLean}, {Benford}, {Lonsdale}, {Blain}, {Mendez}, {Irace}, {Duval}, {Liu}, {Royer}, {Heinrichsen}, {Howard}, {Shannon}, {Kendall}, {Walsh}, {Larsen}, {Cardon}, {Schick}, {Schwalm}, {Abid}, {Fabinsky}, {Naes}, \& {Tsai}}]{WISE10}
{Wright}, E.~L., {Eisenhardt}, P. R.~M., {Mainzer}, A.~K., {et~al.} 2010, \aj, 140, 1868, \dodoi{10.1088/0004-6256/140/6/1868}

\bibitem[{{Yajima} {et~al.}(2021){Yajima}, {Sorai}, {Miyamoto}, {Muraoka}, {Kuno}, {Kaneko}, {Takeuchi}, {Yasuda}, {Tanaka}, {Morokuma-Matsui}, \& {Kobayashi}}]{YAJIMA21LINES}
{Yajima}, Y., {Sorai}, K., {Miyamoto}, Y., {et~al.} 2021, \pasj, 73, 257, \dodoi{10.1093/pasj/psaa119}

\bibitem[{{Young} \& {Scoville}(1991)}]{YOUNG91SFGAS}
{Young}, J.~S., \& {Scoville}, N.~Z. 1991, \araa, 29, 581, \dodoi{10.1146/annurev.aa.29.090191.003053}

\bibitem[{{Young} {et~al.}(1995){Young}, {Xie}, {Tacconi}, {Knezek}, {Viscuso}, {Tacconi-Garman}, {Scoville}, {Schneider}, {Schloerb}, {Lord}, {Lesser}, {Kenney}, {Huang}, {Devereux}, {Claussen}, {Case}, {Carpenter}, {Berry}, \& {Allen}}]{YOUNG95SFGAS}
{Young}, J.~S., {Xie}, S., {Tacconi}, L., {et~al.} 1995, \apjs, 98, 219, \dodoi{10.1086/192159}

\end{thebibliography}

\bibliographystyle{aasjournal}

\appendix
    
\section{Procedure to replicate these measurements for comparison to numerical simulations or other observations}
\label{sec:stepbystep}

We adopt a measurement scheme that can be easily implemented to allow rigorous, ``apples to apples'' comparison between PHANGS--ALMA and either numerical simulations or observations of other samples of galaxies. 

\begin{enumerate}
\item When working with CO data, first apply a local best-estimate $\alpha_{\rm CO}$ to convert the data into units of molecular gas mass. In a simulation estimate the mass in the CO-bright molecular gas phase likely to enter our analysis. CO bright emission visible to PHANGS--ALMA at 150~pc would be defined as $> 0.6$~K~km~s$^{-1}$, or about $3\times$ the typical rms integrated intensity noise.

\item Convolve the CO or simulated molecular gas data to have a resolution of $150$~pc (FWHM). Estimate the surface density, $\Sigma_{\rm mol}$ and line width, $\sigma_{\rm mol}$ along each line of sight.  Apply inclination corrections as in \citet{SUN22CLOUDS} to calculate face-on values of each quantity. In our framework the line width is the ``effective width'' line width as defined in \citet{SUN22CLOUDS} and references therein. From $\Sigma_{\rm mol}^{\rm cloud}$, $\sigma_{\rm mol}^{\rm cloud}$, and $R_{\rm pix}$ ($\equiv (\ell^2 H_{\rm mol} / (8 \cos i))^{1/3}$ with $H_{\rm mol} = 100$~pc and $\ell$ the resolution, here $150$~pc) calculate $\alpha_{\rm vir}$ and $\tau_{\rm ff}^{\rm mol}$.

\item Divide the target region or galaxy into hexagonal apertures with diameter $1.5$~kpc. Within each aperture, calculate the mass-weighted expectation value of $\Sigma_{\rm mol}^{\rm cloud}$, $\sigma_{\rm mol}^{\rm cloud}$, and $\alpha_{\rm vir}^{\rm cloud}$. To do this, select all sightlines in the region and then for quantity $X$ calculate $\left< X \right> = \sum \Sigma_{\rm mol}^{\rm cloud} X / \sum \Sigma_{\rm mol}^{\rm cloud}$. In the case of $\tau_{\rm ff}^{\rm mol}$, calculate the average via $\left< \tau_{\rm ff}^{\rm cloud} \right> = \left( \sum \Sigma_{\rm mol} \left( \tau_{\rm ff}^{\rm cloud} \right)^{-1} / \sum \Sigma_{\rm mol} \right)^{-1}$ so that for a fixed efficiency per free fall time $\Sigma_{\rm SFR} = \epsilon_{\rm ff}^{\rm mol} \left< \tau_{\rm ff}^{\rm cloud} \right> \Sigma_{\rm mol}$.

\item Estimate the molecular gas depletion time by convolving the $\Sigma_{\rm SFR}$ and $\Sigma_{\rm mol}$ maps to 1.5~kpc (FWHM) resolution. Divide them to estimate $\tau_{\rm dep}^{\rm mol}$ and sample the map at the center of each region to obtain the local $\tau_{\rm dep}^{\rm mol}$.

\item Divide $\tauff / \taudep$ to estimate the mean star formation efficiency per free fall time, \eff , in the region.

\item Select regions with mass-weighted average molecular gas surface density $\sdmol > 20$~M$_\odot$~pc$^{-2}$ to match our completeness cut.
\end{enumerate}

\section{PHANGS--ALMA at higher resolution}
\label{sec:resn}

\begin{figure*}
\centering
\includegraphics{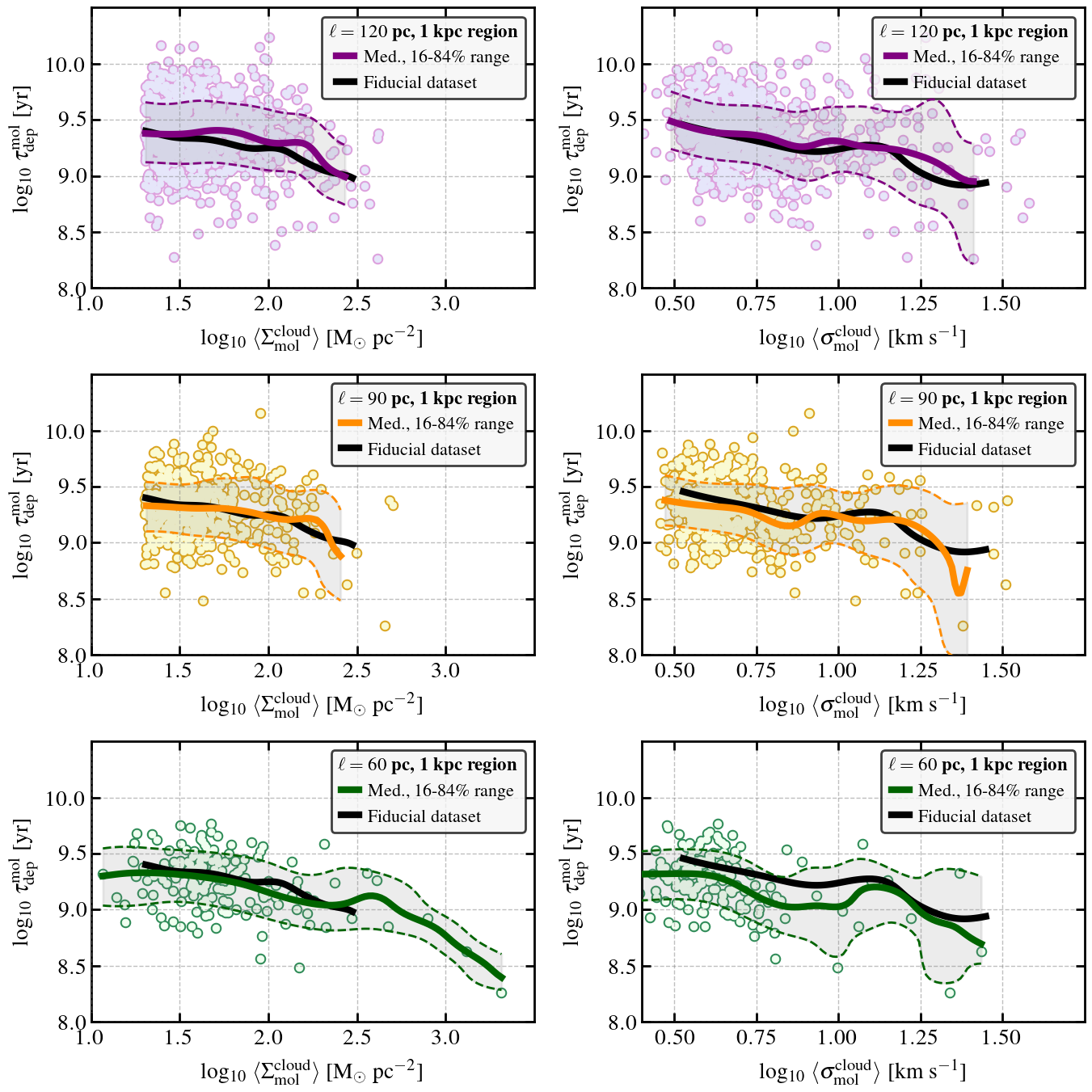}
\caption{Molecular gas depletion time as a function of cloud scale mean (\textit{left}) surface density and (\textit{right}) line width measured for sharper cloud scale resolutions than the $\ell = 150$~pc used in the main text, \textit{top} to \textit{bottom} $\ell = 120$~pc, $\ell = 90$, $\ell = 60$~pc. We show the running median and 16{-}84\% range for each quantity in each panel and plot the median trend for the data set used in the main paper  ($\ell = 150$~pc over 1.5 kpc regions) in black. The sample of galaxies that reach a given resolution varies, but overall the same trends are evident at all four resolutions. We provide these higher resolution, lower sample-size data sets as additional data products.
\label{fig:props_vs_res}
}
\end{figure*}

\begin{deluxetable*}{lccccccccccccc}[t!]
\tabletypesize{\footnotesize}
\tablecaption{Measurements for individual regions at $\ell = 120$~pc in 1.0~kpc diameter regions \label{tab:data120}}
\tablewidth{0pt}
\tablehead{
\colhead{Galaxy} & 
\colhead{Radius} & 
\colhead{$\tau_{\rm dep}^{\rm mol}$} & 
\colhead{$\frac{\tau_{\rm dep}^{\rm mol}}{\left< \tau_{\rm dep}^{\rm mol} \right>_{\rm gal}}$} & 
\colhead{$\alpha_{\rm CO}^{2-1}$} & 
\colhead{$i$} & 
\colhead{$\log_{10} M_\star$} & 
\colhead{$\log_{10} {\rm SFR}$} & 
\colhead{$\left< \Sigma_{mol}^{\rm cloud} \right>$} & 
\colhead{$\left< \tau_{ff}^{\rm cloud} \right>$} & 
\colhead{$\left< \sigma_{mol}^{\rm cloud} \right>$} & 
\colhead{$\left< \alpha_{vir}^{\rm cloud} \right>$} & 
\colhead{$\epsilon_{\rm ff}^{\rm mol}$} & 
\colhead{$\frac{\epsilon_{\rm ff}^{\rm mol}}{\left< \epsilon_{\rm ff}^{\rm mol} \right>_{\rm gal}}$}
\\
\colhead{} & 
\colhead{(kpc)} & 
\colhead{(Gyr)} & 
\colhead{(norm.)} & 
\colhead{($\frac{{\rm M}_\odot~{\rm pc}^{-2}}{{\rm K~km~s}^{-1}}$)} & 
\colhead{($^\circ$)} & 
\colhead{(M$_\odot$)} & 
\colhead{(M$_\odot$~yr$^{-1}$)} & 
\colhead{(M$_\odot$~pc$^{-2}$)} & 
\colhead{(Myr)} & 
\colhead{(km s$^{-1}$)} & 
\colhead{} & 
\colhead{} & 
\colhead{(norm.)}
}
\startdata
IC1954 & 0.0 & 0.7 & 0.51 & 4.36 & 57.1 & 9.7 & -0.4 & 33.0 & 8.03 & 6.9 & 4.11 & 0.0115 & 1.66 \\
IC1954 & 1.22 & 1.44 & 1.06 & 6.57 & 57.1 & 9.7 & -0.4 & 26.0 & 9.08 & 4.9 & 2.37 & 0.0063 & 0.91 \\
IC1954 & 1.22 & 1.27 & 0.94 & 6.8 & 57.1 & 9.7 & -0.4 & 37.0 & 7.82 & 5.3 & 2.35 & 0.0062 & 0.89 \\
IC1954 & 1.31 & 1.61 & 1.18 & 7.05 & 57.1 & 9.7 & -0.4 & 35.0 & 7.95 & 5.2 & 2.41 & 0.0049 & 0.71 \\
IC1954 & 1.31 & 1.55 & 1.14 & 7.1 & 57.1 & 9.7 & -0.4 & 30.0 & 8.69 & 4.9 & 2.54 & 0.0056 & 0.81 \\
IC1954 & 1.74 & 1.97 & 1.45 & 8.23 & 57.1 & 9.7 & -0.4 & 21.0 & 10.13 & 4.1 & 2.07 & 0.0051 & 0.74 \\
IC1954 & 1.84 & 1.33 & 0.98 & 7.63 & 57.1 & 9.7 & -0.4 & 26.0 & 9.35 & 4.6 & 2.63 & 0.0071 & 1.02 \\
IC1954 & 2.95 & 1.23 & 0.91 & 9.53 & 57.1 & 9.7 & -0.4 & 23.0 & 9.61 & 4.4 & 2.08 & 0.0078 & 1.13 \\
IC5273 & 0.0 & 0.38 & 0.51 & 3.87 & 52.0 & 9.7 & -0.3 & 62.0 & 6.66 & 7.6 & 4.0 & 0.0177 & 1.1 \\
\enddata
\tablecomments{This is a stub. The full table is available online as part of a machine readable table. Columns: (1) galaxy name, (2) galactocentric radius, (3) molecular gas depletion time, (4) molecular gas depeletion time normalized to galaxy average, (5) CO (2-1) to H$_2$ conversion factor, (6) galaxy inclination, (7) galaxy integrated stellar mass, (8) galaxy integrated star formation rate, region-averaged mass weighted molecular gas: (9) surface density, (10) gravitational free-fall time, (11) line width, and (12) virial parameter, (13) star formation efficiency per free fall time, (14) star formation efficiency per free fall time normalized to galaxy average.}
\end{deluxetable*}

\begin{deluxetable*}{lccccccccccccc}[t!]
\tabletypesize{\footnotesize}
\tablecaption{Measurements for individual regions at $\ell = 90$~pc in 1.0~kpc diameter regions \label{tab:data90}}
\tablewidth{0pt}
\tablehead{
\colhead{Galaxy} & 
\colhead{Radius} & 
\colhead{$\tau_{\rm dep}^{\rm mol}$} & 
\colhead{$\frac{\tau_{\rm dep}^{\rm mol}}{\left< \tau_{\rm dep}^{\rm mol} \right>_{\rm gal}}$} & 
\colhead{$\alpha_{\rm CO}^{2-1}$} & 
\colhead{$i$} & 
\colhead{$\log_{10} M_\star$} & 
\colhead{$\log_{10} {\rm SFR}$} & 
\colhead{$\left< \Sigma_{mol}^{\rm cloud} \right>$} & 
\colhead{$\left< \tau_{ff}^{\rm cloud} \right>$} & 
\colhead{$\left< \sigma_{mol}^{\rm cloud} \right>$} & 
\colhead{$\left< \alpha_{vir}^{\rm cloud} \right>$} & 
\colhead{$\epsilon_{\rm ff}^{\rm mol}$} & 
\colhead{$\frac{\epsilon_{\rm ff}^{\rm mol}}{\left< \epsilon_{\rm ff}^{\rm mol} \right>_{\rm gal}}$}
\\
\colhead{} & 
\colhead{(kpc)} & 
\colhead{(Gyr)} & 
\colhead{(norm.)} & 
\colhead{($\frac{{\rm M}_\odot~{\rm pc}^{-2}}{{\rm K~km~s}^{-1}}$)} & 
\colhead{($^\circ$)} & 
\colhead{(M$_\odot$)} & 
\colhead{(M$_\odot$~yr$^{-1}$)} & 
\colhead{(M$_\odot$~pc$^{-2}$)} & 
\colhead{(Myr)} & 
\colhead{(km s$^{-1}$)} & 
\colhead{} & 
\colhead{} & 
\colhead{(norm.)}
}
\startdata
IC1954 & 0.0 & 0.7 & 0.51 & 4.36 & 57.1 & 9.7 & -0.4 & 35.0 & 7.04 & 6.2 & 3.82 & 0.0101 & 1.66 \\
IC1954 & 1.22 & 1.44 & 1.06 & 6.57 & 57.1 & 9.7 & -0.4 & 28.0 & 7.94 & 4.5 & 2.3 & 0.0055 & 0.91 \\
IC1954 & 1.22 & 1.27 & 0.94 & 6.8 & 57.1 & 9.7 & -0.4 & 38.0 & 6.92 & 5.0 & 2.31 & 0.0054 & 0.9 \\
IC1954 & 1.31 & 1.61 & 1.18 & 7.05 & 57.1 & 9.7 & -0.4 & 37.0 & 7.02 & 4.9 & 2.37 & 0.0044 & 0.72 \\
IC1954 & 1.31 & 1.55 & 1.14 & 7.1 & 57.1 & 9.7 & -0.4 & 31.0 & 7.66 & 4.6 & 2.39 & 0.005 & 0.82 \\
IC1954 & 1.74 & 1.97 & 1.45 & 8.23 & 57.1 & 9.7 & -0.4 & 22.0 & 8.97 & 3.7 & 1.99 & 0.0046 & 0.75 \\
IC1954 & 1.84 & 1.33 & 0.98 & 7.63 & 57.1 & 9.7 & -0.4 & 30.0 & 7.94 & 4.2 & 2.4 & 0.006 & 0.99 \\
IC1954 & 2.95 & 1.23 & 0.91 & 9.53 & 57.1 & 9.7 & -0.4 & 25.0 & 8.41 & 4.0 & 1.94 & 0.0068 & 1.13 \\
NGC0628 & 0.0 & 1.53 & 0.56 & 2.63 & 8.9 & 10.3 & 0.2 & 22.0 & 11.62 & 4.6 & 6.65 & 0.0076 & 1.82 \\
\enddata
\tablecomments{This is a stub. The full table is available online as part of a machine readable table. Columns: (1) galaxy name, (2) galactocentric radius, (3) molecular gas depletion time, (4) molecular gas depeletion time normalized to galaxy average, (5) CO (2-1) to H$_2$ conversion factor, (6) galaxy inclination, (7) galaxy integrated stellar mass, (8) galaxy integrated star formation rate, region-averaged mass weighted molecular gas: (9) surface density, (10) gravitational free-fall time, (11) line width, and (12) virial parameter, (13) star formation efficiency per free fall time, (14) star formation efficiency per free fall time normalized to galaxy average.}
\end{deluxetable*}

\begin{deluxetable*}{lccccccccccccc}[t!]
\tabletypesize{\footnotesize}
\tablecaption{Measurements for individual regions at $\ell = 60$~pc in 1.0~kpc diameter regions \label{tab:data60}}
\tablewidth{0pt}
\tablehead{
\colhead{Galaxy} & 
\colhead{Radius} & 
\colhead{$\tau_{\rm dep}^{\rm mol}$} & 
\colhead{$\frac{\tau_{\rm dep}^{\rm mol}}{\left< \tau_{\rm dep}^{\rm mol} \right>_{\rm gal}}$} & 
\colhead{$\alpha_{\rm CO}^{2-1}$} & 
\colhead{$i$} & 
\colhead{$\log_{10} M_\star$} & 
\colhead{$\log_{10} {\rm SFR}$} & 
\colhead{$\left< \Sigma_{mol}^{\rm cloud} \right>$} & 
\colhead{$\left< \tau_{ff}^{\rm cloud} \right>$} & 
\colhead{$\left< \sigma_{mol}^{\rm cloud} \right>$} & 
\colhead{$\left< \alpha_{vir}^{\rm cloud} \right>$} & 
\colhead{$\epsilon_{\rm ff}^{\rm mol}$} & 
\colhead{$\frac{\epsilon_{\rm ff}^{\rm mol}}{\left< \epsilon_{\rm ff}^{\rm mol} \right>_{\rm gal}}$}
\\
\colhead{} & 
\colhead{(kpc)} & 
\colhead{(Gyr)} & 
\colhead{(norm.)} & 
\colhead{($\frac{{\rm M}_\odot~{\rm pc}^{-2}}{{\rm K~km~s}^{-1}}$)} & 
\colhead{($^\circ$)} & 
\colhead{(M$_\odot$)} & 
\colhead{(M$_\odot$~yr$^{-1}$)} & 
\colhead{(M$_\odot$~pc$^{-2}$)} & 
\colhead{(Myr)} & 
\colhead{(km s$^{-1}$)} & 
\colhead{} & 
\colhead{} & 
\colhead{(norm.)}
}
\startdata
NGC0628 & 0.0 & 1.53 & 0.55 & 2.63 & 8.9 & 10.3 & 0.2 & 24.0 & 9.07 & 4.2 & 7.21 & 0.0059 & 1.95 \\
NGC0628 & 1.0 & 2.79 & 1.0 & 3.61 & 8.9 & 10.3 & 0.2 & 28.0 & 8.5 & 3.8 & 4.68 & 0.0031 & 1.0 \\
NGC0628 & 1.0 & 2.09 & 0.75 & 3.51 & 8.9 & 10.3 & 0.2 & 24.0 & 8.98 & 4.0 & 5.7 & 0.0043 & 1.41 \\
NGC0628 & 1.01 & 2.18 & 0.78 & 3.51 & 8.9 & 10.3 & 0.2 & 35.0 & 7.75 & 4.6 & 5.55 & 0.0036 & 1.17 \\
NGC0628 & 1.01 & 2.31 & 0.83 & 3.61 & 8.9 & 10.3 & 0.2 & 27.0 & 8.66 & 4.3 & 5.26 & 0.0038 & 1.23 \\
NGC0628 & 1.01 & 2.95 & 1.06 & 3.71 & 8.9 & 10.3 & 0.2 & 21.0 & 9.94 & 3.5 & 5.02 & 0.0034 & 1.11 \\
NGC0628 & 1.01 & 2.24 & 0.8 & 3.72 & 8.9 & 10.3 & 0.2 & 30.0 & 8.48 & 4.4 & 5.19 & 0.0038 & 1.24 \\
NGC0628 & 1.74 & 2.67 & 0.96 & 4.64 & 8.9 & 10.3 & 0.2 & 20.0 & 10.02 & 3.3 & 3.77 & 0.0038 & 1.23 \\
NGC0628 & 1.74 & 2.47 & 0.88 & 4.67 & 8.9 & 10.3 & 0.2 & 22.0 & 9.47 & 3.2 & 4.28 & 0.0038 & 1.26 \\
\enddata
\tablecomments{This is a stub. The full table is available online as part of a machine readable table. Columns: (1) galaxy name, (2) galactocentric radius, (3) molecular gas depletion time, (4) molecular gas depeletion time normalized to galaxy average, (5) CO (2-1) to H$_2$ conversion factor, (6) galaxy inclination, (7) galaxy integrated stellar mass, (8) galaxy integrated star formation rate, region-averaged mass weighted molecular gas: (9) surface density, (10) gravitational free-fall time, (11) line width, and (12) virial parameter, (13) star formation efficiency per free fall time, (14) star formation efficiency per free fall time normalized to galaxy average.}
\end{deluxetable*}

In the main text, we present results at $\ell = 150$~pc resolution. PHANGS--ALMA also represents the largest survey of CO from galaxies at $120$~pc, $90$~pc, and $60$~pc resolution, though we have fewer galaxies at these higher resolutions. Tables \ref{tab:data120}, \ref{tab:data90}, and \ref{tab:data60} report results that measure cloud-scale properties at these higher resolutions. We apply the same completeness and surface density cuts, but work with $1$~kpc diameter hexagonal regions rather than the $1.5$~kpc diameter used for the main results. This is possible because the galaxies with higher physical resolution are also closer, meaning that our $\Sigma_{\rm SFR}$ estimates (which use lower resolution WISE data) can reach $1$~kpc resolution. Given this, our sample at $\ell = 120$~pc consists of $966$ 1~kpc regions in $41$ galaxies. At $\ell = 90$~pc we have 552 regions in $28$ galaxies, and at $\ell = 60$~pc we have 170 regions in $9$ galaxies. Figure \ref{fig:props_vs_res} show \taudep\ as a function of \sdmol\ and \sigmol\ for each of these resolutions, with our fiducial results overplotted as a black line. Overall all three resolutions show consistent results, though the number of galaxies centers with extreme conditions is lower for the smaller samples at higher linear resolution. Because radial variations in gas properties drive our dynamic range (\S \ref{sec:byrad}), especially extreme conditions at galaxy centers, the trends are noisier for these high resolution data sets.

\suppressAffiliationsfalse
\allauthors


\end{document}